\documentclass[aps,pra,onecolumn,superscriptaddress,10pt]{revtex4-2}

\usepackage[english]{babel}
\usepackage{graphicx}  
\usepackage{dcolumn}   
\usepackage{bm}        
\usepackage{bbm}        
\usepackage{amssymb}   
\expandafter\let\csname equation*\endcsname\relax
\expandafter\let\csname endequation*\endcsname\relax
\usepackage{amsmath}
\usepackage{mathrsfs}
\usepackage{mathtools}
\usepackage[utf8]{inputenc}
\usepackage{makecell}

\usepackage{wrapfig}

\usepackage{changes}

\definecolor{myurlcolor}{rgb}{0,0,0.7}
\definecolor{myrefcolor}{rgb}{0.8,0,0}
\usepackage{hyperref}
\hypersetup{colorlinks, linkcolor=myrefcolor,
citecolor=myurlcolor, urlcolor=myurlcolor}

\usepackage{cleveref}
\crefrangeformat{equation}{#3(#1 #4--#5 #2)#6}

\usepackage{comment}
\usepackage{soul} 
\newcommand{\ignore}[1]{}
\usepackage{enumitem}

\newcommand{\eg}{\emph{e.g. }}				
\newcommand{\ie}{\emph{i.e. }}				
\newcommand{\etal}{\emph{et al. }}			
\newcommand{\var}[2]{\ensuremath{(\Delta #1)^2_{#2}}}

\renewcommand{\Re}{\text{Re}}
\renewcommand{\Im}{\text{Im}}

\newcommand{\cov}{\ensuremath{\, \text{Cov}}}

\newcommand{\op}[1]{\hat{#1}}				
\newcommand{\vect}[1]{{\bm{#1}}}			
\newcommand{\avg}[1]{\left\langle #1 \right\rangle}		
\newcommand{\dd}{\text{d}}				

\newcommand{\scN}{\mathcal{N}}

\newcommand{\bra}[1]{\ensuremath{\langle#1|}}
\newcommand{\ket}[1]{\ensuremath{|#1\rangle}}

\newcommand{\Tr}{\ensuremath{\, \mathrm{Tr}}}

\newcommand{\abs}[1]{\ensuremath{\left\vert #1\right\vert}}

\newcommand{\RHO}{\ensuremath{\hat{\rho}}}
\newcommand{\PSI}{\ensuremath{\hat{\psi}}}
\newcommand{\GEN}{\ensuremath{\hat{G}}}
\newcommand{\MM}{\ensuremath{\hat{M}}}
\newcommand{\qq}{\ensuremath{\hat{r}}}
\newcommand{\bMM}{\ensuremath{\mathbf{\hat{M}}}}
\newcommand{\bqq}{\ensuremath{\mathbf{\hat{r}}}}
\newcommand{\ID}{\ensuremath{\hat{\mathbb{I}}}}
\renewcommand{\aa}{\ensuremath{\hat{a}}}
\newcommand{\aad}{\ensuremath{\hat{a}^\dagger}}
\newcommand{\xx}{\ensuremath{\hat{x}}}
\newcommand{\pp}{\ensuremath{\hat{p}}}
\newcommand{\nn}{\ensuremath{\hat{n}}}

\newcommand{\Gam}{\ensuremath{\Gamma}}

\newcommand{\nT}{\ensuremath{n_T}}

\newcommand{\NN}{\ensuremath{\mathcal{N}}}
\newcommand{\DD}{\ensuremath{\hat{D}}}

\newcommand{\PP}{\ensuremath{\hat{\Pi}}}

\usepackage[squaren]{SIunits}

\begin{document}

\title{Quantum metrology with a continuous-variable system}

\author{Matteo Fadel}
\email[]{fadelm@phys.ethz.ch} 
\affiliation{Department of Physics ETH Z\"urich - 8093 Z\"urich - Switzerland} 
\author{Noah Roux}
\affiliation{Department of Physics ETH Z\"urich - 8093 Z\"urich - Switzerland} 
\author{Manuel Gessner}
\email[]{manuel.gessner@uv.es} 
\affiliation{Instituto de Física Corpuscular (IFIC), CSIC‐Universitat de València and Departament de Física Teòrica, UV, C/Dr Moliner 50, E-46100 Burjassot (Valencia), Spain}

\date{\today}

\begin{abstract}
As one of the main pillars of quantum technologies, quantum metrology aims to improve measurement precision using techniques from quantum information. The two main strategies to achieve this are the preparation of nonclassical states and the design of optimized measurement observables. We discuss precision limits and optimal strategies in quantum metrology and sensing with a single mode of quantum continuous variables. We focus on the practically most relevant cases of estimating displacements and rotations and provide the sensitivities of the most important classes of states that includes Gaussian states and superpositions of Fock states or coherent states. Fundamental precision limits that are obtained from the quantum Fisher information are compared to the precision of a simple moment-based estimation strategy based on the data obtained from possibly sub-optimal measurement observables, including homodyne, photon number, parity and higher moments. Finally, we summarize some of the main experimental achievements and present emerging platforms for continuous-variable sensing. These results are of particular interest for experiments with quantum light, trapped ions, mechanical oscillators, and microwave resonators. 
\end{abstract}

\maketitle

\clearpage
\newpage

\tableofcontents

\clearpage
\newpage

\section{Introduction}
The common goal of all quantum technologies is to perform tasks that are inaccessible classically or to improve their performance beyond the limits of classical strategies through the dedicated use of quantum resources. The central figure of merit in quantum metrology is the precision in estimating an unknown parameter~\cite{BraunsteinPRL1994, Paris2009, PhysRevLett.96.010401, Giovannetti2011}. Strategies inspired by quantum information to improve this precision include the preparation of nonclassical states, such as squeezed~\cite{Caves1981,PhysRevA.50.67, PhysRevA.46.R6797} or entangled states~\cite{PhysRevLett.102.100401,PezzeBook,Toth2014}, and the optimization of measurement observables~\cite{Giovannetti2011}. Quantum metrology offers a wide range of potential applications, including in fundamental physics~\cite{schnabel_squeezed_2017,TsePRL2019, PhysRevLett.123.231108, chou2023quantum}, biology~\cite{DegenRMP}, quantum information~\cite{PezzeBook,Toth2014}, optics~\cite{Rafal_quantumlimits_2015}, astronomy and microscopy~\cite{PhysRevX.6.031033,TsangReview}.

Quantum systems that are used for quantum information applications can generally be classified into two main categories based on the degree of freedom being controlled: discrete variables (DV) and continuous variables (CV). DV systems are characterized by finite-dimensional Hilbert spaces, such as qubits, \ie two-level or equivalently spin-1/2 systems, $d$-level systems or qudits, and multipartite systems composed of these elemental constituents. The fundamental building block of CV systems is the quantum harmonic oscillator, whose Hilbert space is unbounded and allows for the construction of phase-space observables with a continuous spectrum.

Metrology with DV has been largely investigated in the context of sensing with spin states of trapped atoms and atomic ensembles, such as vapour cells, cold atoms, Bose-Einstein condensates, and trapped ions~\cite{pezze_quantum_2018}, but also with superconducting qubits~\cite{PhysRevLett.128.150501}. Sensitivity beyond classical limits has been demonstrated through the preparation of spin-squeezed and other families of entangled states, including Schrödinger cat states. For an extensive review of the DV case, see Ref.~\cite{pezze_quantum_2018}.

Quantum metrology has an even longer tradition in CV systems since quantum optical modes, which are described by continuous variables, were the first to be prepared in squeezed states~\cite{slusher_observation_1985, PhysRevLett.57.691, Wu_generation_86}. Today, the well established field of optical interferometry is mature enough to lead to applications beyond proof-of-principle experiments, such as gravitational wave detectors~\cite{schnabel_squeezed_2017,TsePRL2019, PhysRevLett.123.231108}. The recent experimental development in the control of massive systems in the quantum regime (as opposed to photons that are massless), raised interest in considering quantum sensing with motional states of harmonic-oscillator-like systems. First examples of high-fidelity quantum control over such systems were the external degrees of freedom of trapped ions~\cite{RevModPhys.75.281}, but recent advances have made a similar degree of control possible also for solid-state mechanical oscillators \cite{Satzinger18,mason_continuous_2019,wollack_quantum_2022,vonLupke22,catSCI23,marti2023,youssefi_squeezed_2023}.

The purpose of this review is to provide the theoretical description of the metrological sensitivity of CV systems, and to discuss the metrological properties of widely-used classes of single-mode CV states for parameter estimation tasks of particular interest. In particular, we focus on the metrology of phase-space displacements and rotations, \ie phase shifts. For each task and state, we discuss
\begin{enumerate}
    \item[(i)] the fundamental metrological sensitivity in terms of the quantum Fisher information, corresponding to an optimal estimation strategy based on the data obtained from an optimally designed observable;
    \item[(ii)] the achievable precision from a simple estimation strategy known as the method of moments, based on the measurement of the expectation value of a single, possibly non-optimal observable.
\end{enumerate}
Analytical expressions for these figures of merit allow us to identify optimal strategies under various constraints, such as with or without \textit{a priori} knowledge of the displacement direction. By providing a comprehensive overview of known results in single-mode CV sensing and filling in open gaps in the existing literature with original contributions, this work aims to serve as a valuable resource for both theoretical and experimental research in the field.

Other evolutions for the generation of the parameter dependence, including nonlinear Hamiltonians with the capability of generating squeezing, or non-unitary evolutions in the presence of noise, will not be discussed in this manuscript. Nonetheless, these situations are of high interest and, in particular, for the case of Gaussian states under Gaussian evolutions, analytical results for the sensitivity of arbitrary parameters and their optimization can be found in the literature~\cite{Monras2007,PinelPRA2013,safranek_optimal_2016,RevModPhys.90.035006,oh_optimal_2019}.

Furthermore, multimode approaches, despite their evident interest and relevance, are beyond the scope of the present work. This means that we will not discuss the CV metrology of ancilla-assisted setups~\cite{PhysRevLett.98.090401,PhysRevLett.113.250801,PhysRevA.101.012124,matsubara_optimal_2019,RevModPhys.90.035006}, superresolution imaging~\cite{Albarelli2020,PhysRevResearch.4.L032022}, multiparameter estimation~\cite{Safranek2015,Nichols2018,Gessner2020,PhysRevResearch.2.023030,Albarelli2020}, metrological CV entanglement detection~\cite{Adesso_2007,BosonicSqueezing,Qin2019}, or multiple frequency modes that are needed to discuss the noise budget of gravitational wave detectors~\cite{schnabel_squeezed_2017,TsePRL2019, PhysRevLett.123.231108}. The single-mode description employed here can be understood as the limit of a two-mode interferometer with a strongly populated coherent state in one of the modes acting as a phase reference. Our description focuses on the quantum aspects of the other mode, which implies that the macroscopic number of photons that populate the reference laser are not considered as part of the resource. For descriptions of two-mode interferometers that explicitly consider the beam-splitter mixing with a coherent state, see Refs.~\cite{PezzePRL2008,PinelPRA2012,CavesLangPRL2013}.

\subsection*{Outline}
We begin by introducing the formalism of (single-mode) CV systems in Section~\ref{sec:cv}, along with a definition of the states that are particularly relevant both theoretically and experimentally. In Section~\ref{sec:qm}, we provide an overview of quantum metrology, with a focus on the key concepts that will be essential for our analysis, such as the classical and quantum Fisher information, the Cramér-Rao bound, and the method of moments. 
In Section~\ref{sec:disp}, we analyze the QFI for displacements of the states considered, along with the sensitivity achievable using experimentally practical measurement strategies. Specifically, we examine scenarios where the phase of the displacement is unknown and discuss both best- and worst-case scenarios. In Section~\ref{sec:rot}, we turn to the case of phase estimation, which corresponds to a rotation in phase space.

\section{CV systems}\label{sec:cv}
In this Section, we briefly summarize the formalism of CV systems, focusing on the tools that will be used in this paper. For more detailed introductions and accounts of quantum CV systems, we refer to extensive review articles, tutorials and lecture notes on the topic~\cite{Paris2005,Adesso_2007,Wang2007,RevModPhys.77.513,Weedbrook,Adesso2014,Serafini2017,brask2022gaussian,PRXQuantum.2.030204}.

\subsection{Quadrature operators and Wigner function}
The quantum description of a single-mode continuous variable system is equivalent to that of a quantum harmonic oscillator. The position and momentum operators $\xx$ and $\pp$ satisfy the canonical commutation relations $[\xx,\pp]=i$, where henceforth we will set $\hbar=1$. We introduce bosonic creation and annihilation operators $\aa$ and $\aad$, satisfying $[\aa,\aad]=1$, such that
\begin{equation}
    \xx =\dfrac{1}{\sqrt{2}}(\aa+\aad) \;,\qquad  \pp =\dfrac{1}{i\sqrt{2}}(\aa-\aad) \;.
    \label{eq:def_x_p}
\end{equation}

An important tool for the representation of quantum states in phase space is the Wigner function. For an arbitrary state $\hat{\rho}$ it can be defined as
\begin{equation}
    W(x,p) = \frac{1}{2\pi}\int_{-\infty}^{\infty}\mathrm{d}q\bra{x-q/2}\hat{\rho}\ket{x+q/2}e^{ipq},
    \label{WignerDef}
\end{equation}
where $\ket{x}$ are the eigenstates of the position operator $\xx$. Normalization implies $\int\dd x \, \dd p \, W(x,p) =1$. The information contained in the Wigner function is equivalent to that of the full quantum state $\hat{\rho}$ and therefore provides a complete description of the CV system's state. We will often switch between the phase space coordinates $(x,p)\in\mathbb{R}^2$ and $\alpha\in \mathbb{C}$ using the mapping $\alpha=\frac{x+ip}{\sqrt{2}}$. The Wigner function can be written as the expectation value of a displaced parity measurement~\cite{Cahill1969b,gerry_introductory_2004}, namely 
\begin{equation}
    W(x,p) = \frac{1}{\pi}\Tr\left[\hat{\rho}\hat{D}(x,p)\hat{\Pi}\hat{D}^{\dagger}(x,p)\right],
    \label{eq:wigner_parity_measurement}
\end{equation}
where $\hat{D}(x,p)=\hat{D}\left(\frac{x+ip}{\sqrt{2}}\right)$ with
\begin{equation}\label{eq:dispOp}
\hat{D}(\alpha) \equiv e^{\alpha \aad - \alpha^\ast \aa} 
\end{equation}
is the displacement operator, and 
\begin{equation}
    \hat{\Pi} = (-1)^{\hat{n}} = \sum_n (-1)^{n} \ket{n}\bra{n}
\end{equation}
is the parity operator, which measures whether the number of excitations is even or odd. In complex coordinates, the Wigner function reads
\begin{equation}\label{eq:defWignerAlpha}
    W(\alpha) = \frac{2}{\pi}\Tr\left[\hat{\rho}\hat{D}(\alpha)\hat{\Pi}\hat{D}^{\dagger}(\alpha)\right] \;,
\end{equation}
where the factor 2 comes from the Jacobian of the coordinate transformation, such that $\int \dd^2\alpha\, W(\alpha)=1$.
Equation~\eqref{WignerDef} provides a prescription to experimentally measure the Wigner function when parity measurements are available, such as in cavity QED and trapped ions \cite{LutterbachPRL97}. In other cases, however, parity measurements can be challenging and one needs to rely on other observables. For example, typical measurement performed on propagating optical fields are homodyne measurements, from which the Wigner function can be reconstructed via tomographic approaches \cite{Lvovsky2009}.

\subsection{Quantum states of interest}

\begin{figure}[h!]
    \centering
    \includegraphics[width=\textwidth]{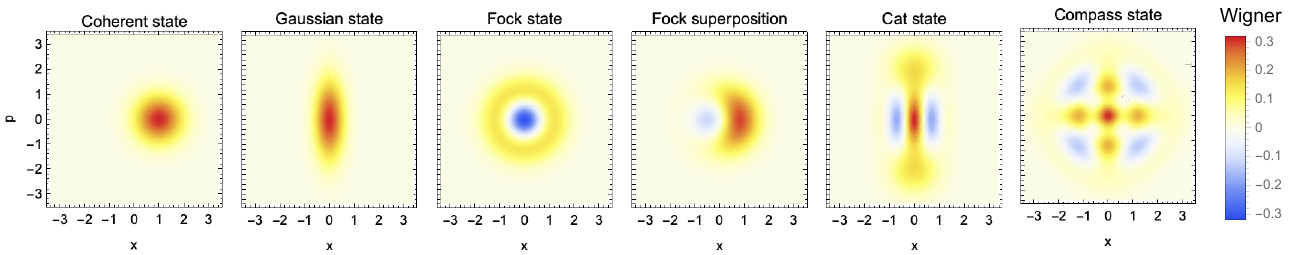}
    \caption{\textbf{Wigner functions of a set of states investigated in this work.} From left to right: coherent state $\ket{\alpha=1/\sqrt{2}}$, pure Gaussian state $\ket{\alpha=0,\xi=0.5}$, Fock state $\ket{n=1}$, superposition of Fock states $(\ket{0}+\ket{1})/\sqrt{2}$, cat state $\NN_1(\ket{\alpha}+\ket{-\alpha})$ with $\alpha=i/\sqrt{2}$, compass state $\NN_2 \left( |\alpha\rangle+|-\alpha\rangle+|i\alpha\rangle+|-i\alpha\rangle \right)$ with $\alpha=1/\sqrt{2}$. The color scale indicates the value of the Wigner function. See main text for details.}
    \label{fig:states}
\end{figure}

In the following we introduce specific families of single-mode CV states, with particular focus on those that will feature prominently below in the context of quantum metrology.

\textbf{Fock states.} We begin by introducing a natural basis for the quantum harmonic oscillator that will allow us to express all other states. Such a basis is given by \textbf{Fock states}, which are defined for $n\in \mathbb{N}$ as
\begin{equation}
\ket{n}= \dfrac{(\op{a}^\dagger)^n}{\sqrt{n!}} \ket{0} \,.
\end{equation}
These are eigenstates of the Hamiltonian of the harmonic oscillator $\hat{H}=\frac{1}{2}(\hat{x}^2+\hat{p}^2)=\hat{a}^{\dagger}\hat{a}+\frac{1}{2}$, with the vacuum state $\ket{0}$ being the ground state. The action of the associated bosonic ladder operators is $\aa\ket{n}=\sqrt{n}\ket{n-1}$, $\aad\ket{n}=\sqrt{n+1}\ket{n+1}$. 

\textbf{Coherent states.} A displacement of the vacuum to phase space coordinates $\alpha$ produces a coherent state
\begin{align}
        \ket{\alpha}=\hat{D}(\alpha)\ket{0}.
\end{align}
In the basis of Fock states, this leads to the expression
\begin{equation}
    \ket{\alpha} = e^{-\frac{1}{2}\abs{\alpha}}\sum_{n=0}^{\infty}\frac{\alpha^{n}}{\sqrt{n!}}\ket{n} \;.
\end{equation}
Since perfectly coherent laser light is described by a coherent state, these states have played an important role in the development of the quantum theory of light, and they are the experimentally most easily accessible family of states. In the context of quantum metrology, coherent states determine the limits of classical measurement strategies.

\textbf{Gaussian states.} Coherent states are part of a larger family of states known as Gaussian states. These can be defined as those states that have a Gaussian Wigner function. As the most readily accessible family of states in quantum optical experiments, these states play a crucial role in continuous-variable quantum information processing~\cite{Paris2005,Adesso_2007,Wang2007,RevModPhys.77.513,Weedbrook,Adesso2014,Serafini2017,brask2022gaussian}. Any pure Gaussian state can be expressed as a displaced squeezed vacuum state
\begin{equation}
\ket{\alpha,\xi}=\hat{D}(\alpha)\hat{S}(\xi)\ket{0} \;,
\end{equation}
where $\hat{S}(\xi)$ is the squeezing operator \cite{Infeld1955,Plebanski1956,StolerPRD70,WallsNat83}
\begin{equation}
    \hat{S}(\xi) \equiv e^{\frac{1}{2} (\xi^{*}\aa^{2}-\xi \op{a}^{\dagger 2} )},
\end{equation}
with $\xi = r e^{i\gamma}$, $r\equiv|\xi|$. This definition can be generalised to mixed Gaussian states that can be written as 
\begin{equation}\label{eq:rhoG}
    \RHO(\alpha,\xi,n_T) = \hat{D}(\alpha)\hat{S}(\xi)\RHO_{T} \hat{S}^{\dagger}(\xi)\hat{D}^{\dagger}(\alpha) \;,
\end{equation}
with $\RHO_{T}$ the thermal state with average number of thermal excitations $n_T \equiv \text{Tr}[\op{\rho_T}\aad\aa]$. Here, note that the total average number of excitation in the state is $\overline{n} \equiv \text{Tr}[\RHO(\alpha,\xi,n_T)\aad\aa]=n_T\cosh(2r) + |\alpha|^2 + \sinh^2 r$, meaning that for $|\alpha|^2=0$ we have $\cosh(2r)=(1+2\overline{n})/(1+2n_T)$ and $\sinh^2 r=(\overline{n}-n_T)/(1+2n_T)$. Gaussianity implies that these states can be fully described by first and second moments of quadrature operators. For this reason, a Gaussian state is fully determined by its first-moment vector $\langle \bqq\rangle_{\RHO}$ and its covariance matrix $\Gamma[\RHO,\bqq]$, with $\bqq=(\xx,\pp)^T$, $\langle \hat{M}\rangle_{\RHO}=\Tr\left[\hat{M}\RHO\right]$, and 
\begin{align}
    \Gamma[\RHO,\bMM]_{ij}=\frac{1}{2}\langle \hat{M}_i\hat{M}_j+\hat{M}_j\hat{M}_i\rangle_{\RHO}-\langle\hat{M}_i\rangle_{\RHO}\langle\hat{M}_j\rangle_{\RHO}.
\end{align}
For the state~(\ref{eq:rhoG}), we obtain the first-moment vector
\begin{align}
    \langle \bqq\rangle_{\RHO(\alpha,\xi,n_T)}=\begin{pmatrix}
        \langle \xx\rangle_{\RHO(\alpha,\xi,n_T)}\\
        \langle \pp\rangle_{\RHO(\alpha,\xi,n_T)},
    \end{pmatrix} =\begin{pmatrix}
        \sqrt{2}\Re[\alpha]\\
        \sqrt{2}\Im[\alpha]
    \end{pmatrix}  \;,
\end{align}
and the covariance matrix
\begin{equation}
    \Gamma[\RHO(\alpha,\xi,n_T),\bqq] = \dfrac{2 n_T + 1}{2} \begin{pmatrix}
\cosh(2r)-\sinh(2r)\cos(\gamma) & -\sinh(2r)\sin(\gamma) \\
-\sinh(2r)\sin(\gamma) & \cosh(2r)+\sinh(2r)\cos(\gamma)
\end{pmatrix} \;.
\end{equation}

A necessary and sufficient condition for a matrix $\Gamma$ to describe a Gaussian quantum state is that it satisfies the uncertainty relation $\Gamma+\frac{i}{2}\Omega\geq 0$, where $\Omega_{ij}=[\qq_i,\qq_j]$ is the symplectic form with entries
\begin{align}
    \Omega=\begin{pmatrix}
        0 & 1\\ -1 & 0
    \end{pmatrix}.
\end{align}
For a single-mode system, the uncertainty relation is equivalent to the two conditions $\Gamma\geq 0$ and $\det\Gamma\geq 1/4$~\cite{Serafini2017}, with pure Gaussian states satisfying $\det\Gamma = 1/4$.

\textbf{Cat states.} Based on these fundamental classes of states, we can build superposition states of particular interest. A paradigmatic case are  quantum superpositions of two opposite coherent states, typically referred to as cat states~\cite{gerry_quantum_1997}
\begin{equation}
    \ket{\Psi_{\text{cat},\gamma}} = \NN_1 (\ket{\alpha} + e^{i\gamma}\ket{-\alpha}) \;,
    \label{eq:general_cat_state}
\end{equation}
where $\NN_1^{-1}=\sqrt{2+2\cos(\gamma) e^{-2\abs{\alpha}^2}}$ is the normalization constant. Specific real choices for the phases are referred to as even cat state $(\gamma=0)$ and odd cat state $(\gamma=\pi)$. 

\textbf{Compass states.} A superposition of four coherent states is known as a compass state, which have been suggested as particularly sensitive probes of phase space~\cite{WojciechNature2001}. These states are given by
\begin{align}
|\Psi_\text{comp}\rangle=\NN_2 \left( |\alpha\rangle+|-\alpha\rangle+|i\alpha\rangle+|-i\alpha\rangle \right) \;,
\end{align}
where $\alpha\in\mathbb{R}$ and $\NN_2^{-1}=2\sqrt{1+e^{-2\alpha^2}+2e^{-\alpha^2}\cos(\alpha^2)}$ is a normalization constant. 

\textbf{Superpositions of two Fock states.} Another way to create highly nonclassical states is to superpose two Fock states. We will also consider the metrological properties of states of the type $(\ket{m} + e^{i\gamma} \ket{n})/\sqrt{2}$ with $m\neq n$.

The first and second moments of all classes of states mentioned in this section are provided in Appendix~\ref{app:avg} and~\ref{app:cov}, respectively. The Wigner functions of examples of these families are shown in Fig.~\ref{fig:states}.

\subsection{Gaussian evolutions}
\textbf{Unitary evolutions.} Of particular interest for Gaussian states are Gaussian evolutions, since they maintain the Gaussianity of the state at all times. For a detailed discussion we refer to the extensive literature on the topic~\cite{Paris2005,Adesso_2007,Wang2007,RevModPhys.77.513,Weedbrook,Adesso2014,Serafini2017,brask2022gaussian}. An example of a Gaussian evolution is a unitary evolution generated by a quadratic Hamiltonian 
\begin{align}\label{eq:quadraticH}
    \hat{U}(\theta)=\exp\left[-i\frac{1}{2}\hat{\bm{r}}^TH\hat{\bm{r}}\theta\right].
\end{align}
The state $\RHO(\theta)=\hat{U}(\theta)\RHO\hat{U}^{\dagger}(\theta)$ remains Gaussian for all $\theta$ if the initial state $\RHO$ is Gaussian. This implies that for any value of $\theta$, the state is fully characterized by its first and second moments, which under this evolution transform as
\begin{align}\label{eq:symplectic}
    \langle\hat{\bm{r}}\rangle_{\RHO(\theta)}&=\mathcal{S}(\theta)\langle\hat{\bm{r}}\rangle_{\RHO}\notag\\
    \Gam[\RHO(\theta),\bqq]&=\mathcal{S}(\theta)\Gam[\RHO,\bqq]\mathcal{S}(\theta)^T,
\end{align}
where
\begin{align}\label{eq:Ssymp}
    \mathcal{S}(\theta)=\exp[\Omega H \theta]
\end{align}
describes a symplectic transformation with the defining property $\mathcal{S}(\theta)\Omega \mathcal{S}(\theta)^T=\Omega$.

Another example of a Gaussian unitary evolution is the one generated by a linear Hamiltonian, \ie a displacement $\hat{D}(\alpha)$ leading to the displaced state $\RHO(\alpha)=\hat{D}(\alpha)\RHO\hat{D}^{\dagger}(\alpha)$. The displacement only shifts the first moments while leaving the second moments invariant:
\begin{align}\label{eq:disptransformation}
    \langle\hat{\bm{r}}\rangle_{\RHO(\alpha)}&=\langle\hat{\bm{r}}\rangle_{\RHO}+\sqrt{2}\begin{pmatrix}
        \Re[\alpha]\\\Im[\alpha]
    \end{pmatrix}\notag\\
    \Gam[\RHO(\alpha),\bqq]&=\Gam[\RHO,\bqq].
\end{align}

\textbf{Gaussian channels.} When the system of interest interacts with an inaccessible environment, its effective evolution is no longer unitary but instead governed by a completely positive trace-preserving map, commonly referred to as a quantum channel~\cite{BreuerPetruccione2006}. Gaussian channels preserve the Gaussianity of the initial state and are described by the evolution~\cite{Serafini2017}
\begin{align}\label{eq:channel}
    \langle\hat{\bm{r}}\rangle_{\RHO(\theta)}&=\mathcal{X}(\theta)\langle\hat{\bm{r}}\rangle_{\RHO}+\bm{d}(\theta)\notag\\
    \Gam[\RHO(\theta),\bqq]&=\mathcal{X}(\theta)\Gam[\RHO,\bqq]\mathcal{X}(\theta)^T+\mathcal{Y}(\theta),
\end{align}
where $\mathcal{X}(\theta)$ and $\mathcal{Y}(\theta)$ satisfy
\begin{align}\label{eq:open}
    \mathcal{Y}(\theta)+\frac{i}{2}\Omega\geq \frac{i}{2}\mathcal{X}(\theta)\Omega\mathcal{X}(\theta)^T.
\end{align}
These channels will play a crucial role when analyzing the metrological performances of CV systems subject to common decoherence channels, such as losses or heating.

\clearpage
\newpage

\section{Quantum metrology}\label{sec:qm}

\begin{figure}[h!]
    \centering
    \includegraphics[width=10cm]{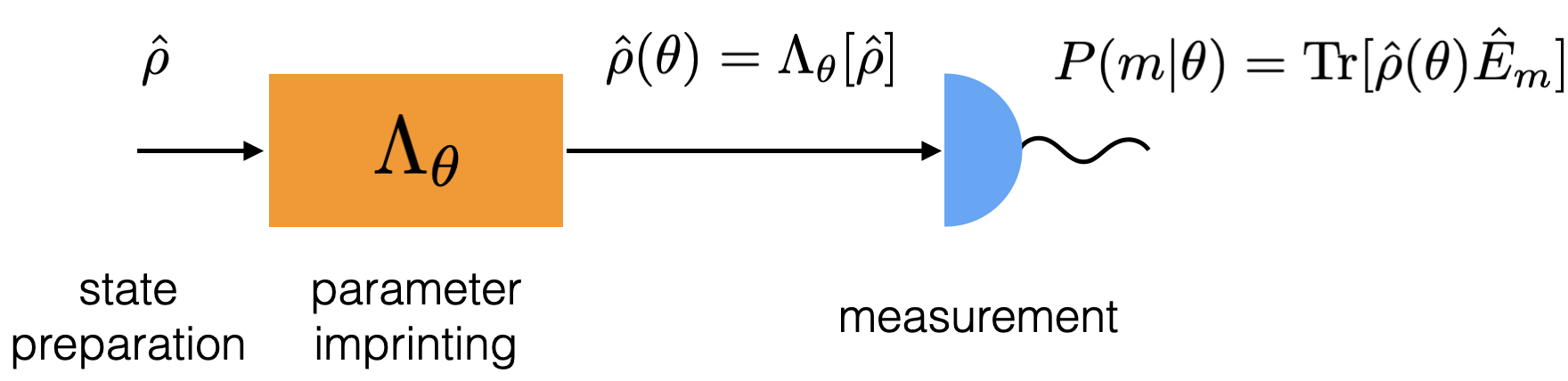}
    \caption{\textbf{Schematic illustration of a metrological protocol.} A parameter $\theta$ is imprinted on an initial quantum state $\hat{\rho}$ through the evolution described by the quantum channel $\Lambda_\theta$. This results in the state $\hat{\rho}(\theta)$, on which a POVM measurement $\{E_m \}_m$ is performed to estimate the unknown $\theta$.
    }
    \label{fig:scheme}
\end{figure}

\subsection{Quantum Fisher information and Cram\'er-Rao bound}
\textbf{Classical bound.} We consider the task of estimating the value of an unknown parameter that determines the state $\RHO(\theta)$ of a quantum system, see Fig.~\ref{fig:scheme}. To infer the value of $\theta$, we perform repeated measurements on the system and construct an estimator function $\theta_{\mathrm{est}}$ that maps the measurement outcomes onto a possible value for $\theta$. The Cram\'er-Rao (CR) bound defines a fundamental limit for the variance of an unbiased estimator:
\begin{equation}
    \var{\theta_{\mathrm{est}}}{} \geq \frac{1}{\mu F[\RHO(\theta),\{\hat{E}_m\}]} \;,
\end{equation}
where $\mu$ is the number of independent measurements and
\begin{equation}\label{eq:classicalFI}
    F[\RHO(\theta),\{\hat{E}_m\}] = \sum_{m}\frac{1}{P(m|\theta)}\left(\frac{\partial P(m|\theta)}{\partial\theta}\right)^2 \;
\end{equation}
is the Fisher information. This quantity describes the information that can be retrieved about the parameter $\theta$ from measurement results $m$ that are distributed according to the probability distribution $P(m|\theta)$, conditioned on $\theta$. A measurement on a quantum system is described by a positive operator-valued measure (POVM), \ie a set $\{\hat{E}_m\}_m$ of positive semidefinite operators $\hat{E}_m\geq 0$ such that $\sum_m\hat{E}_m=\ID$. The probability of obtaining the result $m$ given a phase $\theta$ is then given, according to quantum theory, by  $P(m|\theta)=\Tr[\RHO(\theta)\hat{E}_{m}]$, where $\RHO(\theta)$ is the quantum state of the system. It can be shown that a maximum likelihood analysis of the measurement data saturates the CR bound when a large enough sample has been recorded~\cite{Fisher1922,Fisher1925}.

\vspace{2mm}
\textbf{Quantum bound.} 
Since the Fisher information is a function of the measurement setting, the achievable precision still depends crucially on the chosen POVM. A POVM of particular interest is the one that minimizes the estimation error and gives rise to the quantum Fisher information (QFI):
\begin{equation}\label{eq:maxF}
    F_{\mathrm{Q}}[\RHO(\theta)]=\max_{\{\hat{E}_m\}}F[\RHO(\theta),\{\hat{E}_m\}] \;,
\end{equation}
The QFI can be expressed as
\begin{align}\label{eq:QFIspectral}
    F_{\mathrm{Q}}[\RHO(\theta)] = \sum_{k,l}  \frac{2}{\lambda_k+\lambda_l}\vert\bra{k}\frac{\partial\RHO(\theta)}{\partial\theta}\ket{l}\vert^2 \;,
\end{align}
where $\lambda_k$ and $\ket{k}$ are the eigenvalues and eigenvectors of $\RHO(\theta)$, respectively, and the summation includes all $k,l$ such that $\lambda_k+\lambda_l>0$. The optimal POVM that achieves the maximum in Eq.~(\ref{eq:maxF}) can be obtained as the projectors onto the eigenvalues of the symmetric logarithmic derivative operator, i.e., if $\hat{L}=\sum_ml_m\hat{E}_m$, where $\hat{L}$ is defined as the operator satisfying the relation~\cite{Helstrom1969,BraunsteinPRL1994}
\begin{align}\label{eq:SLD}
    \frac{\partial \RHO(\theta)}{\partial \theta}=\frac{1}{2}\left(\hat{L}(\theta)\RHO(\theta)+\RHO(\theta)\hat{L}(\theta)\right) \;.
\end{align}
The QFI can be expressed using $\hat{L}(\theta)$ as $F_{\mathrm{Q}}[\RHO(\theta)]=\Tr{\left[\RHO(\theta)\hat{L}(\theta)^2\right]}$.

\vspace{2mm}
\textbf{Unitary evolution.} 
A relevant special scenario is that of a unitarily imprinted parameter. A unitary evolution $\hat{U}(\theta)=\exp[-i \GEN \theta]$, generated by $\GEN$ transforms the initial state $\RHO$ into the parameter dependent state $\RHO(\theta)=\hat{U}(\theta)\RHO\hat{U}(\theta)^{\dagger}$. For unitary evolutions, the QFI is independent of $\theta$ and only depends on the initial state $\RHO$ and the generator $\GEN$, which we express by writing $F_{\mathrm{Q}}[\RHO(\theta)]=F_{\mathrm{Q}}[\RHO,\GEN]$. In this case, Eq.~(\ref{eq:QFIspectral}) reduces to
\begin{equation}
    F_{\mathrm{Q}}[\RHO,\GEN] = 2 \sum_{k,l}  \frac{(\lambda_k-\lambda_l)^2}{\lambda_k+\lambda_l}\vert\bra{k}\GEN\ket{l}\vert^2 \;.
\end{equation}
The QFI is bounded from above by the variance of the generator:
\begin{equation}
    F_{\mathrm{Q}}[\RHO, \GEN] \leq 4 \var{\GEN}{\RHO} \;.
    \label{eq:QFI_var}
\end{equation}
Moreover, this bound is saturated for all pure states $\PSI=|\psi\rangle\langle\psi|$:
\begin{align}\label{eq:QFI_varpure}
    F_{\mathrm{Q}}[\PSI, \GEN]=4\var{\GEN}{\PSI} \;.
\end{align}

\vspace{2mm}
\textbf{Gaussian states.} 
In addition, an analytical expression for the QFI can also be obtained for arbitrary Gaussian states. For a single-mode Gaussian state $\RHO(\theta)$ under unitary evolution, the QFI is given in terms of their first-moment vector $\langle\bqq\rangle_{\RHO(\theta)}$ and covariance matrix $\Gam[\RHO(\theta),\bqq]$ as~\cite{PinelPRA2013,monras2013phase,Jiang2014,Serafini2017}
\begin{align}\label{eq:QFIgauss}
    F_Q[\RHO(\theta)]=\frac{1}{2}\frac{\Tr\left[(\Gam^{-1}[\RHO(\theta),\bqq]\frac{\partial}{\partial\theta}\Gam[\RHO(\theta),\bqq])^2\right]}
    {1+\mathcal{P}(\RHO(\theta))^2}+\left(\frac{\partial\langle\bqq\rangle_{\RHO(\theta)}}{\partial\theta}\right)^T\Gam^{-1}[\RHO(\theta),\bqq]\left(\frac{\partial\langle\bqq\rangle_{\RHO(\theta)}}{\partial\theta}\right),
\end{align}
where
\begin{align}\label{eq:purityGauss}
    \mathcal{P}(\RHO)=\Tr\left[\RHO^2\right]=\frac{1}{2\sqrt{\det\Gam[\RHO,\bqq]}}
\end{align}
is the purity. 
For an evolution generated by a single-mode quadratic Hamiltonian of the form~(\ref{eq:quadraticH}), we obtain
\begin{align}
    F_Q[\RHO,\frac{1}{2}\hat{\bm{r}}^TH\hat{\bm{r}}]&=\frac{\Tr\left[\Gam[\RHO,\bqq] H\right]^2-4\det\Gam[\RHO,\bqq]\det{H}}
    {\frac{1}{4}+\det\Gam[\RHO,\bqq]}+\frac{\langle\hat{\bm{r}}\rangle_{\RHO}^TH\Gam[\RHO,\bqq] H\langle\hat{\bm{r}}\rangle_{\RHO}}{\det\Gam[\RHO,\bqq]}.
\end{align}
A derivation of this result is given in App.~\ref{app:gaussrotqfi}. For mixed states under non-unitary Gaussian channels~(\ref{eq:channel}), the purity-changing term $2\left(\frac{\partial\mathcal{P}(\RHO(\theta))}{\partial\theta}\right)^2/(1-\mathcal{P}(\RHO(\theta))^4)$ is added to the QFI~(\ref{eq:QFIgauss})~\cite{PinelPRA2013,Serafini2017}.

\vspace{2mm}
\textbf{Convexity.}
The QFI is a convex function of the quantum state, \ie for any probability distribution $p_k$ and family of quantum states $\RHO_k$, we have
\begin{align}
    F_{\mathrm{Q}}[\sum_kp_k\RHO_k, \GEN]\leq\sum_kp_kF_{\mathrm{Q}}[\RHO_k, \GEN] \;.
\end{align}
This implies that the maximum QFI over all quantum states is achieved for a pure state:
\begin{align}\label{eq:optstate}
    \max_{\RHO}F_{\mathrm{Q}}[\RHO, \GEN]=\max_{\PSI}F_{\mathrm{Q}}[\PSI, \GEN]=\max_{\PSI}4\var{\GEN}{\PSI} \;.
\end{align}

\subsection{Method of moments}

\begin{figure}[h!]
    \centering
    \includegraphics[width=14cm]{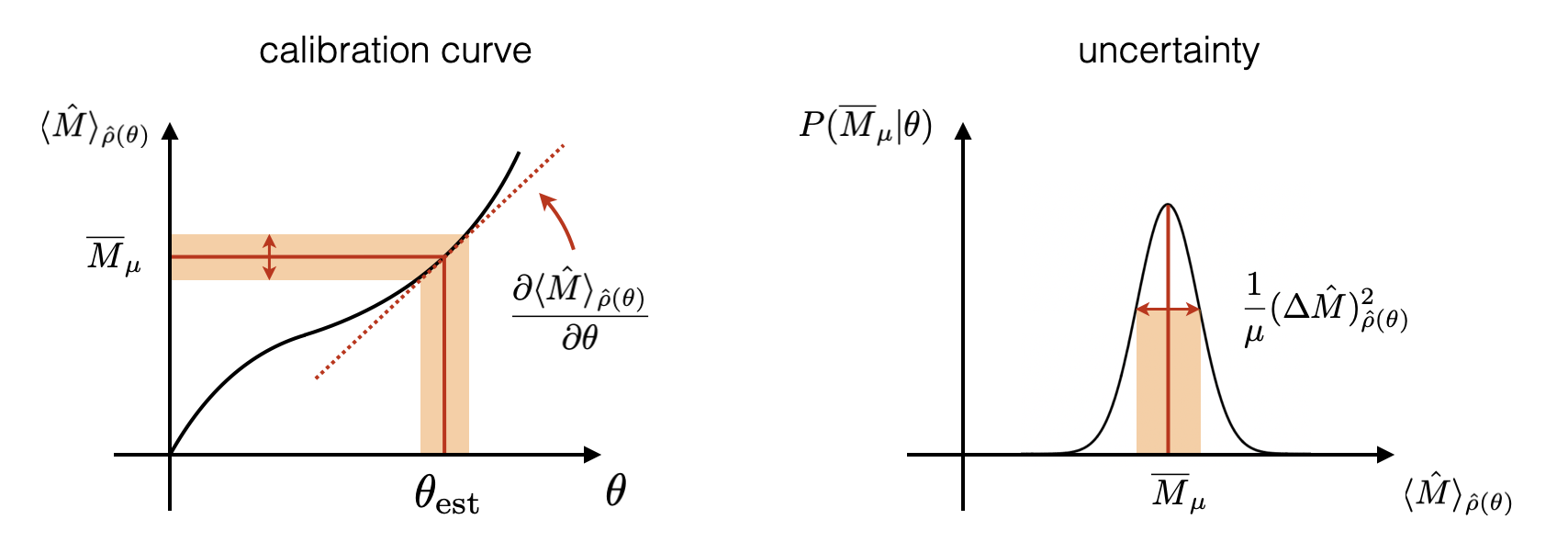}
    \caption{\textbf{Illustration of the method of moments.} First, the calibration curve $f(\theta)$ relating $\langle \MM\rangle_{\RHO(\theta)}$ to $\theta$ is recorded or predicted from theory. Then, from $\mu$ measurements of the observable $\hat{M}$ the sample average $\overline{M}_\mu$ is computed and subsequently used to estimate the unknown parameter by inverting the calibration curve, \ie $\theta_{\text{est}}=f^{-1}(\overline{M}_\mu)$. In the central limit, when $\mu\gg 1$, the distribution of the sample average becomes Gaussian with variance $\var{\MM}{\RHO(\theta)}/\mu$. The error on the estimator is then determined by maximum likelihood estimation based on the sample average data, or equivalently, by error propagation, and depends on the derivative of the calibration curve.}
    \label{fig:FigMom}
\end{figure}

The QFI describes the maximal metrological sensitivity that can be extracted from the quantum state $\RHO$ using an optimal estimation strategy. 
A practical approach to parameter estimation based on the average value of a single observable $\MM$ is given by the method of moments, see Fig.~\ref{fig:FigMom}. 
This method provides a simple estimator for $\theta$ by comparing the sample average of $\mu$ repeated measurements of $\MM$ with a previously established relation $\langle \MM\rangle_{\RHO(\theta)}$, \ie a ``calibration curve''. 
Once this relation is known, implementation of the method of moments is straightforward: The estimator is simply given by the parameter value that yields equality between the sample average and the calibration curve.

For this fixed choice of measurement that is the sample average of an observable $\MM$, in the asymptotic limit of many repetitions, $\mu\gg 1$, it can be shown that this estimator is optimal, since it becomes equivalent to the maximum likelihood estimator and thus achieves the CR bound. An explicit calculation (see \eg \cite{PezzeBook}) of the FI then yields
\begin{align}\label{eq:mmesterror}
    \var{\theta_{\mathrm{est}}}{}
    =\frac{\var{\MM}{\RHO(\theta)}}{\mu\abs{\frac{\partial\langle\MM\rangle_{\RHO(\theta)}}{\partial\theta}}^2}=\frac{1}{\mu}\chi^2[\RHO(\theta),\MM] \;,
\end{align}
where
\begin{align}\label{eq:mmoments}
    \chi^{-2}[\RHO(\theta),\MM] =\frac{\abs{\frac{\partial}{\partial\theta}\langle\MM\rangle_{\RHO(\theta)}}^2}{\var{\MM}{\RHO(\theta)}}
\end{align}
is the moment-based sensitivity of the state $\RHO(\theta)$ for estimations of $\theta$ from measurements of $\MM$.

The estimation error Eq.~(\ref{eq:mmesterror}) takes the form of an error propagation expression and can be intuitively understood by considering that the sample average will in the central limit be Gaussian distributed with variance $\frac{\var{\MM}{\RHO(\theta)}}{\mu}$. This variance then propagates onto the estimator and the impact it has depends on the gradient of the calibration curve, \ie on $\abs{\frac{\partial\langle\MM\rangle_{\RHO(\theta)}}{\partial\theta}}^2$.

\vspace{2mm}
\textbf{Unitary evolution.} For unitary evolutions generated by $\GEN$, we make use of the von Neumann equation $\frac{\partial\langle \MM \rangle_{\RHO}}{\partial\theta} = -i \langle[\MM,\GEN]\rangle_{\RHO}$ to describe the evolution with respect to the parameter $\theta$. In the following, we focus on estimations of $\theta$ in the vicinity of $\theta=0$ and express the moment-based sensitivity as a function of the initial state $\RHO$, the generator $\GEN$ and the measurement observable $\MM$ as
\begin{align}\label{eq:chiunitary}
    \chi^{-2}[\RHO,\GEN,\MM] =\frac{\abs{\langle[\MM, \GEN]\rangle_{\RHO}}^2}{\var{\MM}{\RHO}} \;.
\end{align}

As a specific measurement strategy, any choice of $\MM$ leads, by virtue of the method of moments, to the lower bound on the full metrological potential
\begin{align}
    \chi^{-2}[\RHO, \GEN, \MM]\leq F_{\mathrm{Q}}[\RHO, \GEN] \;.
\end{align}
A systematic optimization over all observables $\MM$ recovers the QFI~\cite{gessner_metrological_2019}
\begin{equation}
    F_{\mathrm{Q}}[\RHO, \GEN]=\max_{\MM}\chi^{-2}[\RHO, \GEN, \MM] \;.
\end{equation}
This maximum is achieved by the symmetric logarithmic derivative $\hat{L}$, defined in Eq.~(\ref{eq:SLD})~\cite{gessner_metrological_2019}.

\vspace{2mm}
\textbf{Dichotomic measurements.} For dichotomic measurements the method of moments equals the classical Fisher information. Consider an observable $\MM=m_1\hat{\Pi}_1+m_2\hat{\Pi}_2$ with two eigenvalues $m_{i}$ and corresponding projectors $\hat{\Pi}_{i}$, such that $\hat{\Pi}_1+\hat{\Pi}_2=\ID$. The conditional probability is given by $P(m_1|\theta)=\Tr\left[\RHO(\theta)\hat{\Pi}_1\right]=1-\Tr\left[\RHO(\theta)\hat{\Pi}_2\right]=1-P(m_2|\theta)$. The variance of $\MM$ reads
\begin{align}
    \var{\MM}{\RHO(\theta)}&=m_1^2P(m_1|\theta)+m_2^2P(m_2|\theta)-(m_1P(m_1|\theta)+m_2P(m_2|\theta))^2\notag\\
    &=(m_1-m_2)^2P(m_1|\theta)(1-P(m_1|\theta)) \;,
\end{align}
and the derivative of the average value is given by
\begin{align}
    \frac{\partial}{\partial\theta}\langle \MM\rangle_{\RHO(\theta)}&=m_1\frac{\partial P(m_1|\theta)}{\partial\theta}+m_2\frac{\partial P(m_2|\theta)}{\partial\theta}\notag\\
    &=(m_1-m_2)\frac{\partial P(m_1|\theta)}{\partial\theta} \;.
\end{align}
We thus obtain from the method of moments
\begin{align}
    \chi^{-2}[\RHO(\theta),\MM] =\frac{\abs{\frac{\partial}{\partial\theta}\langle\MM\rangle_{\RHO(\theta)}}^2}{\var{\MM}{\RHO(\theta)}}=\frac{\left(\frac{\partial P(m_1|\theta)}{\partial\theta}\right)^2}{P(m_1|\theta)(1-P(m_1|\theta))} \;.
\end{align}
On the other hand, the Fisher information reads in this case
\begin{align}
    F[\RHO(\theta),\{\hat{E}_m\}]&=\frac{1}{P(m_1|\theta)}\left(\frac{\partial P(m_1|\theta)}{\partial\theta}\right)^2+\frac{1}{P(m_2|\theta)}\left(\frac{\partial P(m_2|\theta)}{\partial\theta}\right)^2\notag\\
    &=\frac{1}{P(m_1|\theta)(1-P(m_1|\theta))}\left(\frac{\partial P(m_1|\theta)}{\partial\theta}\right)^2 \;.
\end{align}
Indeed, the two expressions coincide:
\begin{align}\label{eq:momentsdichotomic}
    \chi^{-2}[\RHO(\theta),\MM]=F[\RHO(\theta),\{\hat{E}_m\}] \;.
\end{align}

\vspace{2mm}
\textbf{Optimizing the measurement.} Assume that any linear combination of measurement observables $\bMM=\{\MM_1,\dots,\MM_m\}^T$ can be accessed. In this case, it is convenient to identify the maximal sensitivity that can be reached through the measurement of an optimal linear combination using the method of moments. We obtain
\begin{align}\label{eq:maxchi}
    \max_{\MM\in\mathrm{span}(\bMM)}\chi^{-2}[\RHO, \GEN, \MM]=C^T[\RHO,\GEN,\bMM]\Gam^{-1}[\RHO,\bMM]C[\RHO,\GEN,\bMM] \;,
\end{align}
with $\Gam^{-1}[\RHO,\bMM]$ the inverse of the covariance matrix
\begin{align}
    \Gam[\RHO,\bMM]_{ij}=\cov[\MM_i,\MM_j]_{\RHO} \;,
\end{align}
and the vector of commutators
\begin{align}\label{eq:comvec}
    C[\RHO,\GEN,\bMM]_i=-i\langle [\MM_i,\GEN]\rangle_{\RHO} = \frac{\partial}{\partial\theta}\avg{\MM_i}_{\RHO} \;.
\end{align}
The covariance is given by $\cov[\hat{A},\hat{B}]_{\RHO} = \frac{1}{2} \langle \hat{A}\hat{B}+\hat{B}\hat{A} \rangle_{\RHO} - \langle \hat{A} \rangle_{\RHO}\langle \hat{B} \rangle_{\RHO}$.
The optimal linear combination that achieves this sensitivity is identified as $\MM=\bMM^T\bm{c}=\sum_{i=1}^mc_i\MM_i$ with
\begin{align}\label{eq:maxc}
    \bm{c}=\zeta \Gam^{-1}[\RHO,\bMM]C[\RHO,\GEN,\bMM] \;,
\end{align}
where $\zeta\in\mathbb{R}$ is a normalization constant. 

\subsection{Measurement-after-interaction technique}
A convenient method to access a wider range of measurement observables is to apply an additional unitary evolution $\hat{U}$ before the measurement of $\MM$. This effectively leads to the replacement of $\RHO(\theta)$ by $\hat{U}\RHO(\theta)\hat{U}^{\dagger}$, or equivalently, to the effective measurement of the observable $\hat{U}^{\dagger}\MM\hat{U}$ with the same state $\RHO(\theta)$. Typically, the generator $\op{H}$ of the additional unitary $\op{U}=e^{-i\op{H}t}$ is a Hamiltonian that describes interactions or nonlinear processes, which motivates the name of the measurement-after-interaction (MAI) technique. In certain cases, the sensitivity
\begin{align}\label{eq:chiunitaryMAI}
    \chi^{-2}[\RHO,\GEN,\hat{U}^{\dagger}\MM\hat{U}] =\frac{\abs{\langle[\hat{U}^{\dagger}\MM\hat{U}, \GEN]\rangle_{\RHO}}^2}{\var{(\hat{U}^{\dagger}\MM\hat{U})}{\RHO}} \;
\end{align}
may be better than $\chi^{-2}[\RHO,\GEN,\MM]$, \ie the one obtained from a direct measurement of $\MM$. The MAI technique can be used to access higher-order moments of $\MM$, or to reduce the effect of detection noise.

\subsection{Classical Fisher information}
The method of moments estimates the value of the unknown parameter $\theta$ from the variations of the expectation value of the measured observable $\langle \MM\rangle_{\RHO(\theta)}$. Sometimes, however, for a fixed choice of $\MM$, or more generally a POVM $\{\op{E}_m\}$, it is necessary to consider more than just the first moment to extract the maximum information about $\theta$ that is contained in the data. The complete description of the measurement distribution is given by the probability distribution $P(m|\theta)=\Tr[\RHO(\theta)\hat{E}_m]$. The full information contained in this data is described by the classical Fisher information, Eq.~(\ref{eq:classicalFI}).

As an example, we state the Fisher information of an arbitrary single-mode Gaussian state, characterized by first-moment vector $\langle \bqq\rangle_{\RHO(\theta)}$ and covariance matrix $\Gamma[\RHO(\theta),\bqq]$. A homodyne measurement is a measurement of a quadrature observable $\qq(\varepsilon) =\xx\sin\varepsilon + \pp\cos \varepsilon= \bqq^T\bm{w} $, where $\bm{w}=(\sin\varepsilon,\cos\varepsilon)^T$ and the angle $\epsilon$ determines the direction of the measurement in phase space. When applied to a Gaussian state, this measurement leads to a Gaussian probability distribution $P(x|\theta)$ with first moment $\mu=\langle \bqq\rangle_{\RHO(\theta)}^T\bm{w}$ and variance $\sigma^2=\bm{w}^T\Gamma[\RHO(\theta),\bqq]\bm{w}$. Generally, the Fisher information for Gaussian distributions reads
\begin{align}
    F_{\mathrm{G}}(\theta)=\frac{1}{\sigma^2}\left[\left(\frac{\partial\mu}{\partial \theta}\right)^2+\frac{1}{2\sigma^2}\left(\frac{\partial\sigma^2}{\partial \theta}\right)^2\right].
\end{align}
A homodyne measurement in direction $\bm{w}$ on a single-mode Gaussian state, thus, yields the classical Fisher information
\begin{align}\label{eq:GaussianCFI}
    F[\RHO(\theta),\bm{w}^T \bqq] = \frac{\left(\bm{w}^T\frac{\partial\langle \bqq\rangle_{\RHO(\theta)}}{\partial \theta}\right)^2}{\bm{w}^T\Gamma[\RHO(\theta),\bqq]\bm{w}}+\frac{\left(\bm{w}^T\frac{\partial\Gamma[\RHO(\theta),\bqq]}{\partial \theta}\bm{w}\right)^2}{2(\bm{w}^T\Gamma[\RHO(\theta),\bqq]\bm{w})^2}.
\end{align}
This result can be further generalized to account for arbitrary (including multi-mode) Gaussian measurements, \ie so-called general-dyne measurements that may contain imperfections~\cite{Serafini2017, oh_optimal_2019}.

\clearpage
\newpage
\section{Displacement sensing}\label{sec:disp}
\begin{wrapfigure}[15]{r}{0.30\textwidth}
\centering
    \raisebox{0pt}[\dimexpr\height-1.5\baselineskip\relax]{%
        \includegraphics[width=0.28\textwidth]{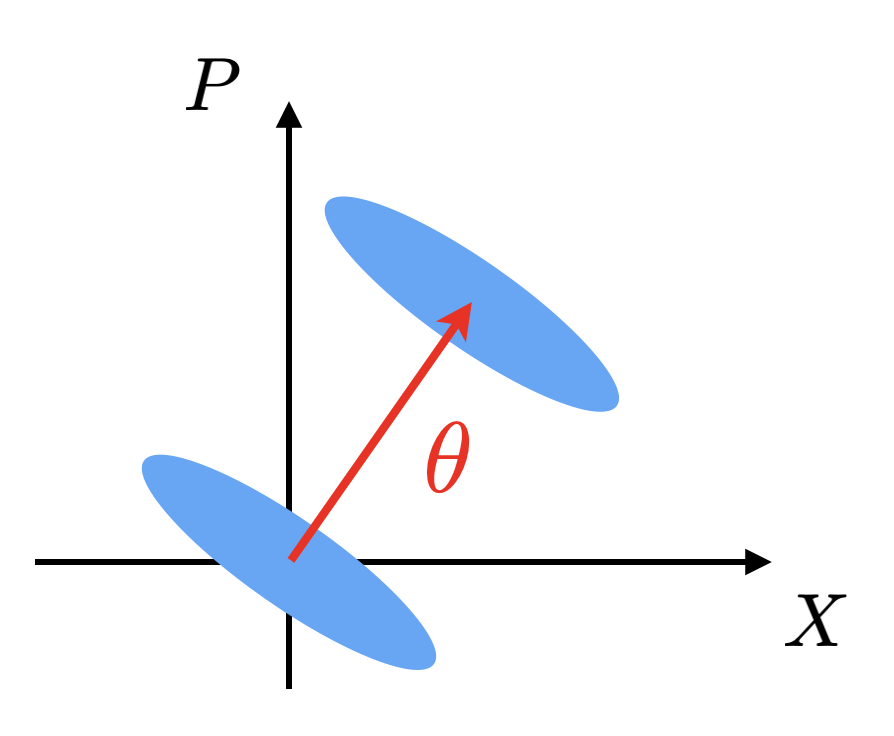}%
    }%
\caption{Phase-space illustration of a displacement. The metrological task considered in this section is to estimate the displacement amplitude $\theta$.}
\label{fig:DispSketch}
\end{wrapfigure}
The first class of perturbations we are going to consider are the one resulting in a phase-space translation of the system's state, see Fig.~\ref{fig:DispSketch}. These can be expressed by the displacement operator Eq.~\eqref{eq:dispOp}, that we rewrite here as
\begin{equation}
\hat{D}(\alpha) \equiv e^{\alpha \aad - \alpha^\ast \aa} = e^{-i\qq(\phi) \theta} \;,
\end{equation}
to identify 
\begin{align}
    \qq(\phi)=\xx\sin\phi + \pp\cos\phi= \bqq^T\bm{u}
\end{align}
as the generator of the perturbation with $\bqq=(\hat{x},\hat{p})^T$, direction $\vect{u}=(\sin\phi,\cos\phi)^T$, and $\alpha=\theta e^{-i \phi}/\sqrt{2}$.
We consider the goal of estimating the amplitude $\theta$ of a displacement. The phase $\phi$ of the displacement may be known or unknown at the time of the state preparation. The task then consists of optimizing the state preparation and the final measurement, in order to maximize the sensitivity.

\subsection{Quantum Fisher information}
As we have seen, the information about the amplitude of the displacement $\theta$ that can be retrieved from the state $\RHO$ is given by the QFI $F_Q[\RHO,\qq(\phi)]$, which here depends on the phase of the displacement $\phi$. In the best-case scenario, this phase is along the most sensitive quadrature, where the QFI is
\begin{equation}\label{eq:Fmax}
    F_Q^{\text{max}}[\RHO] = \max_{\phi} F_Q[\RHO,\qq(\phi)] \;.
\end{equation}
This typically requires some knowledge and control over the phase $\phi$ relative to the initial probe state $\hat{\rho}$. On the contrary, if the phase is unknown or cannot be controlled, an appropriate strategy may be to optimize the performance in the worst-case scenario where the displacement is along the least sensitive quadrature. In this case, the figure of merit is
\begin{equation}\label{eq:Fmin}
    F_Q^{\text{min}}[\RHO] = \min_{\phi} F_Q[\RHO,\qq(\phi)] \;.
\end{equation}
As an alternative strategy in scenarios where $\phi$ is inaccessible, instead of focusing on the worst-case scenario, one may choose to optimize the average sensitivity, namely
\begin{equation}\label{eq:Favg}
    F_Q^{\text{avg}}[\RHO] = \frac{1}{2\pi} \int_0^{2\pi} \dd\phi\; F_Q[\RHO,\qq(\phi)] \;.
\end{equation}

\subsubsection{Pure states}
To evaluate the above expressions for pure states we use that, in this case, the QFI coincides with four times the variance of the generator, Eq.~(\ref{eq:QFI_varpure}). Using the bilinearity of the covariance, this fact can be expressed as
\begin{equation}\label{eq:dispQFIpure}
    F_Q[\PSI,\qq(\phi)]=4\var{\qq(\phi)}{\PSI}=4\vect{u}^T \Gam[\PSI,\bqq] \vect{u} \;,
\end{equation}
where the covariance matrix of an arbitrary state $\RHO$ is defined as
\begin{equation}
    \Gam[\RHO,\bqq] = \begin{pmatrix}
\var{\xx}{\RHO} & \cov[\xx,\pp]_{\RHO} \\
\cov[\xx,\pp]_{\RHO} & \var{\pp}{\RHO}
\end{pmatrix} \;.
\end{equation}
We thus obtain
\begin{align}\label{eq:QFImaxpure}
    F_Q^{\max}[\PSI] &= \max_{\phi}4\var{\qq(\phi)}{\PSI}=4\lambda_{\max}(\Gam[\PSI,\bqq])\notag\\&=2\left( \var{\xx}{\PSI}+\var{\pp}{\PSI}+\sqrt{4\cov[\xx,\pp]_{\PSI}^2+(\var{\xx}{\PSI}-\var{\pp}{\PSI})^2} \right).
\end{align}
Similarly we have
\begin{align}\label{eq:QFIminpure}
    F_Q^{\min}[\PSI] &= \min_{\phi} 4\var{\qq(\phi)}{\PSI}=4\lambda_{\min}(\Gam[\PSI,\bqq])\notag\\&=2\left(\var{\xx}{\PSI}+\var{\pp}{\PSI}-\sqrt{4\cov[\xx,\pp]_{\PSI}^2+(\var{\xx}{\PSI}-\var{\pp}{\PSI})^2} \right) \;,
\end{align}
and
\begin{align}\label{eq:QFIavgpure}
    F_Q^{\text{avg}}[\PSI] &= \frac{1}{2\pi} \int_0^{2\pi} \dd\phi\; 4\var{\qq(\phi)}{\PSI}\notag\\
    &=2\left(\var{\xx}{\PSI}+\var{\pp}{\PSI} \right) \;,
\end{align}
where $\lambda_{\max}(A)$ and $\lambda_{\min}(A)$ are the respective maximal and minimal eigenvalues of the matrix $A$.

\subsubsection{State-independent limits}
It is interesting to identify state-independent upper bounds on these quantities, as they set the ultimate limit for displacement sensing. Despite our focus on pure states here, these bounds set the limits for any state according to Eq.~(\ref{eq:optstate}). A natural constraint for such an optimization is to maintain a fixed average number of photons $\bar{n}$, \ie constant energy across all states.

An upper limit on the covariance of arbitrary observables $\hat{A}$ and $\hat{B}$ can be derived from the Robertson-Schr\"odinger uncertainty relation:
\begin{align}
    \var{\hat{A}}{\RHO}\var{\hat{B}}{\RHO}\geq \cov[\hat{A},\hat{B}]_{\RHO}^2 + \left|\frac{1}{2i}\avg{[\hat{A},\hat{B}]}_{\RHO}\right|^2.
\end{align}
Thus, using $\cov[\xx,\pp]_{\RHO}^2\leq \var{\xx}{\RHO}\var{\pp}{\RHO}-\frac{1}{4}$, we obtain from Eq.~(\ref{eq:QFImaxpure}):
\begin{align}\label{eq:QFImaxpureupper}
    F_Q^{\max}[\PSI] &\leq 2\left(\var{\xx}{\PSI}+\var{\pp}{\PSI}+\sqrt{(\var{\xx}{\PSI}+\var{\pp}{\PSI})^2-1}\right)\notag\\
    &\leq 2(\avg{\xx^2}_{\PSI}+\avg{\pp^2}_{\PSI}+\sqrt{(\avg{\xx^2}_{\PSI}+\avg{\pp^2}_{\PSI})^2-1})\notag\\
    &= 2(1+2\bar{n}+2\sqrt{\bar{n}(\bar{n}+1)}).
\end{align}
In the last step, we used that $\avg{\xx^2}_{\PSI}+\avg{\pp^2}_{\PSI}=1+2\bar{n}$, where $\overline{n}=\avg{\aad\aa}_{\PSI}$ is the average number of particles in the mode. As we will see below, this upper bound is tight and can be saturated by optimally aligned squeezed vacuum states, see Tab.~\ref{tab:dispQFI}. This conclusion continues to be valid in a two-mode interferometer where in the initial step, the CV state is mixed by a beam splitter with a coherent state, see Refs.~\cite{PezzePRL2008,CavesLangPRL2013}. Cat states, on the other hand, do not saturate the upper bound~(\ref{eq:QFImaxpureupper}), even though they show the same scaling with $\bar{n}$ (see App.~\ref{app:catdisp}).

Since the last term in Eq.~(\ref{eq:QFIminpure}) is always negative, it follows that the worst-case sensitivity cannot be larger than 
\begin{align}\label{eq:QFImindisp}
F_Q^{\min}[\PSI]\leq 2(\var{\xx}{\PSI}+\var{\pp}{\PSI})\leq 2(1+2\bar{n}) \;.
\end{align}
A set of sufficient conditions reaching the limit in Eq.~(\ref{eq:QFImindisp}) is $\avg{\xx}_{\PSI}=\avg{\pp}_{\PSI}=0$, $\cov[\xx,\pp]_{\PSI}=0$, and $\var{\xx}{\PSI}=\var{\pp}{\PSI}$. These conditions are fulfilled, \eg by Fock states (see Tab.~\ref{tab:dispQFI}), and it is interesting to note that no Gaussian state would be able to satisfy them~\cite{wolf_motional_2019}. Finally, the average sensitivity can achieve the same maximum value 
\begin{align}\label{QFIavgupper}
F_Q^{\text{avg}}[\PSI]\leq 2(1+2\bar{n}),    
\end{align}
and any pure state with vanishing first moments optimizes the average sensitivity.

\subsubsection{Arbitrary Gaussian states}
The general expression for the single-mode Gaussian QFI is given in Eq.~(\ref{eq:QFIgauss}). Since displacements leave the covariance matrix invariant, this expression reduces in the present case to
\begin{align}\label{eq:QFIGauss}
    F_Q[\RHO,\qq(\phi)] &= \left(\frac{\partial\langle\hat{\bm{r}}\rangle_{\RHO}}{\partial\theta}\right)^T\Gam^{-1}[\RHO,\bqq]\left(\frac{\partial\langle\hat{\bm{r}}\rangle_{\RHO}}{\partial\theta}\right)\notag\\&=\frac{\bm{u}^T\Gam[\RHO,\bqq]\bm{u}}{\det\Gam[\RHO,\bqq]} \;,
\end{align}
where $\det\Gam[\RHO,\bqq]=\var{\xx}{\RHO}\var{\pp}{\RHO}-\cov[\xx,\pp]_{\RHO}^2$ is the determinant of the covariance matix. Eq.~\eqref{eq:QFIGauss} is obtained using the fact that, for a $2\times 2$ matrix $A$, it holds $A ^{-1}={\frac {1}{\det A }}\Omega A^T\Omega^T$ with $\Omega = \bigl( \begin{smallmatrix} 0 & 1\\ -1 & 0\end{smallmatrix}\bigr)$ the symplectic form. Then, for unitary transformations $\frac{\partial\langle \MM \rangle_{\RHO}}{\partial\theta} = -i \langle[\MM,\GEN]\rangle_{\RHO}$, the evolution of $\hat{\bm{r}}$ under displacements is given by
\begin{align}\label{eq:derivVectR}
    \frac{\partial \langle\hat{\bm{r}}\rangle_{\RHO}}{\partial\theta} = -i \begin{pmatrix}
        \langle [\xx,\xx\sin\phi + \pp\cos\phi]\rangle_{\RHO}\\
        \langle [\pp,\xx\sin\phi + \pp\cos\phi]\rangle_{\RHO}
    \end{pmatrix}=\begin{pmatrix}
        \cos\phi\\
        -\sin\phi
    \end{pmatrix}=\Omega \bm{u}
\end{align}
where we recall that $\bm{u}=(\sin\phi,\cos\phi)^T$ defines the direction of the displacement.

The denominator in Eq.~(\ref{eq:QFIGauss}) can be written in terms of the purity $\mathcal{P}(\RHO) = \frac{1}{2}(\var{\xx}{\RHO}\var{\pp}{\RHO}-\cov[\xx,\pp]_{\RHO}^2)^{-1/2}$ of a Gaussian state $\RHO$ using Eq.~(\ref{eq:purityGauss}).
The sensitivity of mixed single-mode Gaussian states can thus be conveniently determined by amending the correction factor $\mathcal{P}(\RHO)^2$ to the pure-state expression~(\ref{eq:QFI_varpure}), namely
\begin{align}\label{eq:QFIGaussPurity}
    F_Q[\RHO,\qq(\phi)] &=4\mathcal{P}(\RHO)^2\bm{u}^T\Gam[\RHO,\bqq]\bm{u}=4\mathcal{P}(\RHO)^2\var{\qq(\phi)}{\RHO} \;.
\end{align}
In particular, for pure Gaussian states, \ie $\mathcal{P}(\RHO)=1$, we find that Eq.~(\ref{eq:QFIGaussPurity}) reproduces the result Eq.~(\ref{eq:dispQFIpure}), as expected. 

In analogy to Eqs.~(\ref{eq:QFImaxpure}), (\ref{eq:QFIminpure}), and~(\ref{eq:QFIavgpure}), we obtain
\begin{align}\label{eq:qfigaussmax}
    F_Q^{\max}[\RHO] &= 2\mathcal{P}(\RHO)^2\left(\var{\xx}{\RHO}+\var{\pp}{\RHO}+\sqrt{4\cov[\xx,\pp]_{\RHO}^2+(\var{\xx}{\RHO}-\var{\pp}{\RHO})^2} \right) \;,\\\label{eq:qfigaussmin}
    F_Q^{\min}[\RHO] &=2\mathcal{P}(\RHO)^2\left(\var{\xx}{\RHO}+\var{\pp}{\RHO}-\sqrt{4\cov[\xx,\pp]_{\RHO}^2+(\var{\xx}{\RHO}-\var{\pp}{\RHO})^2} \right) \;,
\end{align}
and
\begin{align}\label{eq:qfigaussavg}
    F_Q^{\text{avg}}[\RHO
    ] &=2\mathcal{P}(\RHO)^2\left(\var{\xx}{\RHO}+\var{\pp}{\RHO} \right) \;.
\end{align}

Note that, since the Gaussian QFI for displacement sensing is a function only of the covariance matrix $\Gam$, including a displacement in the state preparation does not increase the state's sensitivity, as a translation does not change the second moments that define $\Gam$. For this reason, in the following we will consider only Gaussian states that are centered at the origin (\ie squeezed vacuum/thermal states). An exception are the coherent states whose QFI for displacements is the same as the one of the vacuum state $\ket{0}$.

\subsubsection{Performance of different quantum states}\label{sec:performanceQFIdisp}
We summarize in Table~\ref{tab:dispQFI} the QFI for the state we considered, and in Table~\ref{tab:dispQFImaxminavg} the corresponding maximum, minimum and average QFI when variations of the displacement directions are taken into account. For a direct comparison, we present the results in terms of the average number of photons $\overline{n}=\text{Tr}[\RHO\aad\aa]$. 
Note that in the case of Gaussian states, which are not necessarily pure, this quantity include both ``coherent'' photons $n_c=\sinh^2 r + |\alpha|^2$, \ie those contributing to the state's squeezing or displacement, as well as ``incoherent'' photons $n_T$ that contribute to the thermal distribution.
Because the sensitivity to displacements is invariant under translations (\ie it does not depend on the position of the state in phase space since one can always redefine the origin), we will consider Gaussian states that are centered in the origin, namely $\op{\rho}(0,\xi,n_T)$.

The standard quantum limit (SQL) is set by the sensitivity of an optimally used coherent state. For displacement sensing, this yields a classical limit of $F_Q[\ket{\alpha},\qq(\phi)]=2$, independent of $\bar{n}$. Any value $F_Q[\RHO,\qq(\phi)]>2$ thus indicates a quantum enhancement in sensing displacements. therefore signifies quantum-enhanced displacement sensitivity. It also certifies the nonclassicality of the state $\RHO$ as such performance cannot be reproduced by a convex mixture of coherent states, i.e. by a classical Glauber–Sudarshan $P$-representation~\cite{rivas_precision_2010}.

It is worth recalling that in this discussion of the optimization of displacement sensitivity, we only focused on the single mode that experiences the displacement. This means that in a quantum optical setting, the energy and number of photons that are used to produce the necessary local oscillator to realize the displacement and eventual homodyne measurements, are not part of our consideration. It is important to keep this in mind when comparing results with other approaches that take the local oscillator explicitly into consideration since optimal solutions may differ. In particular, under optimizations with a fixed total number of photons one finds that quantum strategies such as squeezing cannot lead to any scaling enhancement, since the macroscopic number of photons of the local oscillator eclipses the contribution of the nonclassical state. Instead, the sensitivity can only be improved by a constant factor in this case~\cite{PinelPRA2012}.

\begin{table}[h!]
    \centering
    \begin{tabular}{|c|c|}
    \hline
    \textbf{Quantum state} \bm{$\RHO$} & $\bm{F_Q[\RHO,\qq(\phi)]}$ \\\hline  
    Coherent $\ket{\alpha}$ & $2$  \\\hline
    Gaussian state & $\dfrac{2(1 + 2\overline{n} + 2 \sqrt{\overline{n}(1+\overline{n})-n_T(1+n_T)} \cos{[\gamma + 2\phi]}) }{ (1 + 2 n_T)^2 }$ \\\hline
    Fock $\ket{n}$ & $2(1+2n)$ \\\hline
    Fock superposition $(\ket{m}+e^{i\gamma}\ket{n})/\sqrt{2}$, $n>m$  & $2( n + m + 1 - \sqrt{n(n-1)} \cos[\gamma+2\phi] \delta_{n,m+2} - (m+1)\sin^2[\gamma+\phi] \delta_{n,m+1})$ \\ \hline
    Cat $\scN(\ket{\alpha}+e^{i\gamma}\ket{-\alpha})$, $\alpha\in\mathbb{C}$, $\abs{\alpha}\gtrsim 2$ & $2(1+4\overline{n}\cos[\phi]^2)$  \\\hline
    Compass $\scN(\ket{\alpha}+\ket{-\alpha}+\ket{i\alpha}+\ket{-i\alpha})$ & $2(1 + 2\overline{n})$  \\ \hline
    \end{tabular}
    \caption{\textbf{QFI for a displacement along the direction specified by $\phi$.} The QFI has been computed using Eq.~\eqref{eq:QFI_varpure}, apart from Gaussian states (that can be mixed) for which we used Eq.~\eqref{eq:QFIGaussPurity}. For simplicity, here and in the following tables the expressions for cat states are written in the limit of ``large'' $\alpha$ (\ie $\alpha\gtrsim 2$), the general case is considered in the Appendix.}
\label{tab:dispQFI}
\end{table}

\begin{table}[h!]
    \centering
    \begin{tabular}{|c|c|c|c|}
    \hline
    \textbf{Quantum state} \bm{$\RHO$} & \bm{$F_Q^{\text{max}}[\RHO]$} & \bm{$F_Q^{\text{min}}[\RHO]$} & \bm{$F_Q^{\text{avg}}[\RHO]$} \\\hline  
    Coherent $\ket{\alpha}$ & $2$ & $2$ & $2$  \\\hline
    Gaussian state & $\dfrac{2(1 + 2\overline{n} + 2 \sqrt{\overline{n}(1+\overline{n})-n_T(1+n_T)}) }{ (2 n_T + 1)^2}$ & $\dfrac{2(1 + 2\overline{n} - 2 \sqrt{\overline{n}(1+\overline{n})-n_T(1+n_T)} \,) }{ (2 n_T + 1)^2}$ & $\dfrac{2 (1 + 2\overline{n}) }{ (2 n_T + 1)^2}$  \\\hline
    Fock $\ket{n}$ & $2(1+2n)$ & $2(1+2n)$ & $2(1+2n)$  \\\hline
    \makecell{Fock superposition \\ $(\ket{m}+e^{i\gamma}\ket{n})/\sqrt{2}$ \\ $n>m$}  & $\begin{cases}
        4n \qquad\qquad\qquad& n=m+1\\
        2((2n-1)+\sqrt{n(n-1)}) &  n=m+2\\
        2(1+n+m) & \text{otherwise}
    \end{cases}$ & $\begin{cases}
        2n \qquad\qquad\qquad& n=m+1\\
        2((2n-1)-\sqrt{n(n-1)}) &  n=m+2\\
        2(1+n+m) & \text{otherwise}
    \end{cases}$ & $\begin{cases}
        3n \qquad\qquad& n=m+1\\
        2(2n-1) \qquad\qquad& n=m+2\\
        2(1+n+m) & \text{otherwise}
    \end{cases}$  \\ \hline
    Cat $\abs{\alpha}\gtrsim2$ & $2(1+4\overline{n})$ & $2$ & $2(1+2\overline{n})$  \\\hline
    Compass & $2(1 + 2\overline{n})$ & $2(1 + 2\overline{n})$ & $2(1 + 2\overline{n})$ \\ \hline
    \end{tabular}
    \caption{\textbf{Maximum, minimum and average QFI for a displacement along the direction specified by $\phi$}. For a definition of these quantities, see Eqs.~(\ref{eq:Fmax},\ref{eq:Fmin},\ref{eq:Favg}).}
\label{tab:dispQFImaxminavg}
\end{table}

\subsubsection{Noisy displacement sensing}\label{sec:dispsensnoise}
\begin{figure}[h!]
    \centering
    \includegraphics[width=\textwidth]{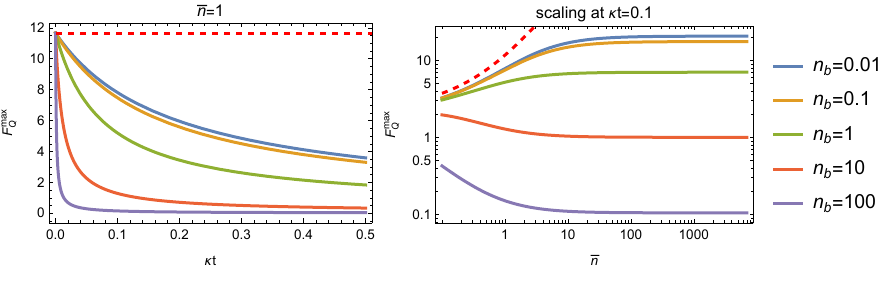}
    \caption{\textbf{Noisy displacement sensitivity of pure Gaussian states.} Left: Sensitivity calculated from Eq.~\eqref{eq:FQNosieDisp} (solid lines) for an initially pure Gaussian state with $\overline{n}=1$ and different thermal baths with average photon number $n_b$ (colors), as a function of the loss $\kappa t$. The red dashed line corresponds to the sensitivity for the ideal unitary evolution. Right: sensitivity scaling at $\kappa t=0.1$.}
    \label{fig:FQDispNoisy}
\end{figure}

If decoherence occurs during interrogation, the QFI, and thus the displacement sensitivity, is typically reduced. The exact dependence on noise and initial state is generally nontrivial, but for Gaussian states under diffusive Gaussian noise, an analytical expression is available.
This is particularly relevant, because it describes the experimentally common scenario of coupling to a thermal bath with average photon number $n_b$ and decay rate $\kappa$ (see Appendix \ref{sec:thermalnoisearbitrary}). For an initially pure Gaussian state, we obtain the maximal QFI as a function of the evolution time $t$:
\begin{equation}\label{eq:FQNosieDisp}
    F_Q^{\text{max}}[\RHO, t] = \dfrac{2\left( 1+2\overline{n}+2\sqrt{\overline{n}(1+\overline{n})} \right)}{\left( (1+2n_b)e^{\kappa t/2} - 2 n_b e^{-\kappa t/2} \right)^2 + 4(1-e^{-\kappa t})(1+2n_b)\overline{n}} \;.
\end{equation}
This expression equals the optimal noiseless QFI Eq.~(\ref{eq:QFImaxpureupper}) multiplied by a reduction factor not exceeding unity for $t,\kappa,n_b \geq 0$. Because this factor depends linearly on $\bar{n}$, it suppresses the ideal linear scaling with $\bar{n}$ and drives the QFI to saturate at a constant value in the large-$\bar{n}$ limit, with the onset of this saturation determined by $n_b$ and $\kappa t$, see Fig.~\ref{fig:FQDispNoisy}.

\clearpage
\newpage

\subsection{Method of moments}

Having established the full potential for displacement sensing, we are now interested in the sensitivity that can be achieved by the convenient estimation procedure that is given by the method of moments for specific measurement settings.

\subsubsection{Homodyne measurements} \label{sec:dispmomentsquadr}
A homodyne measurement can be described in terms of the quadrature vector $\bqq=(\xx,\pp)^T$ and measurement direction $\bm{w}=(\sin\varepsilon,\cos\varepsilon)^T$, giving rise to the general quadrature observable
\begin{align}\label{eq:quadob}
    \qq(\varepsilon) =\xx\sin\varepsilon + \pp\cos \varepsilon= \bqq^T \bm{w}.
\end{align}
The sensitivity that can be achieved for an estimation of $\theta$ using the method of moments based on measurements of the average value of $\qq(\varepsilon)$, according to Eq.~(\ref{eq:chiunitary}), is then given by 
\begin{align}\label{eq:chihom}
\chi^{-2}[\RHO,\qq(\phi), \qq(\varepsilon)]=\frac{\abs{\langle[\qq(\varepsilon),\qq(\phi)]\rangle_{\RHO}}^2}{\var{\qq(\varepsilon)}{\RHO}}.
\end{align}
In the numerator appears the commutator term
\begin{align}\label{eq:chihomnum}
    \langle[\qq(\varepsilon),\qq(\phi)]\rangle_{\RHO}=i\bm w^T\Omega\bm u=i\sin(\varepsilon-\phi).
\end{align}
The denominator of Eq.~(\ref{eq:chihom}) is given by the variance
\begin{align}
    \var{\qq(\varepsilon)}{\RHO}=\bm{w}^T\Gam[\RHO,\bqq]\bm{w}.
\end{align}
We thus obtain
\begin{align}\label{eq:homCHIdef}
    \chi^{-2}[\RHO,\qq(\phi), \qq(\varepsilon)] = \frac{\sin^2(\varepsilon-\phi)}{\bm{w}^T\Gam[\RHO,\bqq]\bm{w}} \;.
\end{align}
For the states we consider, this quantity can be found in Table~\ref{tab:displacementCHI}. Using Eqs.~(\ref{eq:disptransformation}), it can be easily verified that the above expression~(\ref{eq:chihom}) coincides with the corresponding classical Fisher information~(\ref{eq:GaussianCFI}). Therefore, the method of moments extracts the full information about the displacement already from the average value of the quadrature observable.

\begin{table}[h!]
    \centering
    \begin{tabular}{|c|c|}
    \hline
    \textbf{Quantum state} \bm{$\RHO$} & \bm{$\chi^{-2}[\RHO,\qq(\phi), \qq(\varepsilon)]$} \\\hline  
    Coherent $\ket{\alpha}$ & $2 \sin^2(\varepsilon-\phi)$ \\\hline
    Gaussian state & $\dfrac{2 \sin^2(\varepsilon-\phi)}{1 + 2\overline{n} + 2 \sqrt{\overline{n}(1+\overline{n})-n_T(1+n_T)}\cos{\gamma + 2\varepsilon} }$ \\\hline
    Fock $\ket{n}$ & $\dfrac{2\sin^2(\varepsilon-\phi)}{1+2n}$ \\\hline
    Fock superposition $(\ket{m}+e^{i\gamma}\ket{n})/\sqrt{2}$, $n>m$ & $\dfrac{2\sin^2(\varepsilon-\phi)}{n+m+1 - \sqrt{n(n-1)} \cos(\gamma+2\varepsilon) \delta_{n,m+2} - (m+1)\sin^2(\gamma+\varepsilon) \delta_{n,m+1}}$  \\ \hline
    Cat  $\scN(\ket{\alpha}+e^{i\gamma}\ket{-\alpha})$, $\alpha\in\mathbb{C}$, $\abs{\alpha}\gtrsim 2$ & $\dfrac{2 \sin^2(\varepsilon-\phi)}{1 + 4\overline{n} \cos^2{(\varepsilon)} }$ \\\hline
    Compass $\scN(\ket{\alpha}+\ket{-\alpha}+\ket{i\alpha}+\ket{-i\alpha})$  & $\dfrac{2 \sin^2(\varepsilon-\phi)}{1 + 2\overline{n}}$ \\ \hline
    \end{tabular}
    \caption{\textbf{Displacement sensitivity attainable by linear quadrature measurements.} Sensitivity Eq.~\eqref{eq:homCHIdef} with measurement direction specified by the angle $\varepsilon$, see Eq.~\eqref{eq:quadob}.}
    \label{tab:displacementCHI}
\end{table}

As we can see, the sensitivity $\chi^{-2}[\RHO,\qq(\phi), \qq(\varepsilon)]$ in general depends on both the phases $\phi$ and $\varepsilon$. It is therefore relevant to investigate this dependence, and to derive the attainable sensitivities when one or both of these phases can be optimized. To this end, we consider two complementary scenarios that cover experimentally relevant situations. The first consists in having to fix the measurement direction before the generator is known, while the second consists in being able to chose the measurement direction using knowledge about the generator. These two scenarios are analysed in detail below.

\vspace{3mm}
\textit{1) Measurement chosen before the generator is known.---}If the measurement direction $\varepsilon$ must be chosen before the phase $\phi$ of the generator is known, the homodyne sensitivity is directly given by Eq.~\eqref{eq:homCHIdef} with $\varepsilon$ fixed. Suitable figures of merit to gauge the expected precision in the best-case, worst-case and average scenario are then the maximum, minimum and average sensitivity over $\phi$, for a specific choice of $\varepsilon$. We obtain 
\begin{align}\label{eq:chimaxhomfix}
    (\chi^{-2})^{\max}_{\mathrm{hom,fix}}[\RHO,\qq(\varepsilon)]=\max_{\phi}\chi^{-2}[\RHO,\qq(\phi), \qq(\varepsilon)] = \frac{1}{\bm{w}^T\Gam[\RHO,\bqq]\bm{w}} \;,
\end{align}
\begin{align}\label{eq:chiminhomfix}
    (\chi^{-2})^{\min}_{\mathrm{hom,fix}}[\RHO, \qq(\varepsilon)]=\min_{\phi}\chi^{-2}[\RHO,\qq(\phi), \qq(\varepsilon)] = 0 \;,
\end{align}
and
\begin{align}
    (\chi^{-2})^{\mathrm{avg}}_{\mathrm{hom,fix}}[\RHO,\qq(\varepsilon)]=\frac{1}{2\pi} \int_0^{2\pi} \dd\phi\, \chi^{-2}[\RHO,\qq(\phi), \qq(\varepsilon)]=\frac{1}{2}\frac{1}{\bm{w}^T\Gam[\RHO,\bqq]\bm{w}} \;.
\end{align}
In summary, these expressions can be seen as proportional to $(\bm{w}^T\Gam[\RHO,\bqq]\bm{w})^{-1}$. The maximum Eq.~\eqref{eq:chimaxhomfix} is achieved by choosing the displacement direction $\bm u$ orthogonal to the measurement direction $\bm w$. In contrast, if the two directions coincide, the sensitivity yields zero, Eq.~(\ref{eq:chiminhomfix}). 

In order to evaluate the best possible overall sensitivity, we maximize in a second step the above quantities as a function of $\varepsilon$, which determines the measurement direction. The result is determined by the minimum eigenvalue of the covariance matrix $\Gam[\RHO,\bqq]$:
\begin{align}\label{eq:chi2maxhommax}
    (\chi^{-2})^{\max}_{\mathrm{hom,max}}[\RHO]:=\max_{\varepsilon}(\chi^{-2})^{\max}_{\mathrm{hom,fix}}[\RHO, \qq(\varepsilon)]=\frac{1}{\lambda_{\min}(\Gam[\RHO,\bqq])} \;,
\end{align}
and analogously for $\max_{\varepsilon}(\chi^{-2})^{\min}_{\mathrm{hom,fix}}$ and $\max_{\varepsilon}(\chi^{-2})^{\mathrm{avg}}_{\mathrm{hom,fix}}$ with prefactors $0$ and $1/2$, respectively. Since this maximum over $\varepsilon$ is achieved by choosing $\bm{w}$ along the direction of minimal phase-space variance, this means that, if $\phi$ is unknown, the above figures of merit are maximized by the measurement with the smallest variance. Similarly, we could easily obtain the minimum or average over $\varepsilon$. The evaluation of $(\chi^{-2})^{\max}_{\mathrm{hom,fix}}[\RHO,\qq(\varepsilon)]$ and $(\chi^{-2})^{\max}_{\mathrm{hom,max}}[\RHO]$ for a selection of relevant families of states can be found in Table~\ref{tab:displacementCHImaxminavghomfix}.

\begin{table}[h!]
    \centering
    \begin{tabular}{|c|c|c|}
    \hline
    \textbf{Quantum state} \bm{$\RHO$} & \bm{$(\chi^{-2})^{\max}_{\mathrm{hom,fix}}[\RHO,\qq(\varepsilon)]$} & \bm{$(\chi^{-2})^{\max}_{\mathrm{hom,max}}[\RHO]$}  \\\hline  
    Coherent $\ket{\alpha}$ & $2$ &  $2$  \\\hline
    Gaussian state & $\dfrac{2}{1 + 2\overline{n} + 2 \sqrt{\overline{n}(1+\overline{n})-n_T(1+n_T)}\cos{\gamma + 2\varepsilon} }$ &  $\dfrac{2(1+2\overline{n}+2\sqrt{\overline{n}(\overline{n}+1)-n_T(1+n_T)})}{(2 n_T + 1)^2}$ \\\hline
    Fock $\ket{n}$ & $\dfrac{2}{2n+1}$ & $\dfrac{2}{2n+1}$ \\\hline
    Fock superposition    $(\ket{0}+e^{i\gamma}\ket{n})/\sqrt{2}$   & $\begin{cases}
        \frac{1}{1-\frac{1}{2}\sin^2(\gamma+\varepsilon)} \qquad& n=1\\
        \frac{1}{\frac{3}{2}-\frac{1}{\sqrt{2}}\cos(\gamma+2\varepsilon)} & n=2\\
        \frac{2}{1+n} & \text{otherwise}\\
    \end{cases}$ & $\begin{cases}
        2 \qquad& n=1\\
        2 \left(3+\sqrt{2}\right)/7 & n=2\\
       \frac{2}{1+n} & \text{otherwise}\\
    \end{cases}$ \\ \hline    
    Cat  $\scN(\ket{\alpha}+e^{i\gamma}\ket{-\alpha})$, $\alpha\in\mathbb{C}$, $\abs{\alpha}\gtrsim 2$ & $\dfrac{2}{1+4 \overline{n}^2\cos^2 \varepsilon}$ & $2$ \\\hline
    Compass $\scN(\ket{\alpha}+\ket{-\alpha}+\ket{i\alpha}+\ket{-i\alpha})$ & $\dfrac{2}{1+2\overline{n}}$ & $\dfrac{2}{1+2\overline{n}}$  \\ \hline
    \end{tabular}
    \caption{\textbf{Displacement sensitivity attainable by linear quadrature measurements.} Sensitivities computed according to Eqs.~(\ref{eq:chimaxhomfix},\ref{eq:chi2maxhommax}), corresponding to cases where the measurement direction $\varepsilon$ is chosen before the generator is known.}
    \label{tab:displacementCHImaxminavghomfix}
\end{table}

\vspace{3mm}
\textit{2) Measurement chosen after the generator is known.---}If the phase of the homodyne measurement can be chosen after the phase of the displacement is known, then it is natural to choose the $\varepsilon$ that maximizes the sensitivity for a fixed $\phi$. For this reason, we use Eq.~\eqref{eq:maxchi} to introduce the quantity
\begin{align}
    \chi^{-2}_{\mathrm{hom}}[\RHO,\qq(\phi)]:=\max_{\varepsilon}\chi^{-2}[\RHO,\qq(\phi), \qq(\varepsilon)]&=C^T[\RHO,\qq(\phi),\bqq]\Gam^{-1}[\RHO,\bqq]C[\RHO,\qq(\phi),\bqq]\notag\\
    &=\frac{\bm{u}^T \Gam[\RHO,\bqq] \bm{u}}{\det\Gam[\RHO,\bqq]} \;, \label{eq:homCHIcase2}
\end{align}
where we followed the same steps as in the derivation of Eq.~(\ref{eq:QFIGauss}). Notice that the commutator vector was introduced in Eq.~(\ref{eq:comvec}), and is here given by $C[\RHO,\qq(\phi),\bqq]=(\partial/\partial\theta) \langle\hat{\bm{r}}\rangle_{\RHO}=(\cos\phi,-\sin\phi)^T=\Omega\bm{u}$, see Eq.~\eqref{eq:derivVectR}. Using the measurement direction $\bm{w}$, we may express Eq.~(\ref{eq:chihomnum}) as $(C^T[\RHO,\qq(\phi),\bqq]\bm{w})^2=\sin(\varepsilon-\phi)^2$.

The result Eq.~(\ref{eq:homCHIcase2}) is the analytical optimization of Eq.~\eqref{eq:homCHIdef} over $\varepsilon$, where we have to keep in mind that $\bm{w}$ also depends on $\varepsilon$. The optimal quadrature direction is given, according to Eq.~(\ref{eq:maxc}), by the phase
\begin{align}\label{eq:optepsilon}
    \varepsilon&=-\arctan\left(\frac{\var{\pp}{\RHO}\cos\phi+ \cov[\xx,\pp]_{\RHO}\sin\phi}{\var{\xx}{\RHO}\sin\phi+ \cov[\xx,\pp]_{\RHO}\cos\phi}\right).
\end{align}
Note that the optimal homodyne sensitivity coincides with the QFI of a Gaussian state Eq.~(\ref{eq:QFIGauss}), namely, if $\RHO$ is a Gaussian state we have
\begin{align}\label{eq:homoptimalGauss}
    \chi^{-2}_{\mathrm{hom}}[\RHO,\qq(\phi)]=F_Q[\RHO,\qq(\phi)] \;.
\end{align}
This demonstrates that suitably chosen homodyne measurements are optimal for estimating the amplitude of a displacement with arbitrary pure or mixed Gaussian states. We remark that in this scenario, the optimal homodyne measurement direction, defined by $\varepsilon$, Eq.~(\ref{eq:optepsilon}), is not necessarily orthogonal to the generator, which is defined by $\phi$. This is because the fluctuations of the state $\RHO$ in the direction of the measurement have significant impact on the sensitivity, as they appear in the denominator of Eq.~(\ref{eq:homCHIdef}). The optimal direction Eq.~(\ref{eq:optepsilon}) is therefore obtained as a trade-off between the projection of the displacement direction and the fluctuations of the state $\RHO$.

Taking now into account that $\phi$ might be unknown at the time of the state preparation, we can now introduce the maximal, minimal and average sensitivity that can be achieved by an arbitrary state after performing an optimal homodyne measurement. Following closely our previous approach, we have
\begin{align}\label{eq:chi2maxhom}
    (\chi^{-2})^{\max}_{\mathrm{hom}}[\RHO]&=\max_{\phi}\chi^{-2}_{\mathrm{hom}}[\RHO,\qq(\phi)]=\frac{\lambda_{\max}(\Gam[\RHO,\bqq])}{\det\Gam[\RHO,\bqq]}=\frac{1}{\lambda_{\min}(\Gam[\RHO,\bqq])} \;,
\end{align}
\begin{align}\label{eq:chi2minhom}
(\chi^{-2})^{\min}_{\mathrm{hom}}[\RHO]&=\min_{\phi}\chi^{-2}_{\mathrm{hom}}[\RHO,\qq(\phi)]=\frac{\lambda_{\min}(\Gam[\RHO,\bqq])}{\det\Gam[\RHO,\bqq]}=\frac{1}{\lambda_{\max}(\Gam[\RHO,\bqq])} \;,
\end{align}
and
\begin{align}\label{eq:chi2avghom}
(\chi^{-2})^{\mathrm{avg}}_{\mathrm{hom}}[\RHO]&=\frac{1}{2\pi} \int_0^{2\pi} \dd\phi\, \chi^{-2}_{\mathrm{hom}}[\RHO,\qq(\phi)]
    =\frac{\frac{1}{2}\left(\var{\xx}{\RHO}+\var{\pp}{\RHO}\right)}{\var{\xx}{\RHO}\var{\pp}{\RHO}-\cov[\xx,\pp]_{\RHO}^2} \;,
\end{align}
where we have used that the determinant of a $2\times 2$ matrix $A$ is given by $\det A=\lambda_{\min}(A)\lambda_{\max}(A).$ Notice that both quantities $(\chi^{-2})^{\max}_{\mathrm{hom}}[\RHO]$ and $(\chi^{-2})^{\max}_{\mathrm{hom,max}}[\RHO]$ are obtained by maximizing the homodyne sensitivity~(\ref{eq:homCHIdef}) over $\varepsilon$ and $\phi$ but in different order. The above result demonstrates that the sensitivity does not depend on the order of this maximization, \ie Eq.~(\ref{eq:chi2maxhom}) coincides with Eq.~(\ref{eq:chi2maxhommax}). Furthermore, as a consequence of the general result~(\ref{eq:homoptimalGauss}), for Gaussian states the three expressions above coincide with the corresponding Eqs.~(\ref{eq:qfigaussmax}), (\ref{eq:qfigaussmin}), and~(\ref{eq:qfigaussavg}) that were obtained from the QFI. The evaluation of the above quantities for the states of interest can be found in Table~\ref{tab:displacementCHImaxminavg}.

The most relevant quantity above, $(\chi^{-2})^{\max}_{\mathrm{hom}}[\RHO]$, expresses the maximal sensitivity that can be obtained from a quantum state $\RHO$ under displacements with homodyne measurements. The maximization over $\phi$ describes the optimal orientation of the displacement with respect to the fluctuations of the state. This is achieved by choosing $\phi$ such that $\bm u$ coincides with the maximal eigenvector of $\Gam[\RHO,\bqq]$. In the case of a squeezed state, for instance, this means that the displacement is realized in the direction of the squeezed quadrature and, therefore, the generator must coincide with the anti-squeezed quadrature. The optimization over $\varepsilon$, Eq.~(\ref{eq:optepsilon}), assures that the homodyne measurement is realized with an optimal orientation. It is interesting to note that if the displacement direction $\bm u$ coincides with an eigenvector of $\Gam[\RHO,\bqq]$, as is the case for the maximal sensitivity Eq.~(\ref{eq:chi2maxhom}), the optimal homodyne direction $\bm w$ is orthogonal to the displacement direction $\bm u$. To see this, note that from Eq.~(\ref{eq:maxc}) follows that the optimal direction $\bm w=\zeta \Gam^{-1}[\RHO,\bqq]\Omega\bm u_i=\zeta'\Omega\Gam[\RHO,\bqq]\bm u_i=\zeta''\Omega\bm u_i$ is indeed orthogonal to $\bm u_i$, where $\Gam[\RHO,\bqq]\bm u_i=\lambda_i(\Gam[\RHO,\bqq])\bm u_i$, $\zeta''=\lambda_i(\Gam[\RHO,\bqq])\zeta'$, $\zeta'=\zeta/(\det\Gam[\RHO,\bqq])$ and $\zeta$ are irrelevant real constants, and we used that $\Gam[\RHO,\bqq]$ is a $2\times 2$ matrix.

\begin{table}[h!]
    \centering
    \begin{tabular}{|c|c|c|c|}
    \hline
    \textbf{Quantum state} \bm{$\RHO$} & \bm{$(\chi^{-2})^{\max}_{\mathrm{hom}}[\RHO]$} & \bm{$(\chi^{-2})^{\min}_{\mathrm{hom}}[\RHO]$} & \bm{$(\chi^{-2})^{\mathrm{avg}}_{\mathrm{hom}}[\RHO]$} \\\hline  
    Coherent $\ket{\alpha}$ & $2$ & $2$ & $2$  \\\hline
    Gaussian state & $\dfrac{2(1+2\overline{n}+2\sqrt{\overline{n}(\overline{n}+1)-n_T(1+n_T)})}{(2 n_T + 1)^2}$ & $\dfrac{2(1+2\overline{n}-2\sqrt{\overline{n}(\overline{n}+1)-n_T(1+n_T)})}{(2 n_T + 1)^2}$ & $\dfrac{2(1+2\overline{n})}{(2 n_T + 1)^2}$  \\\hline
    Fock $\ket{n}$ & $\dfrac{2}{1+2n}$ & $\dfrac{2}{1+2n}$ & $\dfrac{2}{1+2n}$  \\\hline
    \makecell{Fock superposition \\ $(\ket{0}+e^{i\gamma}\ket{n})/\sqrt{2}$}   & $\begin{cases}
        2 \qquad& n=1\\
        2(3+\sqrt{2})/7 &  n=2\\
        2/(1+n) & \text{otherwise}
    \end{cases}$ & $\begin{cases}
        \dfrac{2}{(3+\sqrt{2})} \qquad& n=2\\
        \dfrac{2}{(1+n)} & \text{otherwise}
    \end{cases}$ & $\begin{cases}
        3/2 \qquad& n=1\\
        6/7 &  n=2\\
        2/(1+n) & \text{otherwise}
    \end{cases}$  \\ \hline
    \makecell{Cat  $\scN(\ket{\alpha}+e^{i\gamma}\ket{-\alpha})$ \\$\alpha\in\mathbb{C}$, $\abs{\alpha}\gtrsim2$} & $2$ & $\dfrac{2}{1+4\overline{n}}$ & $\dfrac{2+4\overline{n}}{1+4\overline{n}}$  \\\hline
    \makecell{Compass\\ $\scN(\ket{\alpha}+\ket{-\alpha}+\ket{i\alpha}+\ket{-i\alpha})$}  & $\dfrac{2}{1 + 2\overline{n}}$ & $\dfrac{2}{1 + 2\overline{n}}$ & $\dfrac{2}{1 + 2\overline{n}}$ \\ \hline
    \end{tabular}
    \caption{\textbf{Displacement sensitivity attainable by linear quadrature measurements.} Sensitivities computed according to Eqs.~(\ref{eq:chi2maxhom},\ref{eq:chi2minhom},\ref{eq:chi2avghom}), corresponding to cases where the measurement direction $\varepsilon$ is chosen after the generator is known.}
    \label{tab:displacementCHImaxminavg}
\end{table}

\subsubsection{Photon number measurements}
A measurement of the average photon number leads with $\MM=\nn$ in Eq.~(\ref{eq:chiunitary}) to the sensitivity
\begin{align}
     \chi^{-2}[\RHO,\qq(\phi),\nn] =\frac{(\bm{u}^T\Omega\langle\hat{\bm{r}}\rangle_{\RHO})^2}{\var{\nn}{\RHO}} \;,
\end{align}
where we used that $[\nn, \xx]=-i\pp$ and $[\nn,\pp]=i\xx$.

The maximal photon number sensitivity under displacements is given by
\begin{align}
    (\chi^{-2})^{\max}_{\mathrm{num}}[\RHO] :=\max_{\phi}\chi^{-2}[\RHO,\qq(\phi),\nn] = \frac{|\langle\hat{\bm{r}}\rangle_{\RHO}|^2}{\var{\nn}{\RHO}}= \frac{\langle\xx\rangle_{\RHO}^2+\langle\pp\rangle_{\RHO}^2}{\var{\nn}{\RHO}} \;,
\end{align}
and this is achieved by choosing $\phi$ such that $\bm{u}$ is parallel to $\Omega\langle\hat{\bm{r}}\rangle_{\RHO}$. Whenever $\bm{u}$ is orthogonal to $\Omega\langle\hat{\bm{r}}\rangle_{\RHO}$, the average photon number does not change in the course of the displacement and therefore this observable provides no information about the parameter. Accordingly, the minimal sensitivity in this case is zero:
\begin{align}
   (\chi^{-2})^{\min}_{\mathrm{num}}[\RHO] = \min_{\phi}\chi^{-2}[\RHO,\qq(\phi),\nn]=0 \;.
\end{align}
The average sensitivity is given by
\begin{align}\label{eq:chi2avgnum}
    (\chi^{-2})^{\mathrm{avg}}_{\mathrm{num}}[\RHO] =\frac{1}{2\pi} \int_0^{2\pi} d\phi\chi^{-2}[\RHO,\qq(\phi),\nn]=\frac{1}{2}\frac{\langle\xx\rangle_{\RHO}^2+\langle\pp\rangle_{\RHO}^2}{\var{\nn}{\RHO}} \;.
\end{align}
These results show that the different definitions of sensitivity are proportional to $(\chi^{-2})^{\max}_{\mathrm{num}}$. For the family of states considered here, this quantity can be found in Table~\ref{tab:chi_max_num}.

Importantly, note that for Fock states $\ket{n}$ we have both $\langle\xx\rangle_{\ket{n}}^2+\langle\pp\rangle_{\ket{n}}^2=0$ and $\var{\nn}{\ket{n}}=0$, even though the ratio of these two quantities is finite. In these cases, we compute the sensitivities $\chi^{-2}$ defined above by considering a displaced Fock state, $\hat{D}(\alpha)\ket{n}$, and then take the limit $\alpha\to 0$.

\begin{table}[h!]
    \centering
    \begin{tabular}{|c|c|}
    \hline
    \textbf{Quantum state} \bm{$\RHO$} & \bm{$(\chi^{-2})^{\mathrm{avg}}_{\mathrm{num}}[\RHO]$} \\\hline
    Coherent $\ket{\alpha}$ & 1\\\hline
    Pure Gaussian states & $\dfrac{1+2\overline{n}}{2\overline{n}(1+\overline{n})} $ \\ \hline
    Fock $\ket{n}$ & $\dfrac{1}{2n+1}$ \\\hline
    Fock superposition  $(\ket{0}+e^{i\gamma}\ket{n})/\sqrt{2}$  & $\begin{cases}
        1\qquad &n=1\\
        0 & \mathrm{otherwise}
    \end{cases}$\\ \hline
    Cat  $\scN(\ket{\alpha}+e^{i\gamma}\ket{-\alpha})$, $\alpha\in\mathbb{C}$, $\abs{\alpha}\gtrsim2$ & 0\\\hline
    Compass $\scN(\ket{\alpha}+\ket{-\alpha}+\ket{i\alpha}+\ket{-i\alpha})$ & 0\\ \hline
    \end{tabular}
    \caption{\textbf{Displacement sensitivity attainable by number measurement.} Sensitivities computed according to Eq.~\eqref{eq:chi2avgnum}. For Gaussian states, we report here the result for pure states, the general expression is more complicated and can be computed from Tab.~\ref{tab:expvalN}.}
\label{tab:chi_max_num}
\end{table}

\clearpage
\newpage

\subsubsection{Measurement of higher-order moments}\label{sec:MeasHighMom}
Comparison between the QFI and the sensitivity obtained from the method of moments reveals that the metrological properties of non-Gaussian states cannot be described by linear and second-order quadrature measurements. The precision of moment-based estimation strategies can be improved by augmenting the set of possible measurement observables $\bMM=\{\MM_1,\dots,\MM_m\}^T$ that determines the maximal sensitivity Eq.~(\ref{eq:maxchi}). A systematic way to obtain corrections beyond Gaussian measurements is to consider, in addition to the linear quadrature observables $\xx$ and $\pp$, all higher moments of order $m$, \ie symmetric $m$-fold products of $\xx$ and $\pp$~\cite{gessner_metrological_2019}. This approach leads to nonlinear squeezing parameter of order $m$, defined as
\begin{align}
    \chi_{(m)}^{-2}[\RHO, \qq(\phi)] &= \max_{\MM\in\mathrm{span}(\hat{\bm{Q}}^{(m)})}\chi^{-2}[\RHO, \qq(\phi), \MM] \notag\\
    & = C^T[\RHO,\qq(\phi),\hat{\bm{Q}}^{(m)}]\Gam^{-1}[\RHO,\hat{\bm{Q}}^{(m)}]C[\RHO,\qq(\phi),\hat{\bm{Q}}^{(m)}] \;,
\end{align}
where $\hat{\bm{Q}}^{(m)}=\bigcup_{i=1}^m\bqq^{(m)}$ contains all quadrature moments up to order $m$, \ie
\begin{equation}\label{eq:Qmdef}
\left\{
\begin{aligned}
    \bqq^{(1)}&=\bqq=\{\xx,\pp\}^T,\notag\\
    \bqq^{(2)}&=\{\xx^2,\frac{\xx\pp+\pp\xx}{2},\pp^2\}^T,\\
    \bqq^{(3)}&=\{\xx^3,\frac{\xx^2\pp+\xx\pp\xx+\pp\xx^2}{3},\frac{\xx\pp^2+\pp\xx\pp+\pp^2\xx}{3},\pp^3\}^T,\notag\\
    &\vdots\notag
\end{aligned} \right.
\end{equation}

\textbf{Fock states.} It was shown in~\cite{gessner_metrological_2019}, that the non-Gaussian metrological properties of Fock states can be entirely captured by third-order moments, \ie $\chi_{(3)}^{-2}[\ket{n}, \qq(\phi)]=F_Q[\ket{n},\qq(\phi)]$. In fact, second-order observables are not useful for Fock-state displacement metrology and the optimal observable Eq.~(\ref{eq:maxc}) is a linear combination of only first and third order
\begin{align}
    \MM_{\mathrm{opt}}=\cos(\phi)\left[-(2n+1)\xx+\xx^3+\frac{\xx\pp^2+\pp\xx\pp+\pp^2\xx}{3}\right]-\sin(\phi)\left[-(2n+1)\pp+\pp^3+\frac{\xx^2\pp+\xx\pp\xx+\pp\xx^2}{3}\right] \;.
\end{align}
Using this observable (up to irrelevant normalization factors), the moment-based sensitivity under displacements generated by $\qq(\phi)=\xx\sin\phi+\pp\cos\phi$ indeed achieves the QFI
\begin{align}
    \chi^{-2}[\ket{n},\qq(\phi),\MM_{\mathrm{opt}}]=\chi_{(3)}^{-2}[\ket{n}, \qq(\phi)]=F_Q[\ket{n},\qq(\phi)] = 2(1 + 2n) \;. \label{eq:ThirdOrdMeasFock}
\end{align}
The sensitivity is independent of the direction $\phi$ of the displacement due to the rotational symmetry of Fock states.

\subsubsection{Measurement-after-interaction technique}\label{sec:mai}

As we have seen, metrological properties of non-Gaussian states are typically encoded in high-order moments. This means that, as illustrated in the previous section, measurements beyond linear and second-order quadrature measurements are required to saturate or approach as closely as possible the quantum Cramér-Rao bound. 

Besides directly accessing high-order moments of quadrature operators, another possibility consists in preceding a linear quadrature measurement $\bMM=\bqq$ by a nontrivial dynamical evolution described by applying Hamiltonian $\op{H}$ for a time $t$. Such evolution can be a squeezing operation which leads to a linear Bogoliubov transformation of quadrature operators, see Eq.~(\ref{eq:bogoliubovsqueezing}), but it could also be a nonlinear transformation, \eg, a Kerr evolution. This so-called Measurement After Interaction (MAI) approach was first proposed for spin systems~\cite{EmilyPRL2016,FlorianPRL2016,PhysRevA.94.010102,SamuelPRL2017} and effectively results in a measurement of the operator $e^{i\op{H}t} \bMM e^{-i\op{H}t}$. In the case of a nonlinear transformation, this observable might involve higher-order moments of quadrature operators and thereby effectively provide access to non-Gaussian measurements. On the other hand, in the case of a linear transformation, the interacting evolution may lead to a rescaled quadrature that yields better measurement precision in the presence of detection noise.

\vspace{2mm}
\textbf{Gaussian operations} (\ie linear transformations of the phase-space coordinates). Even though MAI can be applied in a wide range of settings, we focus here on the task of displacement sensing with homodyne measurements and a moment-based estimation scheme. This means that both the generator $\hat{G}=\qq(\phi)$ and the measurement observables $\hat{M}=\qq(\varepsilon)$ are quadratures. We further assume that the phase of the generator and measurement are chosen optimally and thus maximize the moment-based sensitivity $\chi^{-2}[\RHO,\qq(\phi),\qq(\varepsilon)]$, Eq.~(\ref{eq:chihom}), of the state $\RHO$. As was shown in Sec.~\ref{sec:dispmomentsquadr}, in this case, the quadratures that define the generator $\qq(\phi)=\bqq^T\bm u$ and measurement $\qq(\varepsilon)=\bqq^T\bm w$ correspond to the eigenvectors $\bm u$ and $\bm w$ with maximal and minimal eigenvalues of $\Gam[\RHO,\bqq]$, respectively. In particular, these vectors are orthogonal, \ie they satisfy $|\langle [\qq(\phi),\qq(\varepsilon)] \rangle_{\RHO}|^2=1$.

As we will see below, MAI with linear transformation in CV systems is beneficial in the presence of certain noise processes. Detection noise, for instance, can be effectively modeled by adding a stochastic variable to the measurement observable, $\hat{M} = \qq(\varepsilon) + \Delta M$, where $\Delta M$ is a random variable with mean $\langle \Delta M \rangle =0$ and variance $\langle (\Delta M)^2 \rangle =\sigma^2$. With this, the homodyne sensitivity yields
\begin{align}\label{eq:beforeMAI}
\chi^{-2}[\RHO,\qq(\phi),\MM]&=\frac{|\langle [\qq(\phi),\qq(\varepsilon)] \rangle_{\RHO}|^2}{\var{\qq(\varepsilon)}{\RHO}+\sigma^2 }=\frac{1}{\var{\qq(\varepsilon)}{\RHO}+\sigma^2 } .
\end{align}

The MAI technique replaces the measurement observable $\MM$ by $\hat{U}^{\dagger}\MM\hat{U}$. Here we focus on evolutions $\op{U}$ that produce a linear transformation of the quadrature observable $\MM$. Let us first note that displacements can only lead to constant offsets of the measurement observable which has no effect on the sensitivity. Rotations would only modify the angle $\varepsilon$ of the quadrature observable $\qq(\varepsilon)$ which we already consider to be optimal. We therefore focus our attention on squeezing evolutions, 
\begin{align}
    \MM_{\mathrm{MAI}}&=\hat{S}^{\dagger}(\xi)\MM\hat{S}(\xi)=\hat{S}^{\dagger}(\xi)\qq(\varepsilon)\hat{S}(\xi)+\Delta M,
\end{align}
which lead to the sensitivity
\begin{align}
    \chi^{-2}_{\mathrm{MAI}}[\RHO,\qq(\phi),\qq(\varepsilon)+\Delta M]&=\chi^{-2}[\RHO,\qq(\phi),\MM_{\mathrm{MAI}}]=\frac{|\langle [\qq(\phi),\MM_{\mathrm{MAI}}] \rangle_{\RHO}|^2}{\var{\MM_{\mathrm{MAI}}}{\RHO}+\sigma^2 }=\frac{|\bm{u}^T\Omega \mathcal{S}(r,\gamma)\bm{w}|^2}{\bm{w}^T \mathcal{S}^T(r,\gamma)\Gam[\RHO,\bqq]\mathcal{S}(r,\gamma)\bm w+\sigma^2},
\end{align}
where we have used Eqs.~(\ref{eq:squeezingtransformsquadrature}) and~(\ref{eq:chihomnum}) to introduce the symplectic transformation $\mathcal{S}(r,\gamma)$, see Eq.~\eqref{eq:symplecticsqueezing}. 
By choosing $\gamma=-2\varepsilon$, the anti-squeezing direction is aligned with the measurement direction $\bm w$ and we obtain $\mathcal{S}(r,\gamma)\bm{w}=e^{r}\bm{w}$ such that the signal is effectively amplified. This amplification also affects the variance of the measurement quadrature, but has no impact on the classical fluctuations that model detection noise. The MAI sensitivity
\begin{align}
    \chi^{-2}_{\mathrm{MAI}}[\RHO,\qq(\phi),\qq(\varepsilon)+\Delta M]&=\frac{e^{2r}}{e^{2r}\var{\qq(\varepsilon)}{\RHO}+\sigma^2}=\frac{1}{\var{\qq(\varepsilon)}{\RHO}+e^{-2r}\sigma^2},
\end{align}
thus exhibits effectively suppressed detection noise $\sigma$ compared to the standard protocol~(\ref{eq:beforeMAI}). We show in App.~\ref{app:optmai} that the above-mentioned choice for the squeezing direction $\gamma$ indeed maximizes the sensitivity enhancement due to MAI. We further note that in the considered setting, the MAI technique has no impact in the absence of detection noise. This protocol was experimentally implemented in a trapped-ion system in Ref.~\cite{burd_quantum_2019}. The experiment illustrates how linear transformations on Gaussian states may improve the metrological performance in the presence of detection noise.

\vspace{2mm}
\textbf{Non-Gaussian operations} (\ie non-linear transformations of the phase-space coordinates). In the most general scenario, nonlinear transformations can be implemented before measurement. This class of transformations is however extremely broad, making it difficult to formulate precise statements that are valid in full generality.
For this reason, we will just present some broad observations and suggest to consider each case of interest by itself.

For Gaussian states linear quadrature measurements are optimal, therefore, a MAI protocol with either linear or nonlinear transformations before the measurement provides no advantage in the absence of detection noise (high order moments of the quadratures do not provide any additional information beyond first and second moments).

On the other contrary, the situation is different for non-Gaussian states. There, a nonlinear transformation before the linear quadrature measurement effectively give access to measure high-order moments of the quadrature operators. As we have shown in Section~\ref{sec:MeasHighMom}, these allow for better sensitivities. In principle, it would be possible to find a nonlinear transformation that allows for a saturation of the Cramér-Rao bound with linear quadrature measurements, although this is typically difficult to find and implement experimentally.

\subsubsection{Parity operators and Wigner function measurements}
Of significant interest are parity measurements $\hat{\Pi}=(-1)^{\hat{n}}$ or, more generally, displaced parity measurements
\begin{equation}
    \MM(\beta) = \DD(\beta)\PP\DD(\beta)^\dagger \;.
\end{equation}
Given the link between the Wigner function and a parity measurement Eqs.~(\ref{eq:wigner_parity_measurement}) and (\ref{eq:defWignerAlpha}),
we have $\langle\MM(\beta)\rangle_{\RHO}=\pi W(x,p)$, where $\beta=(x+ip)/\sqrt{2}$. On the other hand, for the variance we use the fact that $\langle\MM(\beta)^2\rangle_{\RHO}=1$ to write $\var{\MM(\beta)}{\RHO}= 1-(\pi W(x,p))^2$. For a displaced state $\RHO(\theta)=\hat{D}^{\dagger}(\alpha)\RHO \hat{D}(\alpha)$, where $\alpha=\theta e^{-i\phi}/\sqrt{2}$, we obtain from $\hat{D}(\beta)^\dagger = \hat{D}(-\beta)$ and $\hat{D}(\beta)\hat{D}(\alpha)=e^{(\alpha^*\beta-\alpha\beta^*)/2}\hat{D}(\alpha+\beta)$ that $\langle\MM(\beta)\rangle_{\RHO(\theta)}=\pi W(x-\theta \cos\phi,p+\theta\sin\phi)$. We obtain the sensitivity from the method of moments using Eq.~(\ref{eq:mmoments}):
\begin{align}\label{eq:xiWfull}
    \chi^{-2}[\RHO(\theta),\qq(\phi),\MM(\beta)] = \dfrac{\left\vert \pi \frac{\partial}{\partial \theta} W(x-\theta\cos\phi,p+\theta\sin\phi) \right\vert^2 }{1-(\pi W(x-\theta\cos\phi,p+\theta\sin\phi))^2} \;.
\end{align}
We will generally be interested in the sensitivity at $\theta=0$, \ie
\begin{align}\label{eq:xiW}
    \chi^{-2}[\RHO,\qq(\phi),\MM(\beta)] = \dfrac{\left\vert \pi \frac{\partial}{\partial \theta} \left.W(x-\theta\cos\phi,p+\theta\sin\phi)\right|_{\theta=0} \right\vert^2 }{1-(\pi W(x,p))^2} \;.
\end{align}
Therefore, from the Wigner function of the state under consideration it is possible to compute Eq.~\eqref{eq:xiW} for arbitrary values of $\phi$ and $\beta$. Note that, since the parity operator $\MM(\beta)$ is dichotomic, namely
\begin{equation}
    \langle{\MM(\beta)}\rangle = P_e(\beta) - P_o(\beta) = \pi W(\beta) \;,
\end{equation}
according to Eq.~(\ref{eq:momentsdichotomic}), the moment-based sensitivity from its average value coincides with the classical Fisher information of the full probability for obtaining the results $1$ or $-1$ respectively,
\begin{align}
    P_o(\beta) &= \dfrac{1}{2}(1-\pi W(\beta)), \\
    P_e(\beta) &= \dfrac{1}{2}(1+\pi W(\beta)).
\end{align}

We note that there are points in phase space where Eq.~\eqref{eq:xiW} is indeterminate. The reason for this is because the Wigner function is a bounded function, and therefore both numerator and denominator of Eq.~\eqref{eq:xiW} are zero at its extremal points. In other words, these indeterminate values happen at points in phase space where a parity measurement gives $\pm 1$, such as for Fock states. However, we note that this indeterminate sensitivity is a mathematical artifact, that cannot be observed in experiments. In practice there will always be some finite amount of noise that prevents from observing $\var{\MM(\beta)}{\RHO}=0$, and thus from having a zero denominator in Eq.~\eqref{eq:xiW}. To describe this effect we add to the denominator of Eq.~\eqref{eq:xiW} a small positive number $\varepsilon$ representing measurement noise, namely we write
\begin{align}\label{eq:xiWerror}
    \chi^{-2}[\RHO,\qq(\phi),\MM(\beta)] = \dfrac{\left\vert \pi \frac{\partial}{\partial \theta} W(x-\theta\cos\phi,p+\theta\sin\phi) \right\vert^2 }{1-(\pi W(x,p))^2 + \varepsilon} \;.
\end{align}
In the following plots we investigate how the sensitivity obtained from Wigner function measurements is affected by this noise level $\varepsilon$.

\begin{figure}[h!]
    \centering
    \includegraphics[width=\textwidth]{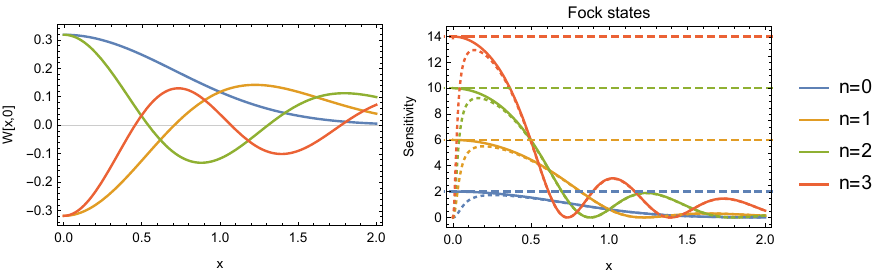}
    \caption{\textbf{Displacement sensitivity of Fock states for parity measurements.} Left panel: slices  for $p=0$ of the Wigner functions for the first four Fock states. Right panel: sensitivity calculated from Eq.~\eqref{eq:xiW} (solid line), compared to the QFI bound $4n+2$ (dashed lines), and to the sensitivity achievable in the presence of noise with $\epsilon=10^{-2}$ (dotted line).}
    \label{fig:FockParityMeas}
\end{figure}

\begin{figure}[h!]
    \centering
    \includegraphics[width=\textwidth]{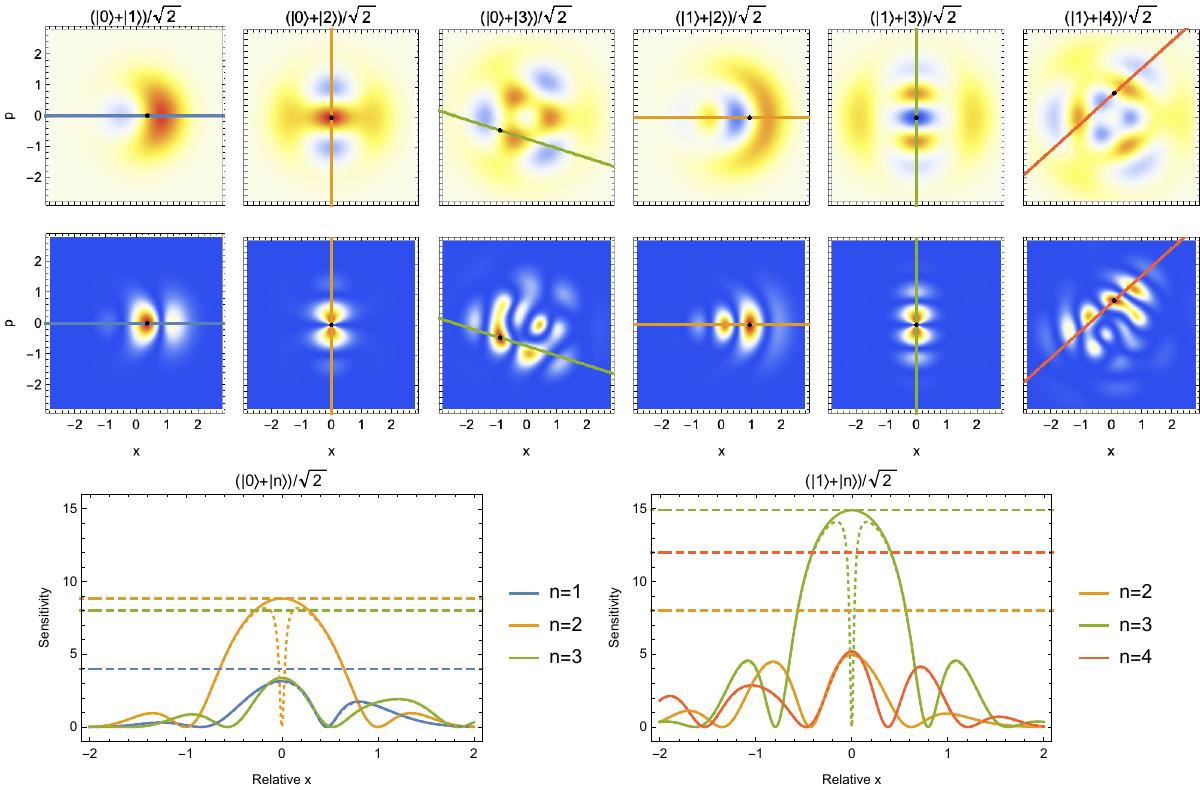}
    \caption{\textbf{Displacement sensitivity of superpositions of two Fock states for parity measurements.} Top row: Wigner functions of different Fock state superpositions. Middle row: sensitivity to displacements obtained through Eq.~\eqref{eq:xiW}. The displacement direction is optimized in order to maximise the sensitivity, and is indicated by coloured lines. The black dot indicate the point of maximum sensitivity (note that states with discrete rotation symmetry have several points with equal maximum sensitivity). Bottom row: sensitivity along the slices indicated by the lines in the two top rows calculated from Eq.~\eqref{eq:xiW} (solid line), compared to the QFI bound $4n+2$ (dashed lines), and to the sensitivity achievable in the presence of noise with $\epsilon=10^{-2}$ (dotted line). Note that these two plots are not symmetric around zero, and that the QFI bounds are saturated for $n$ even or odd respectively, see Fig.~\ref{fig:EvenOddFockSupParityMeas}.}
    \label{fig:FockSupParityMeas}
\end{figure}

\begin{figure}[h!]
    \centering
    \includegraphics[width=\textwidth]{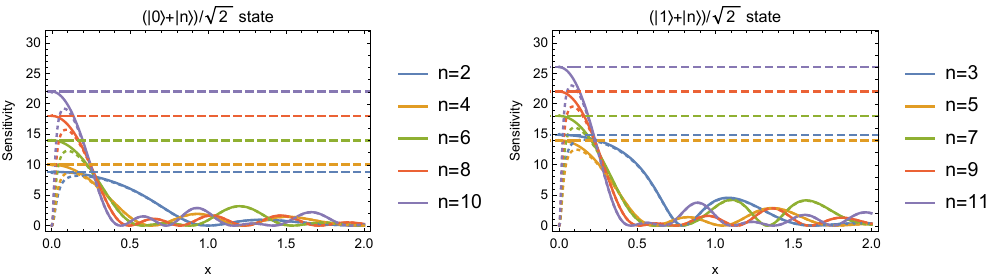}
    \caption{\textbf{Displacement sensitivity of superpositions of two Fock states for parity measurements.} Sensitivity calculated from Eq.~\eqref{eq:xiW} (solid line), compared to the QFI bound $4n+2$ (dashed lines), and to the sensitivity achievable in the presence of noise with $\epsilon=10^{-2}$ (dotted line). Note that the QFI bounds are saturated for $n$ even or odd respectively.}
    \label{fig:EvenOddFockSupParityMeas}
\end{figure}

\begin{figure}[h!]
    \centering
    \includegraphics[width=0.5\textwidth]{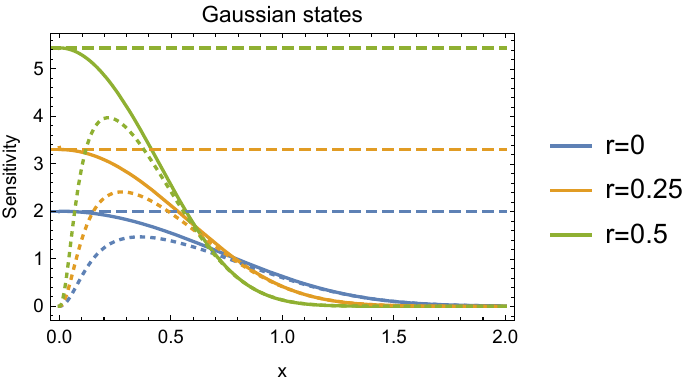}
    \caption{\textbf{Displacement sensitivity of Gaussian states for parity measurements.} Sensitivity calculated from Eq.~\eqref{eq:xiW} (solid line), compared to the QFI bound $2e^{2r}$ (dashed lines), and to the sensitivity achievable in the presence of a thermal population with $n_T=10^{-2}$ (dotted line).}
    \label{fig:GaussianParityMeas}
\end{figure}

\begin{figure}[h!]
    \centering
    \includegraphics[width=\textwidth]{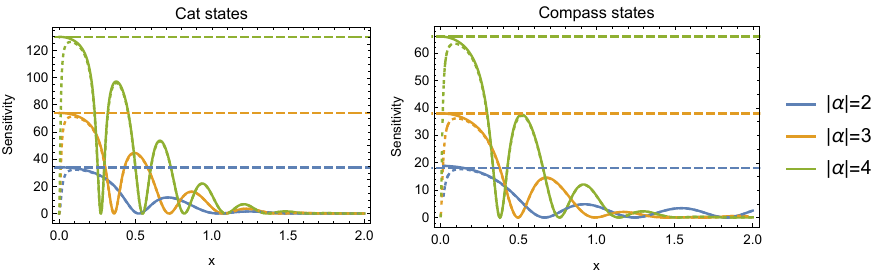}
    \caption{\textbf{Displacement sensitivity of cat and compass states for parity measurements.} Sensitivity calculated from Eq.~\eqref{eq:xiW} (solid line), compared to the QFI bound $2(1+4\abs{\alpha}^2)$ and $2(1+2\abs{\alpha}^2)$ respectively (dashed lines), and to the sensitivity achievable in the presence of noise with $\epsilon=10^{-2}$ (dotted line).}
    \label{fig:CatCompassParityMeas}
\end{figure}

Figure~\ref{fig:FockParityMeas} shows the displacement sensitivity for parity measurements of Fock states, while Figures~\ref{fig:FockSupParityMeas}, \ref{fig:EvenOddFockSupParityMeas} discuss superpositions of two Fock states. Figure~\ref{fig:GaussianParityMeas} is for Gaussian states, while Figure~\ref{fig:CatCompassParityMeas} is for cats and compass states.

\clearpage
\newpage
\subsection{Discussion}

In this section, we have considered the task of sensing displacements along known or unknown directions in phase space. Depending on the situation at hand, appropriate figures of merit may be the maximum, minimum or average sensitivity. The performance of routinely prepared states or of theoretical prototypes is compared in  Tabs.~\ref{tab:dispQFI}, \ref{tab:dispQFImaxminavg}, and put into context with the respective ultimate quantum limits ~(\ref{eq:QFImaxpureupper},\ref{eq:QFImindisp},\ref{QFIavgupper}).

In particular, for a displacement with a known direction an optimally oriented squeezed vacuum state is the only one saturating the upper bound of sensitivity. Neither cat states nor compass states are able to extract the same level of precision from a given average number of photons $\overline{n}$ as a squeezed vacuum state, even though they all show the same scaling with $\overline{n}$ when $\overline{n}$ is large.

When the displacement direction is unknown, the minimum (``worst-case'') sensitivity is optimized by Fock states $\ket{n}$ and compass states, but the latter are significantly more difficult to prepare. The average sensitivity in all directions is optimized by any pure state with vanishing first moments.

We have further provided the sensitivities that become available from an estimation based on error propagation (method of moments) for measurement observables of particular interest, namely linear quadrature measurements (homodyne), Tab.~\ref{tab:displacementCHI}, \ref{tab:displacementCHImaxminavghomfix}, \ref{tab:displacementCHImaxminavg}, average phonon number, Tab.~\ref{tab:chi_max_num}, and Wigner function (displaced parity), Figs.~(\ref{fig:FockParityMeas}-\ref{fig:CatCompassParityMeas}). Linear quadrature measurements are optimal for Gaussian states, while for non-Gaussian states they are in general insufficient and need to be replaced by higher-order observables. For instance, third-order measurements of quadrature operators achieve maximum sensitivity for Fock states; see Eq.~\eqref{eq:ThirdOrdMeasFock}.

Finally, we illustrated the performances of MAI protocols, showing that linear transformations before linear quadrature measurements can provide an advantage only in the presence of detection noise. On the other hand, nonlinear transformations, such as Kerr evolutions, can actually give access to high-order moments of quadrature operators and thus provide an advantage for non-Gaussian states even in the absence of detection noise.

\clearpage
\newpage
\section{Rotation sensing}\label{sec:rot}
\begin{wrapfigure}[16]{r}{0.30\textwidth}
\centering
    \raisebox{0pt}[\dimexpr\height-1.5\baselineskip\relax]{%
        \includegraphics[width=0.28\textwidth]{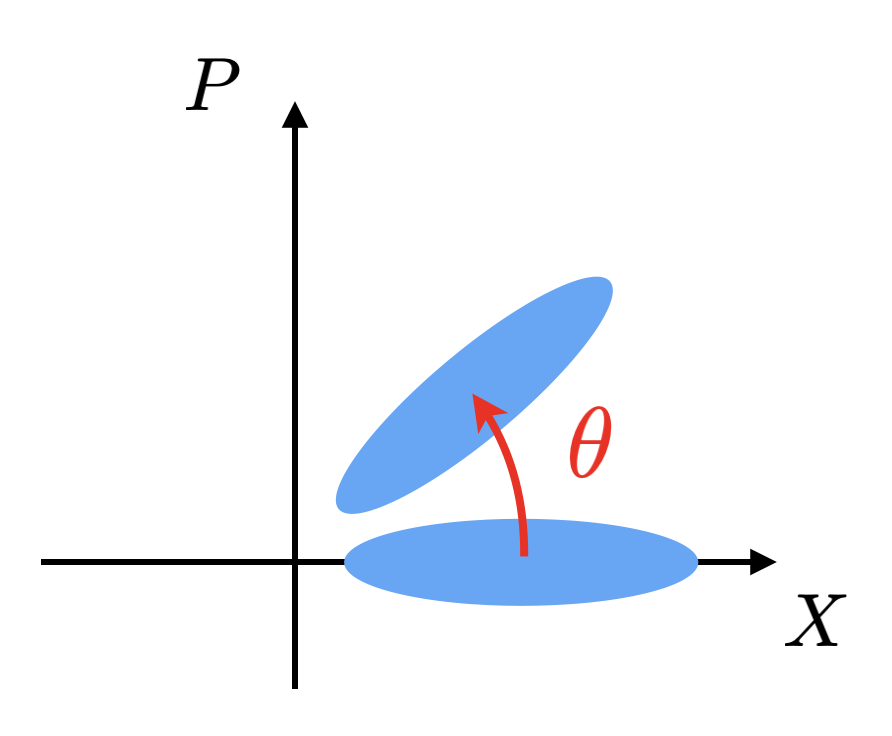}%
    }%
\caption{Phase-space illustration of a rotation. The metrological task considered in this section is to estimate the rotation angle $\theta$.}
\label{fig:RotSketch}
\end{wrapfigure}
Rotation sensing, sometimes referred to simply as phase estimation, is another paradigmatic application of quantum metrology in continuous variable systems. In this case, a phase $\theta$ is imprinted by an evolution generated by the number operator $\nn$, which leads to a rotation of the state in phase space around the origin, see Fig.~\ref{fig:RotSketch}. We discuss the sensitivity properties of a class of relevant states and outline strategies for state preparation and measurement to optimize the metrological precision.

The rotation is described by
\begin{equation}
\hat{R}(\theta) = e^{-i \theta \aad \aa} = e^{-i \theta \nn} \;,
\end{equation}
where $\nn=\aad\aa$ is the generator of the perturbation.

In the following, we compute for different states of interest the sensitivity $\chi^{-2}[\RHO,\GEN,\MM]$ from the method of moments, Eq.~(\ref{eq:chiunitary}), for homodyne and displaced parity measurements and compare to the quantum Fisher information $F_{Q}[\RHO,\GEN]$. Photon number measurements always lead to zero sensitivity here since the measurement observable $\MM=\nn$ commutes with the generator.

\subsection{Quantum Fisher information}
\subsubsection{Pure states}
Since the variance of $\nn$ is independent of $\theta$, for a pure state the QFI is given by
\begin{equation}\label{eq:FQpureRot}
    F_Q[\PSI,\nn] = 4 \var{\nn}{\PSI} \;.
\end{equation}
This also implies that pure states with definite particle number $n$ are insensitive to rotations.

We may wonder if the ultimate quantum limit for phase estimation can be expressed in the form of a state-independent upper limit on Eq.~\eqref{eq:FQpureRot} for states with fixed $\overline{n}$, in analogy to Eq.~(\ref{eq:QFImaxpureupper}) for displacement sensing. However, this is not possible since one can easily construct pure states that have arbitrarily large number fluctuations for any finite~$\overline{n}$~\cite{PhysRevLett.62.2377,Rivas_2012,PhysRevA.88.060101,PhysRevA.92.042115}.
Apparently, such states would enable a sensitivity~(\ref{eq:FQpureRot}) that scales arbitrarily with $\overline{n}$. To put this counter-intuitive phenomenon into the right context, it is crucial to recall that the QFI merely expresses the asymptotic sensitivity that can be achieved after sufficiently many measurement repetitions (see, \eg Ref.~\cite{PhysRevLett.130.260801}). However, the QFI does not provide any information about how many measurements are necessary to achieve this regime. A careful analysis that takes into account the total number of photons leads to a much less advantageous scaling due to finite sampling and the biasedness of the maximum-likelihood estimator~\cite{PhysRevLett.69.2153, PhysRevA.47.1667,PhysRevLett.105.120501,PhysRevA.88.060101,PhysRevLett.108.210404}.

\subsubsection{Arbitrary Gaussian states}
For Gaussian states (both pure and mixed), the QFI for rotation sensing is given by applying the general result~(\ref{eq:QFIgauss}) to rotation sensing, which yields~\cite{monras2013phase,PinelPRA2013}
\begin{align}\label{eq:gaussianQFIrot}
    F_Q[\RHO,\nn]=\frac{4\Tr\left[\Gam[\RHO,\bqq]\right]^2-16\det\Gam[\RHO,\bqq]}
    {4\det\Gam[\RHO,\bqq]+1}+\frac{\langle\hat{\bm{r}}\rangle_{\RHO}^T\Gam[\RHO,\bqq] \langle\hat{\bm{r}}\rangle_{\RHO}}{\det\Gam[\RHO,\bqq]}.
\end{align}
A derivation of this expression can be found in Appendix~\ref{app:gaussrotqfi}.

The SQL, defined by the sensitivity of coherent states, is given by $F_Q[\ket{\alpha},\nn]=4\bar{n}$. Achieving a sensitivity beyond this classical bound, $F_Q[\RHO,\nn]>4\bar{n},$ signals quantum-enhanced performance. Moreover, it certifies the nonclassicality of the state $\RHO$, as captured by the $P$-distribution, since such behavior cannot arise from any convex mixture of coherent states~\cite{rivas_precision_2010}.

\begin{table}[h!]
    \centering
    \begin{tabular}{|c|c|}
    \hline
    \textbf{Quantum state} \bm{$\RHO$} & \bm{$F_Q[\RHO,\nn]$} \\\hline  
    Coherent $\ket{\alpha}$ & $4 \overline{n}$   \\\hline
    Gaussian state & \makecell{$\dfrac{4 \left\lbrace \left[ \left(\Re[\alpha]^2-\Im[\alpha]^2\right)\cos(\gamma) - \Re[\alpha]\Im[\alpha] \sin(\gamma)\right] \sinh (2 r) + \abs{\alpha}^2 \cosh (2
   r) + \frac{(2 n_T+1)^2 \sinh^3(2 r)}{2 (n_T+1)}\right\rbrace}{2(1+2n_T( n_T+1))}$ \\
   \ie\qquad $\dfrac{8(\overline{n}(1+\overline{n})-n_T(1+n_T))}{1+2n_T(1+n_T)}$ \qquad for $|\alpha|=0$
   } \\\hline
    Fock $\ket{n}$ & $0$  \\\hline
    Fock superposition  $(\ket{0}+\ket{n})/\sqrt{2}$ & $n^2 = 4\overline{n}^2$   \\ \hline    
    Cat $\vert\alpha\vert \gtrsim 2$ & $4 \overline{n}$   \\\hline
    Compass & $4 \overline{n}$  \\ \hline
    \end{tabular}
    \caption{\textbf{QFI for a rotation around the origin.} The QFI has been computed using Eq.~\eqref{eq:FQpureRot}, apart from Gaussian states (that can be mixed) for which we used Eq.~\eqref{eq:gaussianQFIrot}.
    }
    \label{tab:rotQFI}
\end{table}

Within the family of Gaussian states the scaling of the variance is limited for fixed $\overline{n}$, which allows us to identify the optimal Gaussian strategy. The optimal QFI for Gaussian states $\RHO_G$ is found by focusing on pure states, \ie $n_T=0$. This is true even within the non-convex set of Gaussian states~\cite{matsubara_optimal_2019}. Assuming a fixed average number of photons $\overline{n}$, the optimal strategy is to use all of this energy for the squeezing and dedicate none of it to the displacement. The optimal Gaussian QFI is therefore achieved by squeezed vacuum states and reads~\cite{Monras_PRA06,Olivares_2009,aspachs_phase_2009,PinelPRA2013,matsubara_optimal_2019}
\begin{align}\label{eq:RotQFIupper}
    F_Q[\RHO_G,\nn]\leq F_Q[\PSI_{G,\mathrm{opt}},\nn]=8\overline{n}(\overline{n}+1).
\end{align}

\subsubsection{Noisy rotation sensing}
\begin{figure}[h!]
    \centering
    \includegraphics[width=\textwidth]{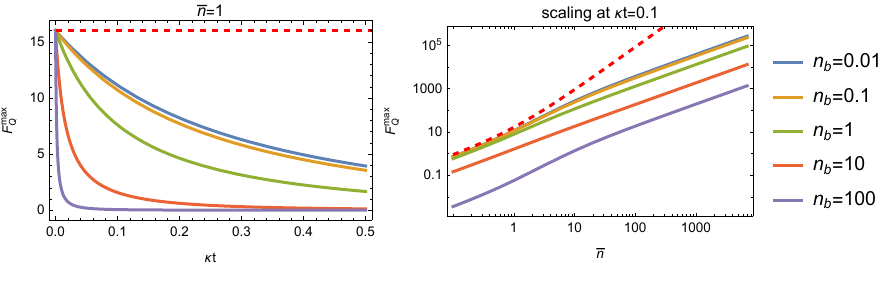}
    \caption{\textbf{Noisy rotation sensitivity of pure Gaussian states.} Left: Sensitivity calculated from Eq.~\eqref{eq:FQNosieRot} (solid lines) for an initially pure Gaussian state with $\overline{n}=1$ and different thermal baths with average photon number $n_b$ (colors), as a function of the loss $\kappa t$. The red dashed line corresponds to the sensitivity for the ideal unitary evolution. Right: sensitivity scaling at $\kappa t=0.1$.}
    \label{fig:FQRotNoisy}
\end{figure}

As in displacement sensing (Sec.~\ref{sec:dispsensnoise}), decoherence also degrades the QFI for rotation sensing. For an initially pure Gaussian state subject to diffusive Gaussian noise from coupling to a thermal bath with average photon number $n_b$ and decay rate $\kappa$, the maximal QFI takes the form (see Appendix~\ref{app:thermalrotsens}):
\begin{equation}\label{eq:FQNosieRot}
    F_Q^{\text{max}}[\RHO, t] = \dfrac{8\overline{n}(1+\overline{n})}{2 n_b^2 - 2 e^{\kappa t} n_b(1+2n_b) + e^{2\kappa t}(1+2n_b(1+n_b)) + 2(e^{\kappa t}-1)(1+2n_b) \overline{n}} \;.
\end{equation}
Again, this equals the maximal noiseless Gaussian QFI~(\ref{eq:RotQFIupper}) multiplied by a suppression factor $\leq 1$. Since this factor scales linearly with $\bar{n}$, the asymptotic scaling is reduced from $\bar{n}^2$ to $\bar{n}$ in the large-$\bar{n}$ limit, with the onset of this regime set by $n_b$ and $\kappa t$, see Fig.~\ref{fig:FQRotNoisy}.

\clearpage
\newpage

\subsection{Method of moments}
\subsubsection{Homodyne measurements}
We consider the sensitivity offered by the method of moments, Eq.~(\ref{eq:chiunitary}), for the quadrature observable Eq.~(\ref{eq:quadob}). We obtain
\begin{align}\label{eq:chi2rothom}
    \chi^{-2}[\RHO,\nn,\qq(\varepsilon)]=\frac{(\bm{w}^T\Omega\langle\bqq\rangle_{\RHO})^2}{\bm{w}^T\Gam[\RHO,\bqq]\bm{w}},
\end{align}
where we used that $i\langle[\qq(\varepsilon),\nn]\rangle_{\RHO}=\bm{w}^TC[\RHO,\nn,\bqq]$, $\bm{w}=(\sin\varepsilon,\cos\varepsilon)^T$ and $C[\RHO,\nn,\bqq]=\Omega\langle\bqq\rangle_{\RHO}$.

Analogously to the definition of Eq.~(\ref{eq:homCHIcase2}) for displacements, an optimally chosen homodyne measurement thus achieves a sensitivity for rotation sensing of 
\begin{align}\label{eq:homrotmaxmm}
    \chi_{\mathrm{hom}}^{-2}[\RHO,\nn] &= \max_{\varepsilon}\chi^{-2}[\RHO,\nn,\qq(\varepsilon)]\notag\\
    &=\langle\bqq\rangle_{\RHO}^T\Omega^T\Gam^{-1}[\RHO,\bqq]\Omega\langle\bqq\rangle_{\RHO}\notag\\
    &=\frac{\langle\bqq\rangle_{\RHO}^T\Gam[\RHO,\bqq]\langle\bqq\rangle_{\RHO}}{\det \Gam[\RHO,\bqq]}.
\end{align}
The minimum homodyne sensitivity is easily found, since choosing $\bm{w}$ orthogonal to $\Omega \langle\bqq\rangle_{\RHO}$ leads to zero sensitivity. Indeterminate values might occur if at the same time also the variance along direction $\bm{w}$ is zero, but this situation would describe a quadrature eigenstate which has unbounded energy and is thus unphysical. The evaluation of $\chi^{-2}[\RHO,\nn,\qq(\varepsilon)]$ and $\chi_{\mathrm{hom}}^{-2}[\RHO,\nn]$ for a selection of relevant families of states can be found in Table~\ref{tab:rotSENS}.

For Gaussian states, the above expression reproduces exactly the second term of the QFI in Eq.~(\ref{eq:gaussianQFIrot}). The first term vanishes if and only if the covariance matrix $\Gam[\RHO,\bqq]$ is proportional to the identity, \ie if the squeezing parameter $r=0$ vanishes, as can easily confirmed using Eq.~(\ref{eq:covgauss}). This describes the family of displaced thermal states. We therefore conclude that the method of moments based on the expectation value of a suitably chosen quadrature is optimal for rotation sensing with displaced thermal states~\cite{Monras_PRA06,oh_optimal_2019}. 

Note that for non-displaced states, the sensitivity~(\ref{eq:homrotmaxmm}) is zero, whereas the QFI~(\ref{eq:gaussianQFIrot}) is generally not. This is because the average value of the quadrature measurement is zero and conveys no information about the parameter. However, there may still be valuable information contained in the data of a homodyne measurement in these cases, but accessing this information requires analyzing more than the first moment that was considered in this simple approach. An analysis of up to the second order will be presented below in Sec.~\ref{sec:MeasHighMomRot}. The classical Fisher information considers the full counting statistics of the measurement data, \ie it accounts for all moments. It is known that the homodyne Fisher information~(\ref{eq:GaussianCFI}) coincides with the QFI~(\ref{eq:gaussianQFIrot}) for low-temperature squeezed thermal states, including the squeezed vacuum states that maximize the QFI among all Gaussian states, as well as for displaced thermal states as is revealed already from an analysis of first moments~\cite{Monras_PRA06,oh_optimal_2019}. Even if the full counting statistics is taken into consideration, for more general single-mode Gaussian states, optimal rotation sensing requires measurements of nonlinear observables to reach the QFI at arbitrary temperatures. Specifically, rotation sensing with arbitrary squeezed thermal states requires measurements of the observable $\hat{x}\hat{p}+\hat{p}\hat{x}$ in order to saturate the QFI~\cite{oh_optimal_2019}.

The above result~(\ref{eq:homrotmaxmm}) further shows that the maximal homodyne rotation sensitivity of any state is achieved when the vector of first moments (up to normalization) coincides with the maximal eigenvector of the covariance matrix. This may be expressed, separating the first-moment vector into modulo and direction, $\langle\bqq\rangle_{\RHO}=\bm{v}|\langle\bqq\rangle_{\RHO}|$, as the ``aligned'' homodyne sensitivity
\begin{align}\label{eq:chi2alghom}
    (\chi^{-2})^{\text{alg}}_{\mathrm{hom}}[\RHO,\nn] = \max_{\bm{v}}\chi_{\mathrm{hom}}^{-2}[\RHO,\nn]&=\max_{\bm{v}}\frac{|\langle\bqq\rangle_{\RHO}|^2\bm{v}^T\Gam[\RHO,\bqq]\bm{v}}{\det \Gam[\RHO,\bqq]}=\frac{|\langle\bqq\rangle_{\RHO}|^2}{\lambda_{\min}(\Gam[\RHO,\bqq])}.
\end{align}

\begin{table}[h!]
    \centering
    \begin{tabular}{|c|c|c|c|}
    \hline
    \textbf{Quantum state} \bm{$\RHO$} & \bm{$\chi^{-2}[\RHO,\nn,\qq(\varepsilon)]$}  & \bm{$\chi_{\mathrm{hom}}^{-2}[\RHO,\nn]$} & \bm{$(\chi^{-2})^{\mathrm{alg}}_{\mathrm{hom}}[\RHO,\nn]$} \\\hline  
    Coherent $\ket{\alpha}$ & $4(\cos\varepsilon\Re[\alpha]-\sin\varepsilon\Im[\alpha])^2$  & $4 \abs{\alpha}^2$ & $4 \abs{\alpha}^2$ \\\hline
    Gaussian state & $\frac{4(\Re[\alpha]\cos\varepsilon-\Im[\alpha]\sin\varepsilon)^2}{(1+2 n_T)(\cosh(2r)+\cos(\gamma+2\varepsilon)\sinh(2r))}$  & $\frac{4 \left\lbrace \left[ \left(\Re[\alpha]^2-\Im[\alpha]^2\right)\cos\gamma - \Re[\alpha]\Im[\alpha] \sin\gamma\right] \sinh (2 r) + \abs{\alpha}^2 \cosh (2
   r) \right\rbrace}{(1 + 2 n_T)}$ & $\frac{4 e^{2r}|\alpha|^2}{(1+2n_T)}$ \\\hline
    Fock $\ket{n}$ & $0$ & $0$ & $0$\\\hline
    \makecell{Fock superposition \\ $(\ket{0}+\ket{n})/\sqrt{2}$} & $\begin{cases} 1-\dfrac{2}{3+\cos(\gamma+2\varepsilon)} \qquad& n=1\\
        0 & \text{otherwise}
    \end{cases}$ & $\begin{cases} \dfrac{1}{2} \qquad& n=1\\
        0 & \text{otherwise}
    \end{cases}$ & $\begin{cases} 1 \qquad& n=1\\
        0 & \text{otherwise}
    \end{cases}$ \\ \hline
    Cat $\vert\alpha\vert \gtrsim 2$ & $0$ & $0$ & $0$ \\\hline
    Compass & $0$ & $0$ & $0$\\ \hline
    \end{tabular}
    \caption{\textbf{Rotation sensitivity attainable by linear quadrature measurements.} Sensitivities Eqs.~(\ref{eq:chi2rothom},\ref{eq:homrotmaxmm},\ref{eq:chi2alghom}) with measurement direction specified by the angle $\varepsilon$, see Eq.~\eqref{eq:quadob}.}
    \label{tab:rotSENS}
\end{table}

\subsubsection{Measurement of higher-order moments}\label{sec:MeasHighMomRot}
In analogy to the method described in Sec.~\ref{sec:MeasHighMom}, we may extract more information from a given observables if also higher-order moments are taken into account. Here, we focus again on the case of homodyne measurements and consider moments up to the order of $2$, \ie $\hat{\bm{Q}}^{(2)}$; see Eq.~(\ref{eq:Qmdef}). The resulting optimized nonlinear squeezing parameter~(\ref{eq:maxchi}) is given by
\begin{align}
    \chi_{(2)}^{-2}[\RHO, \nn] &= \max_{\MM\in\mathrm{span}(\hat{\bm{Q}}^{(2)})}\chi^{-2}[\RHO, \nn, \MM] \notag\\
    & = C^T[\RHO,\nn,\hat{\bm{Q}}^{(2)}]\Gam^{-1}[\RHO,\hat{\bm{Q}}^{(2)}]C[\RHO,\nn,\hat{\bm{Q}}^{(2)}] \;,
\end{align}
To simplify this expression, we write the matrix $\Gam[\RHO,\hat{\bm{Q}}^{(2)}]$ and the vector $C[\RHO,\nn,\hat{\bm{Q}}^{(2)}]$ in block form:
\begin{align}
    \Gam[\RHO,\hat{\bm{Q}}^{(2)}]=\begin{pmatrix}
        \Gam[\RHO,\bqq] & \mathrm{B}[\RHO,\bqq,\bqq^{(2)}] \\
        \mathrm{B}^T[\RHO,\bqq,\bqq^{(2)}] & \Gam[\RHO,\bqq^{(2)}]
    \end{pmatrix},\quad C[\RHO,\nn,\hat{\bm{Q}}^{(2)}]=\begin{pmatrix}
        C[\RHO,\nn,\bqq]\\
        C[\RHO,\nn,\bqq^{(2)}]
    \end{pmatrix},
\end{align}
where
\begin{align}
    \mathrm{B}[\RHO,\bqq,\bqq^{(2)}]=\begin{pmatrix}
        \cov[\xx,\xx^2]_{\RHO} & \cov[\xx,\frac{1}{2}(\xx\pp+\pp\xx)]_{\RHO} & \cov[\xx,\pp^2]_{\RHO}\\
        \cov[\pp,\xx^2]_{\RHO} & \cov[\pp,\frac{1}{2}(\xx\pp+\pp\xx)]_{\RHO} & \cov[\pp,\pp^2]_{\RHO}
    \end{pmatrix}.
\end{align}
Using block inversion techniques, we obtain
\begin{align}\label{eq:rotmm2}
    \chi_{(2)}^{-2}[\RHO, \nn] &= C^T[\RHO,\nn,\bqq]\Gam^{-1}[\RHO,\bqq]C[\RHO,\nn,\bqq]+D^T[\RHO,\nn,\bqq,\bqq^{(2)}]\Sigma^{-1}[\RHO,\bqq,\bqq^{(2)}]D[\RHO,\nn,\bqq,\bqq^{(2)}],
\end{align}
where
\begin{align}
    \Sigma[\RHO,\bqq,\bqq^{(2)}]=\Gam[\RHO,\bqq^{(2)}]-\mathrm{B}^T[\RHO,\bqq,\bqq^{(2)}]\Gam^{-1}[\RHO,\bqq]\mathrm{B}[\RHO,\bqq,\bqq^{(2)}],
\end{align}
and
\begin{align}
    D[\RHO,\nn,\bqq,\bqq^{(2)}]=C[\RHO,\nn,\bqq^{(2)}]-\mathrm{B}^T[\RHO,\bqq,\bqq^{(2)}]\Gam^{-1}[\RHO,\bqq]C[\RHO,\nn,\bqq].
\end{align}
Note that the first term in Eq.~(\ref{eq:rotmm2}) coincides exactly with the moment-based sensitivity~(\ref{eq:homrotmaxmm}) of linear quadrature measurements. The second term therefore describes the sensitivity improvement that is obtained from the second moments.

When we apply the above technique to Gaussian states, we can express all higher-order moments in terms of first and second moments. If, additionally, we restrict to pure Gaussian states, which satisfy $\Gam^{-1}[\RHO,\bqq]=4\Omega^T\Gam[\RHO,\bqq]\Omega$, we find that
\begin{align}\label{eq:mm2rotopt}
    \chi_{(2)}^{-2}[\PSI_G, \nn]=F_Q[\PSI_G,\nn],
\end{align}
meaning that this measurement is optimal. An optimal measurement observable can be identified using Eq.~(\ref{eq:maxc}). For squeezed vacuum states, this observable is a linear combination of $\hat{x}^2$ and $\hat{p}^2$. Further details are provided in Appendix~\ref{app:rotgauss}. Recall from Eq.~(\ref{eq:homrotmaxmm}) that the first moments provide no information about the rotation of non-displaced states. The above result~(\ref{eq:mm2rotopt}) thus demonstrates that for squeezed vacuum states, all the information about the rotation angle is contained in the variation of the second moments.

\subsubsection{Parity operators and measurements of the Wigner function}


Following the derivation of Eq.~\eqref{eq:xiW}, we have the following sensitivity of displaced parity measurements: 
\begin{align}\label{eq:xiWrot}
    \chi^{-2}[\RHO,\nn,\MM(\beta)] = \dfrac{\left\vert \pi \frac{\partial}{\partial \theta} W(x \cos\theta - p \sin\theta, p \cos\theta + x \sin\theta ) \right\vert^2 }{1-(\pi W(x,p))^2} \;.
\end{align}
Therefore, from the Wigner function of the state under consideration it is possible to compute Eq.~\eqref{eq:xiWrot} for arbitrary values of $\beta$.

\begin{figure}[h!]
    \centering
    \includegraphics[width=0.5\textwidth]{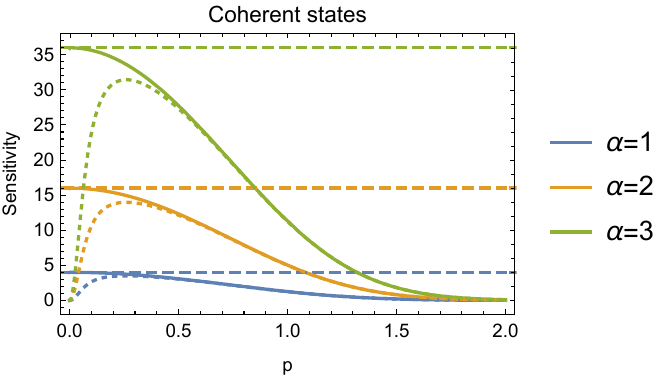}
    \caption{\textbf{Rotation sensitivity of Coherent states for parity measurements.}
    Sensitivity calculated from Eq.~\eqref{eq:xiWrot} (solid line), compared to the QFI bound $4\overline{n}=4|\alpha|^2$ (dashed lines), and to the sensitivity achievable in the presence of noise with $\epsilon=10^{-2}$ (dotted line).
    }
    \label{fig:CohParityMeasRot}
\end{figure}

\begin{figure}[h!]
    \centering
    \includegraphics[width=\textwidth]{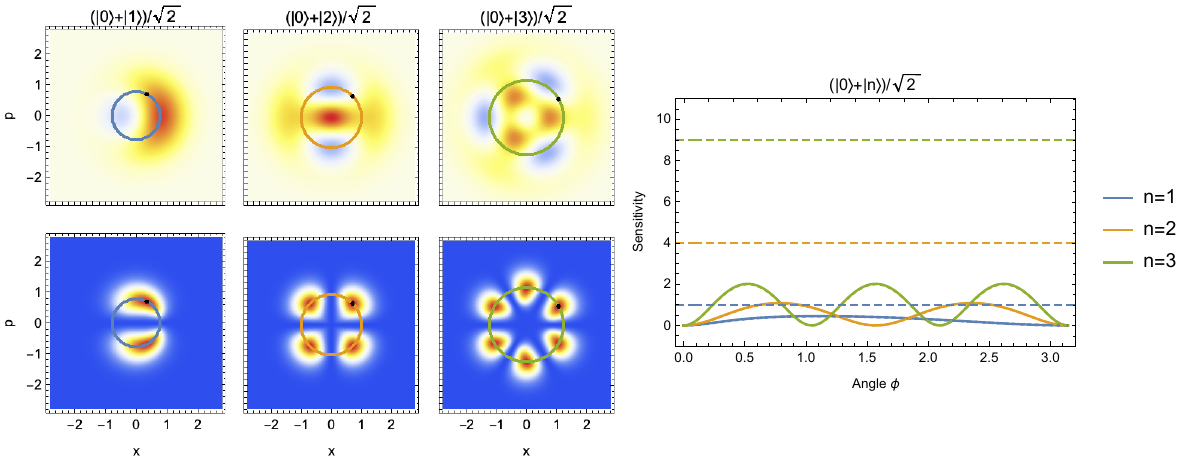}
    \caption{\textbf{Rotation sensitivity of Fock states superpositions for parity measurements.}
    Left panel, top row: Wigner functions of different Fock state superpositions. 
    Left panel, bottom row: sensitivity to rotations obtained through Eq.~\eqref{eq:xiWrot}. The optimal measurement points (\eg black dot) lie on a circumference centered at the origin (coloured circles). 
    Right panel: sensitivity along the circumferences indicated by the circles in the left panel calculated from Eq.~\eqref{eq:xiWrot} (solid line), compared to the QFI bound $n^2=4\overline{n}$ (dashed lines). Note that since these states have discrete rotation symmetry, several points maximize the sensitivity, although the QFI bound is never saturated.}
    \label{fig:Fock0NParityMeasRot}
\end{figure}

\begin{figure}[h!]
    \centering
    \includegraphics[width=0.5\textwidth]{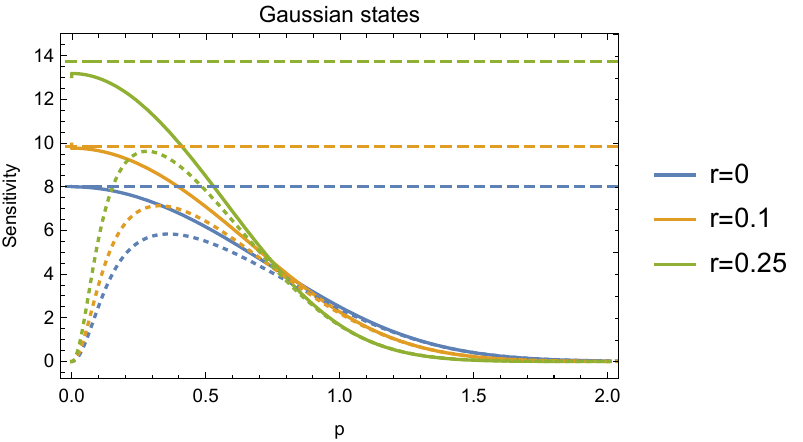}
    \caption{\textbf{Rotation sensitivity of Gaussian states for parity measurements.} The states are $\ket{n_T,\alpha=\sqrt{2},r}$, and the parity measurement is taken at point $(x,p)=(2,p)$. Sensitivity calculated from Eq.~\eqref{eq:xiWrot} for $n_T=0$ (solid line), compared to the QFI bound given in Tab.~\ref{tab:rotQFI} (dashed lines), and to the sensitivity achievable in the presence of a thermal population with $n_T=10^{-2}$ (dotted line).}
    \label{fig:GaussParityMeasRot}
\end{figure}

\begin{figure}[h!]
    \centering
    \includegraphics[width=0.6\textwidth]{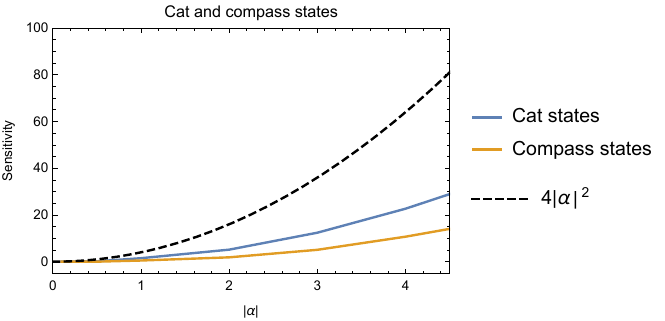}
    \caption{\textbf{Rotation sensitivity of cat and compass states for parity measurements.} Sensitivity calculated from Eq.~\eqref{eq:xiWrot} (solid lines) maximizing over phase space points $(x,p)$, compared to the QFI bound $4\abs{\alpha}^2$ (dashed line).}
    \label{fig:CatCompassParityMeasRot}
\end{figure}

\clearpage

\subsection{Discussion}

In this section, we have considered the task of sensing rotations in phase space. Interestingly, we note that for a fixed number of photons there is no unambiguous state-independent upper bound for the QFI, since pure states with arbitrarily large number fluctuations can be constructed. However, when restricting to Gaussian states we can find the upper bound Eq.~\eqref{eq:RotQFIupper}, which is saturated by squeezed-vacuum states.

The sensitivities of quantum states that are routinely prepared experimentally are compared in Tab.~\ref{tab:rotQFI}. In particular, we find that the sensitivity of a Fock state superposition $(\ket{0}+\ket{n})/\sqrt{2}$ scales quadratically with $\avg{\nn}$, but it does not saturate the upper sensitivity bound of Eq.~\eqref{eq:RotQFIupper}. More complex superposition states such as cat or compass states are not able to achieve similar sensitivity. Actually, for the same average number of photons, their sensitivity for rotations coincides with that of a coherent state.

As discussed for the sensing of displacements, besides computing the sensitivity as given by the QFI, one has also to consider the sensitivity that can be achieved when considering a specific measurement. Having in mind what are the measurements routinely performed, we have calculated the sensitivities achievable by linear quadrature measurements (homodyne), Tab.~\ref{tab:rotSENS}, and Wigner function (displaced parity) measurement, Figs.~(\ref{fig:CohParityMeasRot}-\ref{fig:CatCompassParityMeasRot}). We have found that, generally, these measurements are non-optimal. For Gaussian states, homodyne measurements are optimal in certain regimes, including for squeezed vacuum states and displaced states without squeezing. For squeezed vacuum, comparison between the Fisher information and the method of moments reveals that the crucial information about the parameter is contained in the second moments of the homodyne measurement data.

\clearpage
\newpage

\section{Experimental techniques and applications}

\subsection{Generation of nonclassical states}
\textbf{Massive systems}\\
The preparation of nonclassical CV states has been pioneered in trapped-ion platforms, where the ion's motional state is described by a bosonic mode~\cite{RevModPhys.75.281}. Examples include the preparation of a single trapped ion in coherent states \cite{Meekhof96,kienzler_quantum_2015,alonso_generation_2016}, Fock states \cite{Meekhof96,wolf_motional_2019}, displaced Fock states~\cite{wolf_motional_2019}, squeezed states \cite{Meekhof96,kienzler_quantum_2015,burd_quantum_2019}, squeezed Fock states \cite{ding_quantum_2017}, superposition of two squeezed states \cite{lo_spinmotion_2015}, cat states \cite{monroe_schrodinger_1996,hempel_entanglement-enhanced_2013,kienzler_observation_2016} and superpositions of two Fock states \cite{mccormick_quantum-enhanced_2019}. Besides controlling the motion of a single ion, arrays of trapped ions exhibit normal modes of oscillation that can be prepared collectively in a nonclassical state. Examples include the preparation of squeezed states of motion in a two-dimensional trapped-ion crystal of $\sim 150$ ions \cite{gilmore_quantum-enhanced_2021}. These platforms have been used for the demonstration of sensing protocols based on MAI schemes \cite{burd_quantum_2019,metzner_two-mode_2023}.

A natural extension of these experiments towards more macroscopic systems consists of trapping nanoparticles instead of single atoms. In this case, ground-state cooling becomes especially challenging, as well as the preparation of nonclassical states of motion through light-matter interaction or engineered trapping potentials. The absence of accessible internal degrees of freedom, together with operating in the regime of linear optomechanical interaction, generally limit experiments to the preparation of squeezed thermal states of motion, as it was demonstrated for a levitated nanosphere \cite{rashid_experimental_2016}. Recently, the inevitable nonlinearities in the trapping potential have been exploited for the generation of classical non-Gaussian motional states \cite{setter_characterization_2019,muffato_generation_2025}. To date, in the field of levitodynamics, squeezing below the ground state variance and the preparation of quantum non-Gaussian states still have to be demonstrated.

In the context of solid-state systems, mechanical osccilators such as photonic crystals, nano objects (\eg rods, spheres, disks, toroids), microresonators, cantilevers and membranes have been controlled to exquisite levels using the tools of optomechanics and electromechanics. This allowed for the preparation of their motional degree of freedom in squeezed thermal \cite{Rugar91,pontin_squeezing_2014,bothner_cavity_2020} and vacuum \cite{WollmanSCI15,PirkkalainenPRL15,LecocqPRX15,DelaneyPRL19,youssefi_squeezed_2023} states. As for the case of levitated nanoparticles, challenges in cooling to the ground state and in realizing nonlinear interactions make difficult to prepare quantum non-Gaussian states of motion such as Fock and cat states. A notable exception to this is the use of measurement-based techniques to prepare single-phonon added and subtracted thermal states \cite{enzian_single-phonon_2021,patel_room-temperature_2021}.

A successful approach to circumvent active cooling and the (to first order) linear light-matter interaction consists of coupling high-frequency (gigahertz) acoustic vibration modes in a solid to superconducting qubits. Operating at cryogenic temperatures automatically results in ground state cooling of the acoustic modes, while coupling to the qubit results in a Jaynes-Cummings interaction that is fundamentally nonlinear. Following ideas demonstrated with trapped ions and circuit quantum electrodynamics, recent experiments have demonstrated the preparation of acoustic modes in coherent states, squeezed vacuum states \cite{marti2023}, Fock states or superposition of two Fock states \cite{ChuFockNat18,Satzinger18,wollack_quantum_2022,vonLupke22,macroPRL23} and even cat states \cite{catSCI23}.

\vspace{5mm}
\textbf{Massless systems}\\
Besides massive mechanical systems, CV states also describe massless bosonic modes such as optical and microwave fields.
In optical setups, coherent states are simply obtained as the output of lasers. Besides these, a significant amount of effort has been put in the preparation of squeezed light, motivated by the crucial role played by interferometers (a main example being gravitational wave detectors). Techniques with which this can be achieved involve the interaction of light with nonlinear media, such as atomic ensembles \cite{slusher_observation_1985,mccormick_strong_2007} or solid-state systems \cite{Shelby_generation_86,Wu_generation_86,hirosawa_photon_2005,Vahlbruch_observation_08,Vahlbruch_detection_16}. Alternatively, ponderomotive squeezing of light can be achieved through its interaction with an oscillator, such as a membrane \cite{purdy_strong_2013}, a suspended mirror \cite{aggarwal_room-temperature_2020}, a levitated nanoparticle \cite{magrini_squeezed_2022,militaru_ponderomotive_2022}, or the collective spin of a polarized atomic ensemble \cite{baerentsen_squeezed_2024}.
On the other hand, non-Gaussian states of light are in general challenging to be prepared. A single-photon Fock state $\ket{1}$ can be prepared by deterministic single-photon sources, or by heralding on parametric down-conversion processes. From this elementary state, higher Fock states can be prepared by degenerate parametric down-conversion \cite{yukawa_generating_2013,cooper_experimental_2013}, or through bunching at a beam-splitter \cite{zapletal_experimental_2021}. Optical cat states have been prepared through photon subtraction \cite{ourjoumtsev_generating_2006,wakui_photon_2007,konno_logical_2024} or through homodyne measurement of a Fock state \cite{ourjoumtsev_generation_2007}.

In the microwave regime, the field of circuit and cavity quantum electrodynamics reached control at the single-photon level of electromagnetic resonators and travelling modes. As in the case of trapped ions and of many solid-state mechanical resonators, this level of control is often achieved through the coupling with a two-level system, such as an atom or a superconducting qubit. Experiments demonstrated the preparation of Fock states up to $\ket{100}$ \cite{Bertet_direct_02,varcoe_preparing_2000,hofheinz_generation_2008,deng2023heisenberglimited}, superpositions of two or three Fock states \cite{Hofheinz2009,wang_heisenberg-limited_2019} and cat states \cite{deleglise_reconstruction_2008,Hofheinz2009,kirchmair_observation_2013,pan_realization_2025}.

\clearpage
\newpage

\subsection{Metrology in the presence of noise}

Real-world quantum sensors are inherently open systems that interact with their surrounding environments, leading to decoherence and the gradual loss of quantum properties such as superposition and entanglement~\cite{BreuerPetruccione2006,Serafini2017}. This decoherence often modifies the optimal initial states, measurement strategies, and ultimately the fundamental limits of quantum-enhanced sensitivity~\cite{PhysRevLett.79.3865,Escher2011,Demkowicz2012,Kolodynski2013,PhysRevLett.109.233601,PhysRevLett.119.010403,Sidhu2020Geometric,YoucefPRL2021,YoucefCRP2022}. Exact analytical solutions become elusive in open systems. In this case, several studies have employed channel descriptions to produce upper bounds on the QFI~\cite{Escher2011,Demkowicz2012,Kolodynski2013}. Alternatively, analytically tractable approximations, such as squeezing coefficients, are often still manageable in the presence of noise~\cite{YoucefPRL2021,YoucefCRP2022}.

While many studies of noise have focused on DV systems, revealing that a wide range of noise processes restrict the asymptotic precision scaling to the shot-noise limit~\cite{Demkowicz2012,Kolodynski2013}, significant quantum enhancements may still be present in finite-sized systems and even asymptotically for certain noise types and strengths~\cite{YoucefPRL2021,YoucefCRP2022}. The development of noise-mitigation and error-correction protocols for quantum metrology aims to counter the asymptotic loss of quantum gains~\cite{PhysRevLett.113.250801,PhysRevLett.112.150802,PhysRevLett.129.250503,Shettell2021practical}.

In CV systems, decoherence processes such as photon loss, thermal noise, and amplification can be modeled as Gaussian channels. When combined with Gaussian initial states, these channels permit the use of analytical expressions for the Gaussian QFI. Even in the presence of noise, squeezed vacuum states often remain optimal or near-optimal for metrology, although the achievable enhancement diminishes as decoherence increases~\cite{PinelPRA2013,safranek_optimal_2016,oh_optimal_2019}. A notable non-Gaussian evolution of particular interest is given by the dephasing channel~\cite{PRXQuantum.5.020354}.

\subsection{Gravitational wave detection}

The pursuit of gravitational wave detection in the late 1970s and early 1980s triggered significant advances in quantum-enhanced metrology, notably in the development of squeezed states of light. The challenge of quantum noise in laser interferometry, identified as a major obstracle to achieving the sensitivity required for gravitational wave observatories, motivated early investigations into nonclassical states as a means to surpass the standard quantum limit. Early studies by Hollenhorst~\cite{Hollenhorst1979} and Dodonov~\cite{Dodonov1980} first described the properties of squeezed states, while Caves subsequently demonstrated that injecting squeezed vacuum into the unused port of a Michelson interferometer could mitigate shot noise and enhance sensitivity beyond classical limits~\cite{Caves1980,Caves1981}. These advances inspired extensive research into the potential of squeezed and other nonclassical states for interferometry and gravitational wave detection~\cite{Grishchuk1984,Bondurant1984}. These theoretical studies were experimentally corroborated by Slusher \etal~\cite{slusher_observation_1985} who reported the first observation of squeezed light via four-wave mixing, and Xiao, Wu, and Kimble~\cite{MinPRL1987} who demonstrated squeezed-state generation in an optical parametric oscillator.

Over the ensuing decades, squeezed states became integral to proposals for gravitational wave detectors~\cite{unruh1982quantum,KimblePRD01}. Initial experimental demonstrations of squeezed-light enhancement in prototype interferometers, such as those conducted by Vahlbruch \etal \cite{VahlbruchPRL05} and Chelkowski \etal \cite{Chelkowski05} demonstrated the feasibility of implementing squeezing in full-scale gravitational wave detectors. The the first long-term application of quantum-enhanced metrology in an operational gravitational wave observatory was realized in GEO600~\cite{Grote2013}. Subsequently, the incorporation of squeezed vacuum states into Advanced LIGO has led to significant improvements in sensitivity, increasing its astrophysical reach and enabling the detection of weaker gravitational wave signals that would otherwise remain undetectable \cite{Aasi2013,TsePRL2019}.

\subsection{Quantum-enhanced optical displacement measurements}

Efforts to surpass classical limits in optical displacement and imaging have led to the development of quantum metrology protocols based on multimode nonclassical states of light. Early theoretical analyses showed that single-mode squeezing alone could not overcome the SQL in imaging tasks \cite{fabre_quantum_2000, kolobov_quantum_2000}. This limitation was overcome experimentally by Treps et al., who demonstrated that combining squeezed vacuum in higher-order transverse modes with a coherent fundamental mode enables sub-SQL precision in beam displacement measurements \cite{treps_surpassing_2002}. Extending this approach, a composite beam composed of multiple squeezed modes was shown to reduce pointing noise below the SQL in two transverse directions—the so-called “quantum laser pointer” \cite{treps_quantum_2003}.

These foundational results spurred a broader research program into quantum-enhanced imaging and beam tracking \cite{barnett_ultimate_2003}, including nano-displacement sensing using spatially multimode squeezing \cite{treps_nano-displacement_2004, delaubert_quantum_2008}, quantum-enhanced timing and positioning for clock synchronization \cite{lamine_quantum_2008}, and interferometric timing with squeezed frequency combs \cite{wang_sub-shot-noise_2018, jian_real-time_2012}. Theoretical tools based on multimode Gaussian states and quantum Fisher information were developed to optimize estimation strategies for arbitrary parameters \cite{PinelPRA2012, PinelPRA2013}, leading to experimental demonstrations of quantum-limited distance measurements using spatially structured detection systems \cite{thiel_quantum-limited_2017, cai_quantum_2021}. Optimal observables for small displacement estimation can be implemented via spatial mode-selective detection using multiplane light conversion devices~\cite{labroille_efficient_2014, boucher_spatial_2020}. Remarkably, the same measurements were later shown to be optimal for the problem of resolving two incoherent point sources, revealing a direct quantum route to overcoming the Rayleigh-Abbe limit in far-field passive imaging \cite{PhysRevX.6.031033,TsangReview}.

\subsection{Applications of squeezed states beyond metrology}

Beyond gravitational wave detection and optical precision measurements, squeezed states of light and mechanical systems have found broad applications in other quantum metrology tasks, as well as in quantum computation and quantum communication. 

One example has been the use of squeezed states to exponentially enhance light-matter interactions. By tailoring the quantum fluctuations of electromagnetic fields, it is possible to achieve stronger and more efficient coupling between optical fields and matter. Theoretical proposals and experimental demonstrations have shown that squeezed vacuum states can significantly enhance coupling strengths in cavity quantum electrodynamics (cQED) \cite{Bartkowiak2014}, optomechanical systems \cite{Qin2018}, and trapped ion platforms \cite{Leroux2018,burd_quantum_2019,Burd2021,Burd2024}. This enhancement, in turn, leads to improved quantum sensors that benefit from reduced measurement uncertainty, enabling higher precision in applications such as atomic clocks, magnetometry, and force sensing \cite{Villiers2024,zhang_squeezing-enhanced_2024}. Another striking application of optical squeezing is in the exponentially enhanced dispersive readout of superconducting qubits, where reduced quantum fluctuations in the measurement field enable faster and more accurate qubit state discrimination while reducing backaction \cite{Qin2024}. Moreover, squeezed states of light have been employed to generate spin squeezing, providing a crucial link between quadrature squeezing and entanglement-enhanced quantum sensing \cite{Qin2020,Ma2011}. 

In addition to metrology, squeezed states have recently played a pivotal role in experimental demonstrations of quantum advantage, particularly in photonic quantum computing. One of the most prominent examples is boson sampling, where nonclassical states of light—particularly squeezed vacuum states—serve as input resources for large-scale, high-dimensional interferometers. Recent experimental breakthroughs using squeezed-state-based boson sampling have demonstrated computational tasks that are intractable for classical supercomputers, providing strong evidence of quantum computational advantage \cite{Zhong2020,Zhong2021,Madsen2022}. These results underline the importance of squeezed states in photonic quantum computation and quantum information processing. Furthermore, squeezed states have been leveraged in quantum error correction protocols, where they enable the protection of logical qubits against decoherence. In particular, squeezed cat states — coherent superpositions of squeezed states — have been proposed and demonstrated as a means to enhance fault tolerance in bosonic qubits, extending superposition lifetimes and improving error resilience in quantum processors \cite{LeJeannicPRL18,XiaozhouPRX23,Fluhmann2019,Sivak2023}.

In the context of quantum communication, squeezed states of light play a fundamental role by enabling enhanced information transfer, secure quantum key distribution (QKD), and quantum networking. One of the most prominent applications of squeezed states in quantum communications is in quantum key distribution. Traditional discrete-variable QKD schemes, such as BB84, rely on single-photon states, which can be challenging to generate and detect efficiently. In contrast, continuous-variable QKD (CV-QKD) protocols utilize the quadratures of coherent or squeezed states to encode information, enabling high-speed secure communication with standard telecommunication technology. Early proposals demonstrated that squeezed light could enhance the security and robustness of QKD systems against eavesdropping \cite{HilleryPRA00,Ralph99,RalphPRA00}. More recently, experimental implementations have shown that the use of squeezed states improves the tolerance of CV-QKD protocols to excess noise and loss, extending their practical range in fiber-based and free-space communications \cite{CerfPRA01,Grosshans2003,ScaraniRMP09,Gehring2015}.

Beyond QKD, squeezed states also play a crucial role in quantum teleportation and quantum networking. Quantum teleportation of coherent states was first demonstrated by Furusawa \etal \cite{FurusawaSCI98} using entangled squeezed states to transfer quantum information between distant parties with high fidelity. Subsequent improvements in squeezing levels have enabled near-deterministic teleportation of optical states, with applications in scalable quantum networks and distributed quantum computing \cite{Takeda2013,RevModPhys.77.513}. Furthermore, multimode squeezed states offer a natural route to generate entanglement in continuous-variable networks~\cite{RevModPhys.77.513,Weedbrook,Duan2000,Simon2000,vanLoock2003,Sperling2013,BosonicSqueezing,Qin2019}. They form the basis of continuous-variable cluster states, which constitute a key resource for measurement-based quantum computing as well as for large-scale quantum communication networks~\cite{MenicucciPRL06,Su2007,SuOptLett12,Yokoyama2013,Chen2014,Gerke2015}.
The use of squeezed light in advanced communication protocols, such as continuous-variable quantum secret sharing \cite{LancePRL04} and quantum-enhanced clock synchronization \cite{GiovannettiPRA02}, further highlights the versatility of these nonclassical states.

\clearpage
\newpage

\section{Conclusions}
In summary, we have reviewed the metrological properties of single-mode continuous-variable states, focusing on the quantum Fisher information and its practical approximations based on the method of moments. Gaussian squeezed vacuum states often yield maximal sensitivity when compared to other states with the same average number of photons. For instance, displacement sensing is optimized by squeezed vacuum states when all other parameters (such as the phase of displacement and squeezing) can be controlled. Rotation sensing is also maximized by squeezed vacuum when restricting to Gaussian states, and their sensitivity is unsurpassed by highly nonclassical non-Gaussian states such as Fock, cat, or compass states.

Homodyne measurements are optimal for displacement sensing with Gaussian states, even if only average values (\ie first moments) are considered for the estimation strategy based on the methods of moments. To extract the sensitivity of squeezed vacuum states for rotation sensing, however, second moments of the homodyne measurement data must be considered. Moreover, measurements of displaced parity operators allow us to identify sensitivity bounds directly from the Wigner function, which often turn out to be optimal.

Certain tasks, however, cannot be optimally addressed with only Gaussian resources. This includes the estimation of a displacement with an unknown direction in phase space, whose optimal strategy requires a Fock state. To extract their full sensitivity, information on up to the third moments of homodyne observables is needed.

Our discussion here was limited to the most widely used scenarios, mainly under ideal conditions. For the sake of clarity, our description was limited to the estimation of a single parameter with a single mode. Several of the results presented here for homodyne observables can be generalized to arbitrary Gaussian measurements that include heterodyne and other general-dyne schemes and may account for imperfections~\cite{Serafini2017,oh_optimal_2019}. We limited the detailed discussion to the estimation of a displacement amplitude and of a rotation angle. There are many other parameters of interest, \eg the phase of a displacement, amplitude and phase of a squeezing operation, and even the full characterization of all parameters that define arbitrary Gaussian states (quantum state estimation). Moreover, possibly non-unitary evolutions can be accounted for, \eg in channel estimation; for an overview of results, see~\cite{RevModPhys.90.035006}. More general approaches may incorporate multimode approaches such as ancilla-assisted setups, feedback loops, and the simultaneous estimation of multiple parameter. 

From an experimental perspective, although quantum metrology was initially developed and applied within optical setups, recent advances have enabled the control of a broad range of novel quantum systems, each with unique properties. These systems include microwave photons, motional states of trapped ions, and solid-state oscillators. Notably, the ability to implement quantum metrology protocols on massive systems opens up opportunities to sense previously inaccessible quantities, such as subtle forces, gravity or fundamental decoherence mechanisms. We expect that the results presented in this work will serve to guide future experiments in quantum parameter estimation using continuous variable systems.

\section*{Acknowledgments}
This work is supported by the project PID2023-152724NA-I00, with funding from MCIU/AEI/10.13039/501100011033 and FSE+, by the project CNS2024-154818 with funding by MICIU/AEI /10.13039/501100011033, by the project RYC2021-031094-I, with funding from MCIN/AEI/10.13039/501100011033 and the European Union ‘NextGenerationEU’ PRTR fund, by the project CIPROM/2022/66 with funding by the Generalitat Valenciana, and by the Ministry of Economic Affairs and Digital Transformation of the Spanish Government through the QUANTUM ENIA Project call—QUANTUM SPAIN Project, by the European Union through the Recovery, Transformation and Resilience Plan—NextGenerationEU within the framework of the Digital Spain 2026 Agenda, and by the CSIC Interdisciplinary Thematic Platform (PTI+) on Quantum Technologies (PTI-QTEP+). This work is supported through the project CEX2023-001292-S funded by MCIU/AEI.\\
N.R. acknowledges financial support from the Werner Siemens Foundation. 
M.F. was supported by the Swiss National Science Foundation Ambizione Grant No. 208886, and by The Branco Weiss Fellowship -- Society in Science, administered by the ETH Z\"{u}rich.

\clearpage
\newpage

\bibliography{bibCV.bib}

\clearpage
\newpage

\appendix
\section{Basic identities}\label{app:secA}
The following basic properties will be useful
\begin{align}
    [\aa,\aad] &= 1  \qquad\Rightarrow\qquad \aa\aad = 1 + \aad\aa \\
    \aa\ket{n} &= \sqrt{n}\ket{n-1} \\
    \aad\ket{n} &= \sqrt{n+1}\ket{n+1} \\
    \aad\aa\ket{n} &= n\ket{n} \\
    \aa\ket{\alpha} &= \alpha\ket{\alpha} \\
    \bra{\alpha}\aad &= \bra{\alpha} \alpha^{*} \\
    \bra{\alpha}\aad\aa\ket{\alpha} &= \abs{\alpha}^{2} \\
    \bra{\alpha}(\aad\aa)^2\ket{\alpha} &= \abs{\alpha}^{2} (1 + \abs{\alpha}^{2}) \\
    \langle\beta\ket{\alpha} &= \exp\left[-\dfrac{1}{2}( \abs{\beta}^2 + \abs{\alpha}^2 - 2 \beta^\ast \alpha ) \right]
\end{align}
Moreover, using the fact that quadratures are defined as
\begin{equation}
    \xx =\dfrac{1}{\sqrt{2}}(\hat{a}+\hat{a}^\dagger) \;,\qquad\qquad  \pp =\dfrac{1}{i\sqrt{2}}(\hat{a}-\hat{a}^\dagger) \;,
\end{equation}
we obtain
\begin{align}
\hat{x}^2&=\frac{1}{2}+\hat{a}^{\dagger}\hat{a}+\frac{1}{2}(\hat{a}\hat{a}+\hat{a}^{\dagger}\hat{a}^{\dagger}), \\
\hat{p}^2&=\frac{1}{2}+\hat{a}^{\dagger}\hat{a}-\frac{1}{2}(\hat{a}\hat{a}+\hat{a}^{\dagger}\hat{a}^{\dagger}), \\
\hat{x}\hat{p} + \hat{p}\hat{x} &= i(\hat{a}^{\dagger}\hat{a}^{\dagger}-\hat{a}\hat{a}),
\end{align}
and, thus,
\begin{align}
\avg{\hat{x}^2}&= \frac{1}{2}+\avg{\hat{n}} + \Re[\avg{\hat{a}\hat{a}}]\\
\avg{\hat{p}^2}&= \frac{1}{2}+\avg{\hat{n}} - \Re[\avg{\hat{a}\hat{a}}]\\
\frac{1}{2}\avg{\hat{x}\hat{p} + \hat{p}\hat{x}}&=\Im[\avg{\hat{a}\hat{a}}].
\end{align}

From these relations, we have for Fock states
\begin{align}
    \bra{m}\xx\ket{n} &= \dfrac{1}{\sqrt{2}} \left(  \sqrt{n} \delta_{m,n-1} + \sqrt{n+1} \delta_{m,n+1} \right) \\
    \bra{m}\pp\ket{n} &= \dfrac{1}{i \sqrt{2}} \left(  \sqrt{n} \delta_{m,n-1} - \sqrt{n+1} \delta_{m,n+1} \right) \\
    \bra{m}\xx^{2}\ket{n} &= \dfrac{1}{2} ((1 + 2n) \delta_{m,n} + \sqrt{n+1}\sqrt{n+2} \delta_{m,n+2} + \sqrt{n}\sqrt{n-1} \delta_{m,n-2} ) \\
    \bra{m}\pp^{2}\ket{n} &= \dfrac{1}{2} ((1 + 2n) \delta_{m,n} - \sqrt{n+1}\sqrt{n+2} \delta_{m,n+2} - \sqrt{n}\sqrt{n-1} \delta_{m,n-2})  \\
    \bra{m}\xx\pp + \pp\xx \ket{n} &= i (\sqrt{n+1}\sqrt{n+2} \delta_{m,n+2} - \sqrt{n}\sqrt{n-1} \delta_{m,n-2} )
\end{align}
and for coherent states
\begin{align}
    \bra{\beta}\xx\ket{\alpha} &= \dfrac{1}{\sqrt{2}}\left(\alpha+\beta^\ast\right) e^{-\frac{1}{2}( \abs{\beta}^2 + \abs{\alpha}^2 - 2 \beta^\ast \alpha ) } \\
    \bra{\beta}\pp\ket{\alpha} &= \dfrac{1}{i\sqrt{2}}\left(\alpha-\beta^\ast\right) e^{-\frac{1}{2}( \abs{\beta}^2 + \abs{\alpha}^2 - 2 \beta^\ast \alpha ) } \\
    \bra{\beta}\xx^{2}\ket{\alpha} &= \dfrac{1}{2} \left(\alpha^{2}+1+2\beta^\ast\alpha+(\beta^{*})^{2}\right) e^{-\frac{1}{2}( \abs{\beta}^2 + \abs{\alpha}^2 - 2 \beta^\ast \alpha ) } \\
    \bra{\beta}\pp^{2}\ket{\alpha} &= - \dfrac{1}{2} \left(\alpha^{2}-1-2\beta^\ast\alpha+(\beta^{*})^{2}\right) e^{-\frac{1}{2}( \abs{\beta}^2 + \abs{\alpha}^2 - 2 \beta^\ast \alpha ) } \\
    \bra{\beta}\xx\pp + \pp\xx\ket{\alpha} &= i ((\beta^\ast)^2 - \alpha^2) e^{-\frac{1}{2}( \abs{\beta}^2 + \abs{\alpha}^2 - 2 \beta^\ast \alpha ) }
\end{align}

Pure Gaussian states can be represented as a squeezed and displaced vacuum state, namely
\begin{align}
\ket{\alpha,\xi}=\op{D}(\alpha)\op{S}(\xi)|0\rangle,
\end{align}
where $\op{S}(\xi)=e^{(\xi^{*}\aa^{2}-\xi \op{a}^{\dagger 2} ) /2}$ is the squeezing operator with $\xi=re^{i\gamma}$, $r=|\xi|$, $\op{D}(\alpha)=e^{\alpha \hat{a}^\dagger - \alpha^\ast \hat{a}}$ is the displacement operator, and $|0\rangle$ is the vacuum state.

Using the transformations
\begin{align}
    \op{D}^{\dagger}(\alpha)\hat{a}\op{D}(\alpha) &= \hat{a}+\alpha \\
    \op{D}^{\dagger}(\alpha)\aad \op{D}(\alpha) &= \aad + \alpha^* \\
    \op{S}^{\dagger}(\xi)\hat{a} \op{S}(\xi) &= \mu \hat{a} - \nu \hat{a}^{\dagger} \label{eq:bogoliubovsqueezing}\\
    \op{S}^{\dagger}(\xi)\hat{a}^{\dagger}\op{S}(\xi) &= \mu \hat{a}^{\dagger} - \nu^* \hat{a} \\
    \mu &= \cosh r \\
    \nu &= e^{i\gamma} \sinh r    
\end{align}
it is straightforward to obtain
\begin{align}
    \bra{\xi,\alpha}\aa\ket{\alpha,\xi} &= \alpha \\
    \bra{\xi,\alpha}\aad\ket{\alpha,\xi} &= \alpha^* \\    
    \bra{\xi,\alpha}\aa\aa\ket{\alpha,\xi} &= \alpha^2 - \mu\nu \\
    \bra{\xi,\alpha}\aad\aad\ket{\alpha,\xi} &= (\alpha^*)^2 - \mu\nu^*\\   
    \bra{\xi,\alpha}\aad\aa\ket{\alpha,\xi} &=   \abs{\alpha}^2 + \abs{\nu}^2 \\
    \bra{\xi,\alpha}(\aad\aa)^2\ket{\alpha,\xi} &= \abs{\alpha}^2 ( 1 + \abs{\alpha}^2) + \abs{\nu}^2 (1 + \mu^2 + 4\abs{\alpha}^2 + 2\abs{\nu}^2 ) - (\alpha^*)^2\mu\nu - \alpha^2\mu\nu^*  \;.
\end{align}
From these, in the special case of squeezed vacuum states we have that
\begin{align}
    2 \bra{\xi,0}\op{n}\ket{0,\xi} + 1 &= 2 \sinh^2 r + 1 = \cosh(2r) \\
    \sinh(2r) &= \sqrt{\cosh^2(2r) - 1} = 2\sqrt{\overline{n}(\overline{n}+1)} \;.
\end{align}
Quadratures operators $\qq(\phi)=\xx\sin\phi+\pp\cos\phi=\bqq^T\bm u$ with $\bm u=(\sin\phi,\cos\phi)^T$ transform under displacement and squeezing as
\begin{align}
    \op{D}^{\dagger}(\alpha)\qq(\phi)\op{D}(\alpha) &= \qq(\phi)+\sqrt{2}(\Re[\alpha]\sin\phi+\Im[\alpha]\cos\phi) \\
    \op{S}^{\dagger}(\xi)\qq(\phi) \op{S}(\xi) &= \qq(\phi)\cosh r  + \qq(-\phi-\gamma)\sinh r= \bqq^T\bm{s},\label{eq:squeezingtransformsquadrature} 
\end{align}
with the non-normalized vector $\bm{s}=\mathcal{S}(r,\gamma)\bm{u}$ and
\begin{align}\label{eq:symplecticsqueezing}
    \mathcal{S}(r,\gamma)=\left(
\begin{array}{cc}
 \cosh r - \sinh r \cos \gamma & - \sinh r \sin \gamma \\
 - \sinh r \sin \gamma & \cosh r + \sinh r \cos \gamma   \\
\end{array}
\right)=e^{-r}\mathbf{s}_-\mathbf{s}_-^T+e^{r}\mathbf{s}_+\mathbf{s}_+^T
\end{align}
is the symplectic matrix associated with the squeezing evolution $\hat{S}(\xi)$ with eigenvectors $\mathbf{s}_-=(\cos (\gamma/2),\sin (\gamma/2))^T$ and $\mathbf{s}_+=(-\sin (\gamma/2) ,\cos (\gamma/2))^T$.


\section{First moment of quadrature operators}\label{app:avg}

\begin{table}[h!]
    \centering
    \begin{tabular}{|c|c|c|}
    \hline
    \textbf{Quantum state} \bm{$\RHO$} & \bm{$\avg{\xx}$} & \bm{$\avg{\pp}$}  \\\hline  
    Coherent $\ket{\alpha}$ & $\sqrt{2}\Re[\alpha]$ & $\sqrt{2}\Im[\alpha]$  \\\hline
    Gaussian states & $\sqrt{2}\Re[\alpha]$ & $\sqrt{2}\Im[\alpha]$ \\ \hline
    Fock $\ket{n}$ & $0$ & $0$  \\\hline
    Fock superposition  $(\ket{m}+e^{i\gamma}\ket{n})/\sqrt{2}$, $n>m$  & $\sqrt{\dfrac{n}{2}}\cos[\gamma]\delta_{n,m+1}$ & $\sqrt{\dfrac{n}{2}}\sin[\gamma]\delta_{n,m+1}$ \\ \hline
    Cat $\scN(\ket{\alpha}+e^{i\gamma}\ket{-\alpha})$ & $\dfrac{\sqrt{2}\Im[\alpha]\sin[\gamma]}{e^{2\abs{\alpha}^2}+\cos[\gamma]}$ & $-\dfrac{\sqrt{2}\Re[\alpha]\sin[\gamma]}{e^{2\abs{\alpha}^2}+\cos[\gamma]}$  \\\hline
    Compass $\scN(\ket{\alpha}+\ket{-\alpha}+\ket{i\alpha}+\ket{-i\alpha})$ & $0$ & $0$  \\ \hline
    \end{tabular}
    \caption{\textbf{Expectation values of first moment quadrature operators for the state considered.} These results have been computed using the results given in Appendix~\ref{app:secA}.}
\label{tab:firstmom}
\end{table}

We note that the first moments of cat states can also be expressed as
\begin{align}
    \langle \bqq\rangle_{\text{cat}}=\frac{1}{\sqrt{2}}\left(1-\frac{\avg{\hat{n}}_{\text{cat}}}{|\alpha|^2}\right)\tan[\gamma]\begin{pmatrix}
        \Im[\alpha]\\ -\Re[\alpha]
    \end{pmatrix},
\end{align}
where the average number of photons of the cat state, $\avg{\hat{n}}_{\text{cat}}$, is given in Tab.~\ref{tab:expvalN}.

\section{Covariance matrices}\label{app:cov}
Second moments of quadrature operators can be expressed in terms of the covariance matrix
\begin{equation}\label{eq:suppCM}
    \Gamma[\RHO,\bqq] = \begin{pmatrix}
\var{\xx}{\RHO} & \cov[\xx,\pp]_{\RHO} \\
\cov[\xx,\pp]_{\RHO} & \var{\pp}{\RHO}
\end{pmatrix} \;.
\end{equation}
where the variance is given by $\var{\hat{A}}{\RHO} = \langle \hat{A}^2 \rangle_{\RHO} - \langle \hat{A} \rangle_{\RHO}^2$, and the covariance by $\cov[\hat{A},\hat{B}]_{\RHO} = \frac{1}{2} \langle \hat{A}\hat{B}+\hat{B}\hat{A} \rangle_{\RHO} - \langle \hat{A} \rangle_{\RHO}\langle \hat{B} \rangle_{\RHO}$.

It is often convenient to write Eq.~\eqref{eq:suppCM} in terms of the creation and annihilation operators, which is
\begin{equation}\label{eq:suppCMaadagger}
    \Gamma[\RHO,\bqq] = \begin{pmatrix}
\dfrac{1}{2} + \avg{\aad \aa} + \text{Re}[\avg{\aa\aa}] - \dfrac{1}{2}\avg{\aa+\aad}^2 & \text{Im}[\avg{\aa\aa}] - \dfrac{1}{2}\avg{\aa+\aad}\avg{\aa-\aad} \\
\text{Im}[\avg{\aa\aa}] - \dfrac{1}{2}\avg{\aa+\aad}\avg{\aa-\aad} & \dfrac{1}{2} + \avg{\aad \aa} - \text{Re}[\avg{\aa\aa}] - \dfrac{1}{2}\avg{\aa-\aad}^2
\end{pmatrix} \;,
\end{equation}
where all expectation values are taken on the state $\rho$.

In the following, we provide the covariance matrices for the states considered in the main text. We have for
coherent states
\begin{equation}
    \Gamma[ \ket{\alpha} ,\bqq] = \begin{pmatrix}
\frac{1}{2} & 0\\
0 & \frac{1}{2}
\end{pmatrix} \;,
\end{equation}
arbitrary Gaussian states,
\begin{equation}\label{eq:covgauss}
    \Gamma[ \RHO(\alpha,\xi,n_T) ,\bqq] = \dfrac{2 n_T + 1}{2} \begin{pmatrix}
\cosh(2r)-\sinh(2r)\cos(\gamma) & -\sinh(2r)\sin(\gamma) \\
-\sinh(2r)\sin(\gamma) & \cosh(2r)+\sinh(2r)\cos(\gamma)
\end{pmatrix} \;,
\end{equation}
Fock states,
\begin{equation}
    \Gamma[ \ket{n},\bqq ] = \begin{pmatrix}
\frac{1}{2}+n & 0\\
0 & \frac{1}{2}+n
\end{pmatrix} \;.
\end{equation}

A superposition of two Fock states $(\ket{m}+e^{i\gamma}\ket{n})/\sqrt{2}$ has covariance matrix, for $n=m+1$,
\begin{equation}
    \Gamma[\frac{1}{\sqrt{2}}(\ket{m}+e^{i\gamma}\ket{m+1}),\bqq] = \dfrac{m+1}{4} \begin{pmatrix}
3-\cos(2\gamma) & -\sin(2\gamma)\\
-\sin(2\gamma) & 3+\cos(2\gamma)
\end{pmatrix} \;,
\end{equation}
for $n=m+2$,
\begin{equation}
    \Gamma[\frac{1}{\sqrt{2}}(\ket{m}+e^{i\gamma}\ket{m+2}),\bqq] = \dfrac{1}{2} \begin{pmatrix}
-1+2(m+2)+\sqrt{(m+2)(m+1)}\cos(\gamma) & \sqrt{(m+2)(m+1)}\sin(\gamma)\\
\sqrt{(m+2)(m+1)}\sin(\gamma) & -1+2(m+2)-\sqrt{(m+2)(m+1)}\cos(\gamma)
\end{pmatrix} \;,
\end{equation}
and for $n=m+k$, with $k>2$,
\begin{equation}
    \Gamma[\frac{1}{\sqrt{2}}(\ket{m}+e^{i\gamma}\ket{m+k}),\bqq] = \dfrac{1}{2}\begin{pmatrix}
1+2m+k & 0\\
0 & 1+2m+k
\end{pmatrix} \;.
\end{equation}

\vspace{5mm}

Cat states of the form $\ket{\Psi_{\text{cat},\gamma}}=\mathcal{N}_1 (\ket{\alpha}+e^{i \gamma} \ket{-\alpha})$ have the covariance matrix
\begin{equation} \label{eqsupp:covCat}
    \Gamma[\ket{\Psi_{\text{cat},\gamma}},\bqq] = \begin{pmatrix}
         \frac{1}{2} + \overline{n} + \Re[\alpha^2] - \frac{2 \Im[\alpha]^2 \sin
   ^2(\gamma )}{\left(e^{2 |\alpha|^2}+\cos (\gamma )\right)^2} &  \Im[\alpha^2]
   \left( 1 + \frac{\sin ^2(\gamma )}{\left(e^{2 |\alpha|^2}+\cos (\gamma )\right)^2} \right) \\
 \Im[\alpha^2] \left( 1 + \frac{\sin ^2(\gamma )}{\left(e^{2 |\alpha|^2}+\cos (\gamma )\right)^2} \right) & \frac{1}{2} + \overline{n} - \Re[\alpha^2] - \frac{2 \Re[\alpha]^2 \sin ^2(\gamma
   )}{\left(e^{2 |\alpha|^2}+\cos (\gamma )\right)^2} \\
    \end{pmatrix} \;,
\end{equation}
where $\overline{n}$ can be found in Tab.~\ref{tab:expvalN} and note $\Re[\alpha^2]=\Re[\alpha]^2-\Im[\alpha]^2$ and $\Im[\alpha^2]=2\Re[\alpha]\Im[\alpha]$. Eq.~\eqref{eqsupp:covCat} can be made diagonal by a rotation of the phase-space $xp$ coordinates. In a frame where $\alpha$ is purely imaginary (\ie $\Re[\alpha]=0$, $\Im[\alpha]^2=\abs{\alpha}^2$), such that the cat's interference fringes are vertical, we obtain 
\begin{equation}
\Gamma[\ket{\Psi_{\text{cat},\gamma}},\bqq] =  \begin{pmatrix}
         \frac{1}{2} + \overline{n} - \abs{\alpha}^2 - \frac{2 \abs{\alpha}^2 \sin
   ^2(\gamma )}{\left(e^{2 |\alpha|^2}+\cos (\gamma )\right)^2} &  0 \\
 0 & \frac{1}{2} + \overline{n} + \abs{\alpha}^2  \\
    \end{pmatrix} =
    \begin{pmatrix}
         \frac{1}{2}-\frac{2 |\alpha|^2 \left(\cos (\gamma )+e^{-2 |\alpha|^2}\right)}{e^{2 |\alpha|^2} \left( 1 + e^{-2 |\alpha|^2} \cos (\gamma ) \right)^2} & 0 \\
 0 & \frac{1}{2} + \frac{2 |\alpha|^2}{1 + e^{-2 |\alpha|^2} \cos (\gamma ) }\\
    \end{pmatrix} \;. \label{eq:suppCovCatDiag}
\end{equation}

Specific choices of the superposition phase $\gamma$ are known with specific names in the literature, such as even  ($\gamma=0$) and odd cat states ($\gamma=\pi$) \cite{gerry_quantum_1997}, or Yurke-Stoler cat states ($\gamma=\pi/2$) \cite{YurkeStolerCatPRL}. For these choices of $\gamma$, we have the covariance matrices
\begin{align}
    \text{even cat:}\qquad & \Gamma[\ket{\Psi_{\text{cat},0}},\bqq] =  \begin{pmatrix}
\frac{1}{2} + |\alpha|^2(\tanh(|\alpha|^2)-1) & 0\\
0 & \frac{1}{2} + |\alpha|^2(\tanh(|\alpha|^2)+1)
\end{pmatrix} \;,\\
    \text{odd cat:}\qquad & \Gamma[\ket{\Psi_{\text{cat},\pi}},\bqq] =  \begin{pmatrix}
\frac{1}{2} + |\alpha|^2(\coth(|\alpha|^2)-1) & 0\\
0 & \frac{1}{2} + |\alpha|^2(\coth(|\alpha|^2)+1)
\end{pmatrix} \;,\\
    \text{Yurke-Stoler cat:}\qquad &      \Gamma[\ket{\Psi_{\text{cat},\frac{\pi}{2}}},\bqq] =  \begin{pmatrix}
\frac{1}{2} - 2 |\alpha|^2 e^{-4 |\alpha|^2} & 0\\
0 & \frac{1}{2} + 2 |\alpha|^2 
\end{pmatrix} \;.
\end{align}

\begin{figure}[h]
    \centering
    \includegraphics[width=5.6cm]{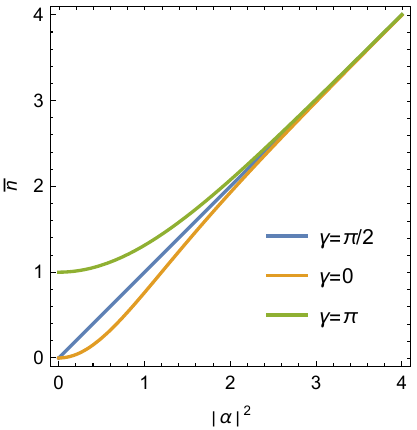}
    \caption{\textbf{Average of the number operator for cat states with different phases.} Value of $\overline{n}$ as given in Tab.~\ref{tab:expvalN}, for different values of $\gamma$ and $|\alpha|^2$.}
    \label{fig:suppNcat}
\end{figure}

So far, our results are exact, as they only involve a rotation of the coordinate system compared to Eq.~\eqref{eqsupp:covCat}. However, we would like to express the above covariance matrices as a function of $\overline{n}$, to compute sensitivities that can be compared between different states for the same amount of resources. To this end, let us note that in the limit of large $|\alpha|^2$ we have $\overline{n}\simeq|\alpha|^2$ independently of $\gamma$, see Fig.~\ref{fig:suppNcat}. This is because $\tanh(|\alpha|^2) \simeq \coth(|\alpha|^2) \simeq 1$ for $|\alpha|^2 \gg 1$. It can be easily demonstrated, \eg by studying first and second derivatives with respect to $\gamma$, that $|\alpha|^2\tanh(|\alpha|^2)\leq \avg{\hat{n}}_{cat}\leq |\alpha|^2\coth(|\alpha|^2)$ and the minimum and maximum are reached for $\gamma=0$ and $\gamma=\pi$, respectively.

Therefore, for values of $|\alpha|^2 \gtrsim 2$ we can approximate $\overline{n} \simeq |\alpha|^2$ and $|\alpha|^2/e^{2|\alpha|^2}\simeq0$, which allows us to approximate the covariance matrix of any ``large'' cat state (\ie $\gamma$ arbitrary) by the much simpler expression (c.f. Eq.~\eqref{eq:suppCovCatDiag})
\begin{equation}
    \text{``Large'' cat, $|\alpha|^2\gg 1$:} \qquad \Gamma[\ket{\Psi_{\text{cat},\gamma}},\bqq] \simeq  \begin{pmatrix}
\frac{1}{2} & 0\\
0 & \frac{1}{2} + 2 \overline{n} 
\end{pmatrix} \;.
\end{equation}
For the calculations presented in the main text we use this expression, as it gives much simpler results without requiring significant approximations.

\vspace{5mm}
For compass states $|\Psi_\text{comp}\rangle=\NN_2 \left( |\alpha\rangle+|-\alpha\rangle+|i\alpha\rangle+|-i\alpha\rangle \right)$, since $\avg{\op{x}}=\avg{\op{p}}=\avg{\aa \aa}=0$, we see from Eq.~\eqref{eq:suppCMaadagger} that the covariance matrix reads
\begin{equation}
    \Gamma[ \ket{\Psi_\text{comp}},\bqq ] = \begin{pmatrix}
         \frac{1}{2} + \overline{n}  & 0 \\
 0 & \frac{1}{2} + \overline{n} \\
    \end{pmatrix} \;,
\end{equation}
where $\overline{n}$ can be found in Tab.~\ref{tab:expvalN}. Note that, in the limit $|\alpha|^2\gg 1$ we have $\overline{n}\simeq |\alpha|^2$.

\section{First and second moments of the number operator}

\begin{table}[h!]
    \centering
    \begin{tabular}{|c|c|c|}
    \hline
    \textbf{Quantum state} \bm{$\RHO$} & \bm{$\avg{\hat{n}}$} & \bm{$\avg{\hat{n}^2}$}  \\\hline  
    Coherent $\ket{\alpha}$ & $\abs{\alpha}^2$ & $\abs{\alpha}^2(1+\abs{\alpha}^2)$  \\\hline
    Gaussian state & $(\mu^2+|\nu|^2) n_T + |\alpha|^2 + |\nu|^2$ & \makecell{$\left( 2|\nu|^4 + 8\mu^2|\nu|^2 + 2\mu^4 \right)\nT^2$ \\
    $\quad + \left( 4|\alpha|^2|\nu|^2 + 3|\nu|^4 - 2\alpha^2\mu\nu^* - 2(\alpha^*)^2\mu\nu + 4|\alpha|^2\mu^2 + 8\mu^2|\nu|^2 + \mu^4 \right)\nT$\\
    $\quad + \left( |\alpha|^4 + 3|\alpha|^2|\nu|^2 + |\nu|^4 - \alpha^2\mu\nu^* - (\alpha^*)^2\mu\nu + |\alpha|^2\mu^2 + 2\mu^2|\nu|^2 \right)$} \\ \hline
    Fock $\ket{n}$ & $n$ & $n^2$  \\\hline
    \makecell{Fock superposition\\ $(\ket{m}+e^{i\gamma}\ket{n})/\sqrt{2}$}  & $\dfrac{m+n}{2}$ & $\dfrac{m^2+n^2}{2}$  \\ \hline
    Cat $\scN(\ket{\alpha}+e^{i\gamma}\ket{-\alpha})$ & $\abs{\alpha}^2  \dfrac{1 - \cos[\gamma]e^{-2\abs{\alpha}^2} }{1 + \cos[\gamma]e^{-2\abs{\alpha}^2} }  $ & $\abs{\alpha}^{2} \dfrac{1+\abs{\alpha}^{2}+\cos[\gamma] e^{-2\abs{\alpha}^{2}}(-1+\abs{\alpha}^{2})}{1+\cos[\gamma] e^{-2\abs{\alpha}^{2}} } $ \\\hline
    \makecell{Compass\\ $\scN(\ket{\alpha}+\ket{-\alpha}+\ket{i\alpha}+\ket{-i\alpha})$} & $\abs{\alpha}^2 \frac{1 - e^{-2\abs{\alpha}^2} - 2e^{-\abs{\alpha}^2}\sin(\abs{\alpha}^2)}{1+e^{-2\abs{\alpha}^2}+2e^{-\abs{\alpha}^2}\cos(\abs{\alpha}^2)}$ & $\abs{\alpha}^2 \frac{1 - e^{-2\abs{\alpha}^2} - 2e^{-\abs{\alpha}^2}\sin(\abs{\alpha}^2)}{1+e^{-2\abs{\alpha}^2}+2e^{-\abs{\alpha}^2}\cos(\abs{\alpha}^2)} + \abs{\alpha}^4 \frac{1 + e^{-2\abs{\alpha}^2} - 2e^{-\abs{\alpha}^2}\cos(\abs{\alpha}^2)}{1+e^{-2\abs{\alpha}^2}+2e^{-\abs{\alpha}^2}\cos(\abs{\alpha}^2)}$  \\ \hline
    \end{tabular}
    \caption{\textbf{Expectation values of first and second moments of the number operator for the state considered.} These results have been computed using the results given in Appendices~\ref{app:secA}, \ref{app:cov}.}
\label{tab:expvalN}
\end{table}

\clearpage
\newpage
\section{Wigner functions}

If the wavefunction $\psi(x)$ of the state is known, the Wigner function can be easily calculated from
\begin{equation}
    W(x,p) = \dfrac{1}{2\pi} \int_{-\infty}^\infty dy\, e^{i p y} \psi^\ast(x+y/2) \psi(x-y/2) \;. 
\end{equation}

\vspace{5mm}
For a coherent state $\ket{\beta}$, with $\beta = (x_0 + i p_0)/\sqrt{2}$, the Wigner function is given by~\cite{gerry_introductory_2004}
\begin{equation}
    W_{\beta}(x,p) = \frac{1}{\pi}e^{-((x-x_0)^2+ (p-p_0)^2)} \;.
\end{equation}

\vspace{5mm}
For a Gaussian state, defining $\vect{X} = (x,p)^T$ and $\avg{\op{\vect{X}}} = (\avg{\op{x}},\avg{\op{p}})^T$, the Wigner function reads~\cite{Paris2005}
\begin{equation}
    W_G(x,p) = \frac{e^{-\frac{1}{2}(\vect{X}-\langle\op{\vect{X}}\rangle)^T\Gamma^{-1}(\vect{X}-\langle\op{\vect{X}}\rangle)}}{2\pi\sqrt{\det[\Gamma]}} \;,
\end{equation}
where $\Gamma$ is the covariance matrix defined in Eq.~\eqref{eq:covgauss}.

\vspace{5mm}
The Wigner function of a Fock state $\ket{n}$ is given by~\cite{gerry_introductory_2004}:
\begin{equation}
    W_{n}(x,p) = \frac{1}{\pi}(-1)^ne^{-(x^2+p^2)}L_{n}[2(x^2+p^2)]
\end{equation}
where $L_n[z]$ is the $n$-th degree Laguerre polynomial.

\vspace{5mm}
Let us consider now the superposition of two Fock states $(\ket{m}+\ket{n})/\sqrt{2}$ with $m\neq n$, whose density matrix is
\begin{equation}
    \begin{split}
        \rho_{mn} &= \frac{1}{2}(\ket{m}\bra{m} + \ket{m}\bra{n} + \ket{n}\bra{m} + \ket{n}\bra{n})
    \end{split} \;.
\end{equation}

To find the corresponding Wigner function, we rely on known results for the Wigner function of $\ket{m}\bra{n}$. Using the formulation from~\cite{leonhardt_measuring_1997}, after adjusting the normalization according to our definitions, we obtain
\begin{equation}
    W_{mn}(x,p) = \begin{cases}
    \frac{1}{\pi}(-1)^m\left(\frac{n!}{m!}\right)^{1/2}e^{-(x^2+p^2)}\left(-\sqrt{2}(x-ip)\right)^{m-n}L_n^{m-n}\left[2(x^2+p^2)\right]\quad &\text{if } m\geq n\\
    \frac{1}{\pi}(-1)^n\left(\frac{m!}{n!}\right)^{1/2}e^{-(x^2+p^2)}\left(-\sqrt{2}(x+ip)\right)^{n-m}L_m^{n-m}\left[2(x^2+p^2)\right] &\text{if } m < n
    \end{cases}.
\end{equation}
Notice that for $n=m$ we recover $W_n(x,p)$, as expected.

For $m=0$, we get the Wigner function for the superposition $(\ket{0}+\ket{n})/\sqrt{2}$ is
\begin{equation}
    \begin{split}
        W_{0n}(x,p) &=\frac{e^{-(x^2+p^2)}}{2 \pi } \left(\frac{2^{n/2}}{\sqrt{n!}} \left((x+i p)^n+(x-i p)^n\right)+(-1)^n L_n\left[2 \left(p^2+x^2\right)\right]+1\right) \;.
    \end{split}
\end{equation}

\vspace{5mm}
The Wigner function of a cat state $\ket{\Psi}=\NN[\ket{\beta}+e^{i\gamma}\ket{-\beta}]$ with $\beta=\beta_r+i\beta_i$ is~\cite{gerry_quantum_1997,BuzekSupCS}
\begin{equation}
    W_{\beta}(x,p) = \dfrac{\left(e^{-([x-\sqrt{2}\beta)^2+(p-\sqrt{2}\beta_i)^2]}+e^{-[(x+\sqrt{2}\beta_r)^2+(p+\sqrt{2}\beta_i)^2]}+2e^{-(x^2+p^2)}\cos[2\sqrt{2}(x \beta_r - p \beta_i) - \gamma]\right)}{2\pi\left(1+ e^{-2\abs{\beta}^2} \cos\gamma \right)} \;.
\end{equation}

\vspace{5mm}
The Wigner function for a compass state can also be computed analytically, but it is extremely lengthy. For this reason, we will not write it here but rather refer the reader to Ref.~\cite{SandersCompass}.

\clearpage
\newpage

\section{Gaussian quantum Fisher information}\label{app:gaussrotqfi}
\subsection{General expression}
For mixed Gaussian states in a general Gaussian channel, the QFI reads
\begin{align}\label{eq:qfiG1}
    F_Q[\RHO(\theta)]=\frac{1}{2}\frac{\Tr\left[(\Gam^{-1}[\RHO(\theta),\bqq]\frac{\partial}{\partial\theta}\Gam[\RHO(\theta),\bqq])^2\right]}
    {1+\mathcal{P}(\RHO(\theta))^2}+\frac{2\left(\frac{\partial\mathcal{P}(\RHO(\theta))}{\partial\theta}\right)^2}{1-\mathcal{P}(\RHO(\theta))^4}+\left(\frac{\partial\langle\bqq\rangle_{\RHO(\theta)}}{\partial\theta}\right)^T\Gam^{-1}[\RHO(\theta),\bqq]\left(\frac{\partial\langle\bqq\rangle_{\RHO(\theta)}}{\partial\theta}\right),
\end{align}
with the Gaussian purity $\mathcal{P}(\RHO)=(2\sqrt{\det\Gam[\RHO,\bqq]})^{-1}$. This expression can be further simplified using the following properties of $2\times 2$ matrices:
\begin{align}\label{eq:2by2rules}
    A ^{-1}&=\frac {1}{\det  {A} }\left[(\Tr{A} ){I} - {A} \right],\notag\\
    (\Tr {A})  {I} - {A} &=\Omega A^T\Omega^T,\notag\\
    \Tr{\left[A^2\right]}&=(\Tr{A})^2-2\det{A},\notag\\
    \det A&=\frac{1}{2}\Tr{\left[A\Omega A^T \Omega^T\right]},
\end{align}
where the last equality reduces to $\Tr{\left[(\Omega H)^2\right]}=-2\det H$ for $H^T=H$. With these identities, we obtain
\begin{align}
    \Gam^{-1}[\RHO,\bqq]&=\frac{\Omega\Gam[\RHO,\bqq]\Omega^T}{\det\Gam[\RHO,\bqq]},
\end{align}
and
\begin{align}\label{eq:f3}
    \Tr\left[(\Gam^{-1}[\RHO(\theta),\bqq]\frac{\partial}{\partial\theta}\Gam[\RHO(\theta),\bqq])^2\right]=\Tr\left[\Gam^{-1}[\RHO(\theta),\bqq]\frac{\partial}{\partial\theta}\Gam[\RHO(\theta),\bqq]\right]^2-2\frac{\det \frac{\partial \Gam[\RHO(\theta),\bqq]}{\partial \theta}}{\det\Gam[\RHO(\theta),\bqq]}.
\end{align}
Moreover, using 
\begin{align}
    \frac{\partial }{\partial \theta}\det \Gam[\RHO(\theta),\bqq]=\det\Gam[\RHO(\theta),\bqq]\Tr\left[\Gam^{-1}[\RHO(\theta),\bqq]\frac{\partial}{\partial\theta}\Gam[\RHO(\theta),\bqq]\right],
\end{align}
the first and second term in Eq.~(\ref{eq:qfiG1}) simplify to
\begin{align}
    \frac{1}{2}\frac{\Tr\left[(\Gam^{-1}[\RHO(\theta),\bqq]\frac{\partial}{\partial\theta}\Gam[\RHO(\theta),\bqq])^2\right]}
    {1+\mathcal{P}(\RHO(\theta))^2}=2\frac{\frac{1}{\det \Gam[\RHO(\theta),\bqq]}\left(\frac{\partial }{\partial \theta}\det \Gam[\RHO(\theta),\bqq]\right)^2-2\det \frac{\partial \Gam[\RHO(\theta),\bqq]}{\partial \theta}}{1+4\det\Gam[\RHO(\theta),\bqq]},
\end{align}
and
\begin{align}
    \frac{2\left(\frac{\partial\mathcal{P}(\RHO(\theta))}{\partial\theta}\right)^2}{1-\mathcal{P}(\RHO(\theta))^4}=2\frac{\frac{1}{\det \Gam[\RHO(\theta),\bqq]}\left(\frac{\partial }{\partial \theta}\det \Gam[\RHO(\theta),\bqq]\right)^2}{16 \det \Gam[\RHO(\theta),\bqq]^2-1},
\end{align}
respectively, leading to the Gaussian QFI for single-mode Gaussian channels
\begin{align}\label{eq:QFIgaussapp}
    F_Q[\RHO(\theta)]=\frac{8\left(\frac{\partial  }{\partial \theta}\det\Gam[\RHO(\theta),\bqq]\right)^2}{16 \det \Gam[\RHO(\theta),\bqq]^2-1}-\frac{4\det \frac{\partial \Gam[\RHO(\theta),\bqq]}{\partial \theta}}{4 \det \Gam[\RHO(\theta),\bqq]+1}+\left(\frac{\partial\langle\bqq\rangle_{\RHO(\theta)}}{\partial\theta}\right)^T\frac{\Omega\Gam[\RHO(\theta),\bqq]\Omega^T}{\det\Gam[\RHO(\theta),\bqq]}\left(\frac{\partial\langle\bqq\rangle_{\RHO(\theta)}}{\partial\theta}\right).
\end{align}
For a general Gaussian channel~(\ref{eq:channel}), the above expression is determined by
\begin{align}\label{eq:appevol}
    \frac{\partial\langle\bqq\rangle_{\RHO(\theta)}}{\partial\theta}&=\frac{\partial\mathcal{X}(\theta)}{\partial\theta}\langle\hat{\bm{r}}\rangle_{\RHO}+\frac{\partial\bm{d}(\theta)}{\partial\theta}\notag\\
     \Gam[\RHO(\theta),\bqq]&=\mathcal{X}(\theta)\Gam[\RHO,\bqq]\mathcal{X}(\theta)^T+\mathcal{Y}(\theta)\notag\\
    \frac{\partial}{\partial\theta}\Gam[\RHO(\theta),\bqq]&=\frac{\partial\mathcal{X}(\theta)}{\partial\theta}\Gam[\RHO,\bqq]\mathcal{X}(\theta)^T+\mathcal{X}(\theta)\Gam[\RHO,\bqq]\frac{\partial\mathcal{X}(\theta)}{\partial\theta}^T+\frac{\partial\mathcal{Y}(\theta)}{\partial\theta}.
\end{align}
These terms can be further simplified using Eq.~(\ref{eq:2by2rules}).

\subsection{Applications}
\subsubsection{Unitary symplectic evolution}
A special case of the Gaussian channel is the unitary Gaussian evolution that is described by a symplectic matrix $\mathcal{X}(\theta)=\mathcal{S}(\theta)=\exp(\Omega H \theta)$ and $\mathcal{Y}(\theta)=0$, with no displacement, $\bm{d}(\theta)=0$; see Eq.~(\ref{eq:symplectic}). In this case, Eqs.~(\ref{eq:appevol}) read,
\begin{align}
    \frac{\partial}{\partial \theta}\langle\hat{\bm{r}}\rangle_{\RHO(\theta)}&=\Omega H\mathcal{S}(\theta)\langle\hat{\bm{r}}\rangle_{\RHO},\notag\\
    \Gam[\RHO(\theta),\bqq]&=\mathcal{S}(\theta)\Gam[\RHO,\bqq]\mathcal{S}(\theta)^T,\notag\\
    \frac{\partial}{\partial \theta}\Gam[\RHO(\theta),\bqq]&=\Omega H\Gam[\RHO(\theta),\bqq]+\Gam[\RHO(\theta),\bqq](\Omega H)^T.
\end{align}
In particular, $\frac{\partial  }{\partial \theta}\det\Gam[\RHO(\theta),\bqq]=0$. Furthermore, we use Eqs.~(\ref{eq:f3}) together with $\Tr\left[\Gam^{-1}[\RHO(\theta),\bqq]\frac{\partial}{\partial\theta}\Gam[\RHO(\theta),\bqq]\right]=0$ and the properties~(\ref{eq:2by2rules}) to show that, in the unitary single-mode case,
\begin{align}\label{eq:unitaryderdgamma}
    \det\frac{\partial\Gam[\RHO(\theta),\bqq]}{\partial \theta}=4\det \Gam[\RHO,\bqq]\det H-\Tr{\left[\Gam[\RHO,\bqq] H\right]}^2.
\end{align}
With this, the QFI~(\ref{eq:QFIgaussapp}) becomes independent of $\theta$ and reduces to 
\begin{align}\label{eq:qfiGaussSympApp}
    F_Q[\RHO,\frac{1}{2}\hat{\bm{r}}^TH\hat{\bm{r}}]
    &=\frac{4\Tr\left[\Gam[\RHO,\bqq] H\right]^2-16\det\Gam[\RHO,\bqq]\det{H}}
    {4\det\Gam[\RHO,\bqq]+1}+\frac{\langle\hat{\bm{r}}\rangle_{\RHO}^TH\Gam[\RHO,\bqq] H\langle\hat{\bm{r}}\rangle_{\RHO}}{\det\Gam[\RHO,\bqq]}.
\end{align}

\subsubsection{Symplectic evolution with thermal noise}\label{sec:symthermalnoisearbitrary}
To account for the effect of diffusive Gaussian noise, caused by weak coupling to a thermal bath, we consider the channel~(\ref{eq:channel}), described by the operators
\begin{align}\label{eq:evolappnrs}
    \mathcal{X}(\theta,t)&=e^{-\kappa t/2}\mathcal{S}(\theta),\notag\\
    \mathcal{Y}(t)&=(1-e^{-\kappa t})\left(n_b+\frac{1}{2}\right)\mathbb{I},\notag\\
    \bm{d}(\theta)&=0,
\end{align}
where $\kappa$ describes the coupling to the bath that has an average number of $n_b$ excitations. In this model, we can interpret the parameter $\theta=\omega t$ as the product of a frequency factor in front of the Hamiltonian and the evolution time $t$, which is consistent with our previous descriptions of unitary evolutions. Equations~(\ref{eq:evolappnrs}) describe the evolution of the system in time $t$, whereas for quantum metrology, we are only interested in the evolution with respect to the parameter $\theta$. To make this distinction explicit, we denote the state at any point in the evolution by $\RHO(\theta,t)$.

Now, Eqs.~(\ref{eq:appevol}) read
\begin{align}
    \frac{\partial}{\partial \theta}\langle\hat{\bm{r}}\rangle_{\RHO(\theta,t)}&=e^{-\kappa t/2}\Omega H\mathcal{S}(\theta)\langle\hat{\bm{r}}\rangle_{\RHO},\notag\\
    \Gam[\RHO(\theta,t),\bqq]&=e^{-\kappa t}\underbrace{\mathcal{S}(\theta)\Gam[\RHO,\bqq]\mathcal{S}(\theta)^T}_{\Gam[\RHO(\theta,0),\bqq]}+(1-e^{-\kappa t})\left(n_b+\frac{1}{2}\right)\mathbb{I},\notag\\
    \frac{\partial}{\partial \theta}\Gam[\RHO(\theta,t),\bqq]&=e^{-\kappa t}\frac{\partial}{\partial \theta}\Gam[\RHO(\theta,0),\bqq],
\end{align}
where $\RHO(\theta,0)$ describes the evolved state in the absence of noise. Using the identity for $2\times 2$ matrices $\det(A+B)=\det A+\det B+\Tr A\Tr B-\Tr\left[AB\right]$, we obtain
\begin{align}\label{eq:detGapp}
    \det\Gam[\RHO(\theta,t),\bqq]&=e^{-2\kappa t}\det \Gam[\RHO,\bqq]+(1-e^{-\kappa t})^2\left(n_b+\frac{1}{2}\right)^2+e^{-\kappa t}(1-e^{-\kappa t})\left(n_b+\frac{1}{2}\right)\Tr{\left[\Gam[\RHO(\theta,0),\bqq]\right]},\\
    \frac{\partial}{\partial \theta}\det\Gam[\RHO(\theta,t),\bqq]&=e^{-\kappa t}(1-e^{-\kappa t})\left(n_b+\frac{1}{2}\right)\Tr{\left[\Omega(H\Gam[\RHO(\theta,0),\bqq]-\Gam[\RHO(\theta,0),\bqq]H)\right]},\label{eq:ddetGapp}
\end{align}
and
\begin{align}
    \det\frac{\partial}{\partial \theta}\Gam[\RHO(\theta,t),\bqq]&=e^{-2\kappa t}\det\frac{\partial}{\partial \theta}\Gam[\RHO(\theta,0),\bqq]\notag\\
    &=e^{-2\kappa t}\left(4\det \Gam[\RHO,\bqq]\det H-\Tr{\left[\Gam[\RHO,\bqq] H\right]}^2\right),
\end{align}
where we used Eq.~(\ref{eq:unitaryderdgamma}) for unitary evolutions. Inserting these expressions in Eq.~(\ref{eq:QFIgaussapp}) yields the QFI for Gaussian evolutions under the impact of thermal noise.

\subsubsection{Displacement with thermal noise}\label{sec:thermalnoisearbitrary}
Alternatively, we may consider the effect of thermal noise on a displacement, described by
\begin{align}\label{eq:evolappnrsdisp}
    \mathcal{X}(t)&=e^{-\kappa t/2}\mathbb{I},\notag\\
    \mathcal{Y}(t)&=(1-e^{-\kappa t})\left(n_b+\frac{1}{2}\right)\mathbb{I},\notag\\
    \bm{d}(\theta)&=\theta\Omega\bm{u},
\end{align}
As above, the parameter $\theta$ is proportional to the displacement amplitude with grows with time $t$, which, however, carries no information about $\theta$.

In this case, Eqs.~(\ref{eq:appevol}) read
\begin{align}
    \frac{\partial}{\partial \theta}\langle\hat{\bm{r}}\rangle_{\RHO(\theta,t)}&=\Omega\bm{u},\notag\\
    \Gam[\RHO(\theta,t),\bqq]&=e^{-\kappa t}\Gam[\RHO,\bqq]+(1-e^{-\kappa t})\left(n_b+\frac{1}{2}\right)\mathbb{I},\notag\\
    \frac{\partial}{\partial \theta}\Gam[\RHO(\theta,t),\bqq]&=0.
\end{align}
We obtain
\begin{align}\label{eq:detGappdisp}
    \det\Gam[\RHO(\theta,t),\bqq]&=e^{-2\kappa t}\det \Gam[\RHO,\bqq]+(1-e^{-\kappa t})^2\left(n_b+\frac{1}{2}\right)^2+e^{-\kappa t}(1-e^{-\kappa t})\left(n_b+\frac{1}{2}\right)\Tr{\left[\Gam[\RHO,\bqq]\right]},\\
    \frac{\partial}{\partial \theta}\det\Gam[\RHO(\theta,t),\bqq]&=0,\label{eq:ddetGappdisp}
\end{align}
and
\begin{align}
    \det\frac{\partial}{\partial \theta}\Gam[\RHO(\theta,t),\bqq]&=0.
\end{align}
From Eq.~(\ref{eq:QFIgaussapp}) we obtain
\begin{align}\label{eq:noisydispQFI}
    F_Q[\RHO(\theta,t)]=\frac{\bm{u}^T\Gam[\RHO(\theta,t),\bqq]\bm{u}}{\det\Gam[\RHO(\theta,t),\bqq]}.
\end{align}

\subsubsection{Rotation sensing: unitary evolution}
Of particular interest is the sensitivity under phase shifts generated by $\nn=\hat{a}^{\dagger}\hat{a}=\frac{1}{2}(\hat{x}^2+\hat{p}^2-1)$. Up to an irrelevant constant, this corresponds to the quadratic Hamiltonian $\frac{1}{2}\hat{\bm{r}}^TH\hat{\bm{r}}$ with Hamiltonian matrix $H=\mathbb{I}$. We obtain
\begin{align}\label{eq:qfiunitaryrotation}
    F_Q[\RHO,\nn]
    &=\frac{4\Tr\left[\Gam[\RHO,\bqq]\right]^2-16\det\Gam[\RHO,\bqq]}
    {4\det\Gam[\RHO,\bqq]+1}+\frac{\langle\hat{\bm{r}}\rangle_{\RHO}^T\Gam[\RHO,\bqq] \langle\hat{\bm{r}}\rangle_{\RHO}}{\det\Gam[\RHO,\bqq]}.
\end{align}

For a Gaussian state with covariance matrix~(\ref{eq:covgauss}) and first moments
\begin{align}\label{eq:fmapgst}
    \langle\hat{\bm{r}}\rangle_{\RHO}=\sqrt{2}|\alpha|\begin{pmatrix}
        \cos\varphi\\
        \sin\varphi
    \end{pmatrix},
\end{align}
we obtain
\begin{align}
  F_Q[\RHO,\nn] &= \frac{4 |\alpha|^2}{2 n_T+1} (\cosh (2 r)-\sinh (2r) \cos (2 \varphi-\gamma))+\frac{2 (2 n_T+1)^2 \sinh
   ^2(2 r)}{2n_T^2+2n_T+1}\notag\\
    &= \frac{4 |\alpha|^2}{2 n_T+1} \left|\mu-e^{-2i\varphi}\nu \right|^2+\frac{8 (2 n_T+1)^2 |\nu|^2\mu^2}{2n_T^2+2n_T+1}.
\end{align}

\subsubsection{Rotation sensing: Thermal noise}\label{app:thermalrotsens}
If the phase shift discussed above is affected by thermal noise, described by~(\ref{eq:evolappnrs}), the results of Sec.~\ref{sec:symthermalnoisearbitrary} apply. Here, $H=\mathbb{I}$, and thus Eqs.~(\ref{eq:detGapp}) and~(\ref{eq:ddetGapp}) yield
\begin{align}
\det\Gam[\RHO(\theta,t),\bqq]&=e^{-2\kappa t}\det \Gam[\RHO,\bqq]+(1-e^{-\kappa t})^2\left(n_b+\frac{1}{2}\right)^2+e^{-\kappa t}(1-e^{-\kappa t})\left(n_b+\frac{1}{2}\right)\Tr{\left[\Gam[\RHO,\bqq]\right]},\\
    \frac{\partial}{\partial \theta}\det\Gam[\RHO(\theta,t),\bqq]&=0,
\end{align}
respectively. The second result reflects the invariance of the state's purity under changes of the parameter $\theta$ that only appears in the Hamiltonian part of the evolution. Hence, using $\det\Gam[\RHO(\theta,t),\bqq]=\det\Gam[\RHO(0,t),\bqq]$, the resulting QFI is again independent of $\theta$ and reads
\begin{align}
    F_Q[\RHO(0,t)]=e^{-2\kappa t}\frac{4\Tr\left[\Gam[\RHO,\bqq]\right]^2-16\det \Gam[\RHO,\bqq]}{4 \det \Gam[\RHO(0,t),\bqq]+1}+e^{-2\kappa t}\frac{\langle\hat{\bm{r}}\rangle_{\RHO}^T\Gam[\RHO,\bqq] \langle\hat{\bm{r}}\rangle_{\RHO}}{\det\Gam[\RHO(0,t),\bqq]}+e^{-\kappa t}(1-e^{-\kappa t})\left(n_b+\frac{1}{2}\right)\frac{\left|\langle\hat{\bm{r}}\rangle_{\RHO}\right|^2}{\det\Gam[\RHO(0,t),\bqq]}.
\end{align}
For $\kappa=0$ we recover the expression~(\ref{eq:qfiunitaryrotation}) of the unitary case.

Considering again a Gaussian state with covariance matrix~(\ref{eq:covgauss}) and first moments~(\ref{eq:fmapgst}), we obtain 
\begin{align}\label{eq:noisyphaseQFI}
    F_Q[\RHO(0,t)]=\frac{4 \alpha ^2 \left[ (2 n_b+1) \left(e^{\kappa  t}-1\right)+(2 n_T+1) (\cosh (2 r)-\sinh (2 r) \cos (2
   \varphi-\gamma))\right]}{2C-e^{2\kappa t}}+\frac{2 (2 n_T+1)^2 \sinh ^2(2 r)}{C},
\end{align}
where
\begin{align}
    C=(2 n_b+1) (2n_T+1) \cosh (2 r) \left(e^{\kappa  t}-1\right)+2n_b (n_b+1) \left(e^{\kappa  t}-1\right)^2+2
   n_T (n_T+1)-e^{\kappa  t}+e^{2 \kappa  t}+1.
\end{align}

\section{Displacement sensing with cat states}\label{app:catdisp}
Squeezed vacuum states were shown to be optimal for the task of displacement sensing with a known displacement direction, as they reach the state-independent upper limit, Eq.~(\ref{eq:QFImaxpureupper}). Cat states, however, are non-optimal and here we explain how far their sensitivity remains from this ultimate limit.

First, from the covariance matrix~(\ref{eqsupp:covCat}) for cat states with an arbitrary phase $\gamma$, we obtain the maximal displacement sensitivity, Eq.~(\ref{eq:QFImaxpure}),
\begin{align}\label{eq:fqmaxcat}
    F_Q^{\max}[\PSI_{\mathrm{cat},\gamma}]=2(1+2\bar{n}+2|\alpha|^2),
\end{align}
where the average number of photons of the cat state $\bar{n}=\avg{\hat{n}}$ is given in Tab.~\ref{tab:expvalN}. It can be easily demonstrated that the sensitivity~(\ref{eq:fqmaxcat}) is always lower than the limit~(\ref{eq:QFImaxpureupper}), $F_Q^{\max}=2(1+2\bar{n}+2\sqrt{\bar{n}(\bar{n}+1)})$. Independently of $\alpha$, the difference is minimal for $\gamma=0$ and maximal for $\gamma=\pi$, see Fig.~\ref{fig:suppNcat}. In the asymptotic limit $|\alpha|\to\infty$, the difference converges to the constant value $2$ for all $\gamma$.

\section{Measurement-after-interaction with squeezing operations in displacement sensing}\label{app:optmai}
We consider the sensitivity for moment-based estimation of a displacement amplitude from a noisy quadrature measurement of $\MM=\qq(\varepsilon)+\Delta M$. As is explained in Sec.~\ref{sec:mai}, the generator $\qq(\phi)$ and measurement quadratures $\qq(\varepsilon)$ are chosen orthogonal and coincide with the quadratures of maximal and minimal variance of the probe state $\RHO$. Without MAI, the sensitivity is given by
\begin{align}
    \chi^{-2}[\RHO,\qq(\phi),\MM]&=\frac{|\langle [\qq(\phi),\qq(\varepsilon)] \rangle_{\RHO}|^2}{\var{\qq(\varepsilon)}{\RHO}+\sigma^2 }=\frac{(\bm u^T\Omega \bm w)^2}{\bm w^T\Gam[\RHO,\bqq]\bm w+\sigma^2}=\frac{1}{\lambda_{\min}(\Gam[\RHO,\bqq])+\sigma^2}.
\end{align}
Using the MAI protocol, we realize the transformation $\MM\to\MM_{\mathrm{MAI}}$ with
\begin{align}
    \MM_{\mathrm{MAI}}=\hat{S}^{\dagger}(\xi)\qq(\varepsilon)\hat{S}(\xi)+\Delta M=\bqq^T\mathcal{S}(r,\gamma)\bm w+\Delta M,
\end{align}
where $\mathcal{S}(r,\gamma)$ is the symplectic matrix~(\ref{eq:symplecticsqueezing}) that describes squeezing. The sensitivity changes to
\begin{align}
    \chi^{-2}_{\mathrm{MAI}}[\RHO,\qq(\phi),\MM]&=\chi^{-2}[\RHO,\qq(\phi),\MM_{\mathrm{MAI}}]=\frac{(\bm u^T\Omega \mathcal{S}(r,\gamma)\bm w)^2}{\bm w^T \mathcal{S}^T(r,\gamma)\Gam[\RHO,\bqq]\mathcal{S}(r,\gamma)\bm w+\sigma^2}.
\end{align}
The $2\times 2$ matrix $\mathcal{S}(r,\gamma)$ has eigenvalues $e^{-r}$ and $e^{r}$ with corresponding eigenvectors $\mathbf{s}_-=(\cos(\gamma/2),\sin(\gamma/2))^T$ and $\mathbf{s}_+=(-\sin(\gamma/2),\cos(\gamma/2))^T$ that are determined by the angle $\gamma$. We now use the fact that $\bm w$ and $\bm u$ span a basis to express the eigenvectors of $\mathcal{S}(r,\gamma)$ as $\mathbf{s}_-=\bm w\sin \chi +\bm u\cos\chi $ and $\mathbf{s}_+=\bm w\cos \chi - \bm u\sin\chi$. This leads to
\begin{align}
    \mathcal{S}(r,\gamma)\bm w
    =\underbrace{(\cosh(r) + \cos(2 \chi) \sinh(r))}_{s_w}\bm w\underbrace{-\sin(2 \chi) \sinh(r)}_{s_u}\bm u,
\end{align}
and the sensitivity
\begin{align}\label{eq:overallsensitivityMAI}
    \chi^{-2}_{\mathrm{MAI}}[\RHO,\qq(\phi),\MM]&=\frac{1}{\lambda_{\min}(\Gam[\RHO,\bqq])+\frac{s_u^2}{s_w^2}\lambda_{\max}(\Gam[\RHO,\bqq])+\frac{1}{s_w^2}\sigma^2}.
\end{align}
The choice $\chi=0$ maximizes $s_w$ and sets $s_u$ to zero, and, thus, maximizes the overall sensitivity~(\ref{eq:overallsensitivityMAI}). In this case, $\bm w$ coincides with the anti-squeezing direction $\bm{s}_+$, which we find by comparison with Eq.~(\ref{eq:quadob}) to amount to the squeezing angle $\gamma=-2\varepsilon$.

\section{Rotation sensing with pure Gaussian states}\label{app:rotgauss}

The wavefunction of a pure Gaussian state can be written in position representation as
\begin{equation}
    \psi(x) = (2 \pi \sigma_x^2)^{-1/4} \exp \left[ -\frac{\left(x-\sqrt{2} \alpha_r \right)^2}{4 \sigma_x^2} + i \sqrt{2} \alpha_i \left(x-\sqrt{2} \alpha_r \right) + i \frac{\text{cov}_{xp} \left(x-\sqrt{2} \alpha_r \right)^2}{2 \sigma_x^2} \right] \;,
\end{equation}
where $\alpha_r=\avg{\op{x}}/\sqrt{2}$, $\alpha_i=\avg{\op{p}}/\sqrt{2}$, $\sigma_x^2 = \var{\hat{x}}{}$ is the variance of $x$ and $\text{cov}_{xp}=\text{Cov}[\op{x},\op{p}]$ is the covariance.

Using this expression, we obtain for the two terms that appear in the expression Eq.~\eqref{eq:rotmm2}:
\begin{align}
    C^T[\RHO,\nn,\bqq]\Gam^{-1}[\RHO,\bqq]C[\RHO,\nn,\bqq] &= 8 \alpha_r^2 \sigma_x^2+\frac{2 \alpha_i^2 \left(4 \text{cov}_{xp}^2+1\right)}{\sigma_x^2}+16 \alpha_i \alpha_r \text{cov}_{xp} \\
    D^T[\RHO,\nn,\bqq,\bqq^{(2)}]\Sigma^{-1}[\RHO,\bqq,\bqq^{(2)}]D[\RHO,\nn,\bqq,\bqq^{(2)}] &= \frac{\left(4 \text{cov}_{xp}^2+1\right)^2}{8 \sigma_x^4}+4 \text{cov}_{xp}^2+2 \sigma_x^4-1.
\end{align}
Importantly, the sum of these two terms coincides with the QFI $F_Q[\PSI,\nn] = 4 \var{\nn}{\PSI}$, see Eq.~\eqref{eq:FQpureRot}.
Note here that the first term vanishes for a state centered at the origin, meaning that a moment-based estimation of the rotation angle using the first moments of quadratures yields zero sensitivity, while measurements of second moments are necessary and sufficient to saturate the QFI.

An optimal measurement observable for the method of moments can be identified, up to irrelevant normalization, using Eq.~(\ref{eq:maxc}) as
\begin{align}
    \MM_{\mathrm{opt}}&= -\frac{\left(4 \text{cov}_{xp}^2+1\right) \left(-4 \alpha_r \sigma_x^4+\alpha_r+4 \alpha_r \text{cov}_{xp}^2-8 \alpha_i \text{cov}_{xp} \sigma_x^2\right)}{\sqrt{2} \sigma_x^4}\hat{x}-2 \sqrt{2} \left(-4 \alpha_i \sigma_x^4+\alpha_i+4 \alpha_i \text{cov}_{xp}^2+8 \alpha_r \text{cov}_{xp} \sigma_x^2\right)\hat{p}\notag\\
 & \quad + \left( \frac{\left(4 \text{cov}_{xp}^2+1\right)^2}{4 \sigma_x^4}+4 \text{cov}_{xp}^2-1\right)\hat{x}^2-\left(4 \text{cov}_{xp}^2+4
   \sigma_x^4-1\right)\hat{p}^2.
\end{align}
For non-displaced pure Gaussian states (\ie squeezed vacuum states), only the second-order terms contribute and we obtain
\begin{align}
    \MM_{\mathrm{opt}}=  \left( \frac{\left(4 \text{cov}_{xp}^2+1\right)^2}{4 \sigma_x^4}+4 \text{cov}_{xp}^2-1\right)\hat{x}^2-\left(4 \text{cov}_{xp}^2+4
   \sigma_x^4-1\right)\hat{p}^2.
\end{align}

\end{document}